\documentclass[twocolumn,english,pra,aps,superscriptaddress]{revtex4-2}
\usepackage[T1]{fontenc}
\usepackage{color}
\usepackage{babel}
\usepackage{mathtools}
\usepackage{bm}
\usepackage{graphicx}
\usepackage{esint}
\usepackage{amsfonts}
\usepackage{graphicx}
\usepackage{float}
\usepackage{subfigure}
\usepackage{amsthm}
\usepackage{amssymb}
\usepackage{braket}
\usepackage{physics}
\usepackage[pdftex,colorlinks=true, linkcolor = blue, citecolor=blue,urlcolor=blue, 
bookmarksnumbered=true, bookmarksopen=true]{hyperref}
\usepackage{breakurl}

\usepackage{orcidlink}

\def\noon{$\mathrm{\bar{n}00\bar{n}}$ }

\begin{document}

\title{Optimal finite-dimensional probe states for quantum phase estimation}

\author{Jin-Feng Qin\orcidlink{0000-0001-8794-9793}}
\affiliation{School of Physics, Huazhong University of Science and Technology, Wuhan 430074, China}
\affiliation{Center for Theoretical Physics and School of Physics and Optoelectronic Engineering, Hainan University, Haikou 570228, China}
\affiliation{School of Electronic Engineering, Chengdu Technological University, Chengdu 611730, China}

\author{Yuqian Xu\orcidlink{0000-0002-1327-4540}}
\affiliation{School of Physics, Huazhong University of Science and Technology, Wuhan 430074, China}

\author{Jing Liu\orcidlink{0000-0001-9944-4493}}
\email{jing.liu@hainanu.edu.cn}
\affiliation{Center for Theoretical Physics and School of Physics and Optoelectronic Engineering, Hainan University, Haikou 570228, China}
\affiliation{School of Physics, Huazhong University of Science and Technology, Wuhan 430074, China}

\begin{abstract}
Phase estimation is a major mission in quantum metrology, especially in quantum interferometry. A full phase estimation scheme 
usually includes the optimal probe state and measurement. For the finite-dimensional states in Fock basis, the N00N state 
ceases to be optimal when the average particle number is fixed yet not equal to the Fock dimension (Fock number of the highest 
occupied Fock state of one mode), and what is the true optimal finite-dimensional probe state in this case is still undiscovered. 
Hereby we present several theorems to answer this question and provide a complete optimal scheme to realize the ultimate precision 
limit in practice. These optimal finite-dimensional probe states reveal an important fact that the Fock dimension could be treated as 
a metrological resource, and the given scheme is particularly useful in scenarios where weak light or limited particle number is 
demanded. 
\end{abstract}

\maketitle

\section{Introduction}

As a fundamental scenario, phase estimation is undoubtedly a core topic in precision measurement. Many measurement scenarios, 
such as ranging, can be naturally translated or modeled into the problem of phase estimation. In quantum mechanics, optical 
quantum phase estimation is the first scenario revealing the power of quantum resources to beat the standard quantum limit, 
thanks to the pioneer works of Caves~\cite{Caves1980,Caves1981}. After decades of studies, quantum phase estimation has now 
become one of the most fertile fields in quantum metrology~\cite{Yurke1986,Holland1993,Giovannetti2004,Giovannetti2006,Monras2007,
Pezze2007,Pezze2008,Boixo2008,Pezze2009,Paris2009,Genoni2011,Peter2011,Spagnolo2012,Genoni2013,Humphreys2013,Lang2013,
Toth2014,Zhang2015,Pezze2017,Degen2017,Gagatsos2017,Liu2020,Rafal2020,Pezze2021,Qiu2022,Birrittella2012,Berry2012,Demkowicz2012,
Rivas2012,Demkowicz2015,Calsamiglia2016,Branford2021}, and many useful schemes have already been experimentally 
realized~\cite{Xiao1987,Grangier1987,Mitchell2004,Higgins2007,Nagata2007,Kacprowicz2010,Berni2015,Xu2020,Liu2021,Liu2023}. 

In quantum phase estimation, especially optical phase estimation, both linear and nonlinear phase shifts can be used to
encode the phase. In theory, the linear phase accumulation on a bosonic mode $a$ can be described by the operator 
$\exp(i\phi_a a^{\dag}a)$ with $\phi_a$ the accumulated phase. For two modes ($a$ and $b$) with such processes, the 
total phase accumulations can also be written as $\exp(i\phi_{\mathrm{tot}}n/2)\exp(i\phi J_z)$ with 
$\phi_{\mathrm{tot}}=\phi_a+\phi_b$ the total phase and $\phi=\phi_a-\phi_b$ the phase difference. $n=a^{\dag}a+b^{\dag}b$ 
is the operator for the average total particle number and $J_z=(a^{\dag}a-b^{\dag}b)/2$ is a Schwinger operator. Similarly, 
the nonlinear phase accumulation on mode $a$ can be described by $\exp(i\phi_a (a^{\dag}a)^2)$ and for two bosonic modes 
it becomes $\exp(i\phi_{\mathrm{tot}}[(a^{\dag}a)^2+(b^{\dag}b)^2]/2)\exp(i\phi n J_z)$. If both phases $\phi_a$ and $\phi_b$ 
are needed to be estimated, an external reference beam is then required~\cite{Jarzyna2012,Pasquale2015} so that the specific 
values of the phases can be measured. In the meantime, if the value of one phase needs to be measured, the value of the other 
phase has to be known. In this case the precision of the unknown phase would be limited by the precision of the known phase, 
and the quantum advantage on the precision may be fully canceled because of it. Therefore, in this paper we focus on the estimation 
of the phase difference $\phi$, which requires no external reference or the absolute value of any phase between $\phi_a$ and $\phi_b$. 
Both linear and nonlinear phase shifts will be studied. 

Quantum Cram\'{e}r-Rao bound is a well-used tool to depict the ultimate precision limit of the phase difference, in which 
the variance of $\phi$, denoted by $\delta^2\phi$, satisfies~\cite{Helstrom1976,Holevo1982} 
\begin{equation}
\delta^2\phi\geq \frac{1}{\mu I}\geq \frac{1}{\mu F}.
\end{equation}
Here $\mu$ is the number of repetitions, $I$ is the classical Fisher information (CFI), and $F$ is the quantum 
Fisher information (QFI). For a pure state $\ket{\psi}$, the QFI with respect to $\phi$ can be calculated via~\cite{Helstrom1976,Holevo1982} 
\begin{equation}
F=4(\bra{\partial_{\phi}\psi}\ket{\partial_{\phi}\psi}-|\bra{\psi}\ket{\partial_{\phi}\psi}|^2).
\end{equation}
Furthermore, for a set of positive operator valued measure $\{\Pi_i\}$ the CFI reads 
$\sum_i (\partial_{\phi}P_i)^2/P_i$ with $P_i=\bra{\psi}\Pi_i\ket{\psi}$ the conditional probability with respect to 
the $i$th result. 

Finite-dimensional states is an important category of quantum states, especially in quantum metrology and quantum parameter 
estimation. Generally speaking, any state with a finite dimension can be referred to as the finite-dimensional state, which is very 
common in quantum mechanics for finite-dimensional Hamiltonians. Here in this paper, the finite-dimensional state is referred to 
the finite-dimensional state in Fock basis. This category of states includes some famous quantum states, such as the N00N 
state~\cite{Sanders1989,Boto2000} and twin-Fock state~\cite{Holland1993}, and have drawn attentions in many 
aspects~\cite{Vogel1993,Park2017,Boas2019,Lee2019}. 

For the sake of designing an optimal scheme for quantum phase estimation, the optimal probe state is the first step that needs to 
be explored~\cite{Dorner2009,Dobrzanski2009,Jarzyna2015,Liu2013}. Regarding the single-mode finite-dimensional states,  in 2012 
Berry \emph{et al.}~\cite{Berry2012} found that with the vacuum-Fock superposition probe state, the precision limit of the phase can 
be enhanced by increasing the state dimension. Further in 2019, Lee \emph{et al.}~\cite{Lee2019} discussed the optimal 
finite-dimensional state for the estimation the phase with a fixed photon number, and found that it could overcome the performance 
of the squeezed vacuum state with the same average photon number. Regarding the quantum phase estimation in two-mode interferometry, 
notice that a general pure finite-dimensional state in this case can be written as $\sum^N_{i,j=0}c_{ij}\ket{ij}$ with $\ket{ij}$ an element of 
the two-mode Fock basis and $c_{ij}$ the corresponding coefficient. Here $N$ is the Fock number of the highest occupied Fock state of 
each mode. In the following $N$ will be referred to as the Fock dimension of the state (or Fock dimension in short).  When the average 
total particle number $\bar{n}:=\expval{n}\in[0,2N]$ ($\expval{\cdot}$ represents the expectation) is unlimited, the optimal 
finite-dimensional probe state (OFPS) for both linear and nonlinear phase shifts is just the N00N state 
$(\ket{0N}+e^{i\theta}\ket{N0})/\sqrt{2}$ with $\theta\in[0,2\pi)$ the relative phase. However, for a fixed average particle 
number satisfying $\bar{n}\neq N$, the \noon state $(\ket{0\bar{n}}+e^{i\theta}\ket{\bar{n}0})/\sqrt{2}$ may not remain optimal 
anymore, and what is the true OFPS in this case is still an open question. The answer to this question is particularly valuable for 
the scenarios requiring limited particle number of the probe, such as the biological detections where weak light is required to avoid 
damaging the specimen~\cite{Taylor2013}, and cost-effective environments like the satellites~\cite{Lu2022a} or chips~\cite{Stokowski2023}. 
Therefore, locating the OFPS with a fixed average particle number for the phase estimation in quantum interferometry and 
providing a complete estimation scheme accordingly are the major motivations of this paper. 

\section{Optimal finite-dimensional probe states}

For the sake of answering the aforementioned question, several theorems are first given to present the OFPSs for both linear 
and nonlinear phase shifts. 

\emph{Theorem 1.}  Consider the linear phase shifts, a fixed average particle number $\bar{n}$, and a fixed Fock dimension 
$N$. The OFPS is
\begin{equation}
\sqrt{1-\frac{\bar{n}}{N}}\ket{00}+\sqrt{\frac{\bar{n}}{2N}}
\left(e^{i\theta_1}\ket{0N}+e^{i\theta_2}\ket{N0}\right)
\label{eq:lin_optstate1}
\end{equation}
when $\bar{n}\in(0,N]$, and 
\begin{equation}
\sqrt{1-\frac{\bar{n}}{2N}}\left(e^{i\theta_1}\ket{0N}
+e^{i\theta_2}\ket{N0}\right)+\sqrt{\frac{\bar{n}}{N}-1}\ket{NN}
\label{eq:lin_optstate2}
\end{equation}
when $\bar{n}\in[N,2N)$. Here $\theta_1,\theta_2\in [0,2\pi)$ are the relative phases. 

The thorough proof of this theorem is given in Appendix~\ref{sec:theorem1}. In the linear case, the QFIs for the states in 
Eqs.~(\ref{eq:lin_optstate1}) and (\ref{eq:lin_optstate2}) are $\bar{n}N$ and $N(2N-\bar{n})$, respectively.  The optimal state 
is just the N00N state in the case that $\bar{n}=N$. The OFPS in Eq.~(\ref{eq:lin_optstate1}) with $\theta_1=\theta_2=0$ has 
also been discussed in Ref.~\cite{Branford2021}, and in Ref.~\cite{Lu2022} as the optimization of the path-symmetric entangled 
states~\cite{Lee2016}.  For the case of nonlinear phase shifts, the form of the OFPS replies on relations between the values 
of $\bar{n}$ and $N$. Hence, we have the following theorems.

\emph{Theorem 2.}  Consider the nonlinear phase shifts, a fixed average particle number $\bar{n}$, and a fixed Fock dimension 
$N$. When  $\bar{n}\in(0,N]$ the OFPS is also in the form of Eq.~(\ref{eq:lin_optstate1}).

\emph{Theorem 3.}   Consider the nonlinear phase shifts, a fixed average particle number $\bar{n}$, and a fixed Fock dimension 
$N$. In the case that $\bar{n}\in\left[N, \left\lfloor\frac{4N+1}{3}\right\rfloor\right]$ the OFPS reads 
\begin{align}
& \sqrt{\frac{\bar{n}-\lfloor \bar{n}\rfloor}{2}}\!\!\left(\ket{\lfloor \bar{n}\rfloor\!+\!1\!-\!N, N}
\!+\!e^{i\theta_1}\ket{N, \lfloor\bar{n}\rfloor\!+\!1\!-\!N}\right) \nonumber \\
& \!+\!\sqrt{\frac{1\!-\!(\bar{n}\!-\!\lfloor\bar{n}\rfloor)}{2}}\!\!\left(e^{i\theta_2} 
\ket{\lfloor\bar{n}\rfloor\!\!-\!\!N, N}\!+\!e^{i\theta_3}\ket{N,\lfloor\bar{n}\rfloor\!\!-\!\!N}\right).  
\label{eq:OFPS1}
\end{align}
Here $\theta_1,\theta_2,\theta_3\in[0,\ 2\pi)$ are the relative phases, $\lfloor\cdot\rfloor$ is the floor function. 

\emph{Corollary 1}
 If $\bar{n}$ is an integer, then in the regime $\bar{n}\in\left[N, \left\lfloor\frac{4N+1}{3}\right\rfloor\right]$ 
 the OFPS reads 
\begin{equation}
\frac{1}{\sqrt{2}}\left(\ket{\bar{n}-N, N} +e^{i\theta}\ket{N, \bar{n}-N}\right).
\label{eq:non_optstate2}
\end{equation}
Here $\theta$ is the relative phase.  

\emph{Theorem 4.}   Consider the nonlinear phase shifts, a fixed average particle number $\bar{n}$, and a fixed Fock 
dimension $N$. In the regime $\bar{n}\in\left[\left\lfloor\frac{4N+1}{3}\right\rfloor,2N\right)$, the OFPS is 
\begin{equation}
\sqrt{\frac{2N\!-\!\bar{n}}{2\!\left(N\!-\!\zeta\right)}}\!\left(e^{i\theta_1}\!\ket{\zeta N}\!+\!
e^{i\theta_2}\!\ket{N \zeta}\right)\!+\!\sqrt{\frac{\bar{n}\!-\!N\!-\!\zeta}{N\!-\!\zeta}}\!\ket{NN},    
\label{eq:OFPS2}
\end{equation}
where $\zeta:=\left\lfloor\frac{N+1}{3}\right\rfloor\!$. Here $\theta_1,\theta_2\in [0,2\pi)$ are 
the relative phases. 

\emph{Corollary 2}
If $N/3$ is an integer, then the regime in Theorem 4 becomes $\bar{n}\in[4N/3,2N)$ and in this regime the OFPS is 
\begin{equation}
\sqrt{\frac{3(2N\!-\!\bar{n})}{4N}}\!\!\left(\!\!e^{i\theta_1}\!\!\ket{\frac{N}{3},N\!\!}
\!+\!e^{i\theta_2}\!\!\ket{N,\frac{N}{3}\!}\!\!\right)\!+\!\sqrt{\frac{3\bar{n}\!-\!4N}{2N}}\!\!\ket{N\!N}.
\label{eq:non_optstate3}
\end{equation}
Here $\theta_1,\theta_2\in [0,2\pi)$ are the relative phases. 

The thorough proofs of Theorems 2 to 4 and corresponding corollaries are given in Appendix~\ref{sec:theorem2}. 
In the nonlinear case, the QFIs for the states in Eqs.~(\ref{eq:OFPS1}) and (\ref{eq:OFPS2}) are given in  
Appendix~\ref{sec:theorem2}, and those for the states in Eqs.~(\ref{eq:lin_optstate1}), (\ref{eq:non_optstate2}), 
and (\ref{eq:non_optstate3}) are $\bar{n}N^3$, $\bar{n}^2 (2N-\bar{n})^2$, and $32N^3(2N-\bar{n})/27$, respectively. 
Similar to the linear case, here the optimal state is just the N00N state in the case that $\bar{n}=N$. 

In a standard Mach-Zehnder interferometer, a 50:50 beam splitter [usually characterized by $\exp(-i\pi J_x/2)$ 
with $J_x=(a^{\dag}b+ab^{\dag})/2$] exists in front of the phase shifts, and the aforementioned OFPSs need to be 
rotated by $\exp(i\pi J_x/2)$ to cancel the influence of the first beam splitter. The expressions of the OFPSs after 
this rotation can be found in Appendix~\ref{sec:MZstate}. 

To make sure that the measurement of phase difference $\phi$ does not need the information of the nuisance 
parameter, i.e., the phase summation $\phi_{\mathrm{tot}}$, we have calculated the quantum Fisher information matrix 
for $\phi$ and $\phi_{\mathrm{tot}}$ with both linear and nonlinear OFPSs. Details of the calculations can be found in 
Appendix~\ref{sec:QFIM}. The result that all the QFIMs of the OFPSs are diagonal means that the measurement of $\phi$ 
indeed does not require the information of $\phi_{\mathrm{tot}}$. This fact can also be confirmed by the optimal measurements 
discussed in Sec.~\ref{sec:opt_measure}, where no information of $\phi_{\mathrm{tot}}$ is used during the entire measurement 
process. 

\begin{figure}[tp]
\centering\includegraphics[width=8.7cm]{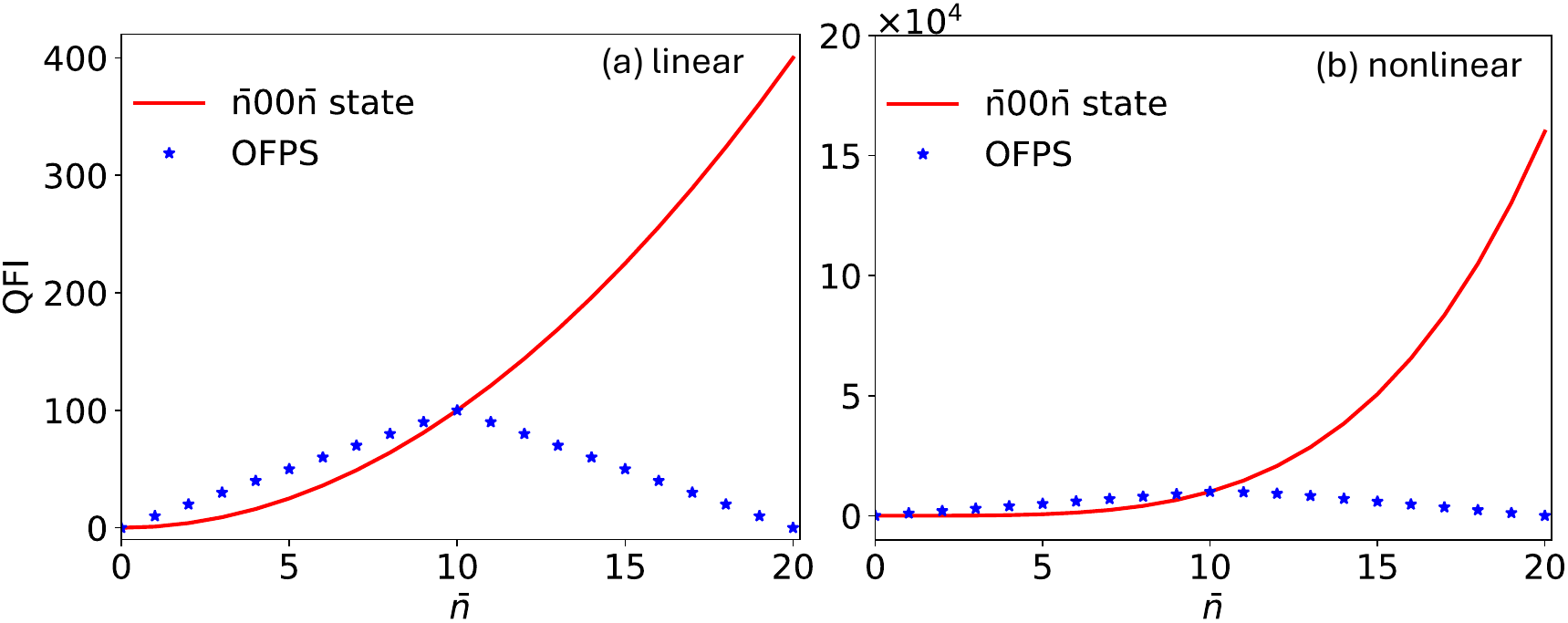}
\caption{Comparison of the QFI between the \noon state (red line) and OFPS (blue stars) for 
(a) linear phase shifts and (b) nonlinear phase shifts. The Fock dimension of the OFPSs is 
$N=10$ in the plots.}
\label{fig:OFPSvsN00N}
\end{figure}

These OFPSs reveal an intriguing fact that the dimension of the OFPS could be a metrological resource in quantum 
interferometry that is different from the particle number, similar to the time and quantum correlations like entanglement. 
This coincides with the Branford and Rubio’s general argument that the average particle number is an insufficient 
metric for interferometry~\cite{Branford2021}. The N00N state $\left[(\ket{N0}+e^{i\theta}\ket{0N})/\sqrt{2}\,\right]$ 
cannot reveal this fact since the average particle number simultaneously increases with the increase of $N$, and thus the 
contribution of Fock dimension and particle number cannot be distinguished. The average particle numbers of the OFPSs given 
in the theorems are fixed and the metrological gain obtained via enlarging $N$ can thus be fully attributed to the growth of the 
Fock dimension. In the meantime, the quantification of entanglement requires dimension independence due to a general belief 
that the same state with different dimensions should have the same amount of entanglement~\cite{Horodecki2009,Eltschka2014}, 
which means the obtained metrological gain can also not be attributed to the entanglement, at least in the current definition. 

To further present the effects of the average particle number and the Fock dimension, the OFPS with a fixed Fock dimension 
($N=10$) and \noon state have been compared for different values of $\bar{n}$, as shown in Fig.~\ref{fig:OFPSvsN00N}. For 
both linear [Fig.~\ref{fig:OFPSvsN00N}(a)] and nonlinear [Fig.~\ref{fig:OFPSvsN00N}(b)] phase shifts, it can be seen that when 
$\bar{n}< N$, the QFI of the OFPS is larger than that of the \noon state, indicating that the theoretical performance of the OFPS 
is better than the \noon state in this regime. In the case that $\bar{n}=N$, the OFPS is nothing but the \noon state, hence the 
values of the QFI are equivalent. These behaviors coincide with the results of the aforementioned theorems. Furthermore, in the 
regime $\bar{n}>N$, the QFI of the \noon state is larger than that of the OFPS. It is important to note that in this case, the Fock 
dimension of the \noon state exceeds that of the OFPS. While this falls outside the scope of the theorems, it also clearly 
demonstrates that a larger Fock dimension could enhance the precision limit. If the same amount of Fock dimension is applied to 
the OFPS,  it just reduces to the \noon state,  which means their performance would then be equivalent. 

The OFPS vividly shows that in the scenarios where limited particle number is required, the precision can still be further improved 
by increasing the Fock dimension without changing the average particle number. Hence, it  would be very useful and promising in 
the scenarios like biological detections. 

\begin{figure*}[tp]
\centering\includegraphics[width=16cm]{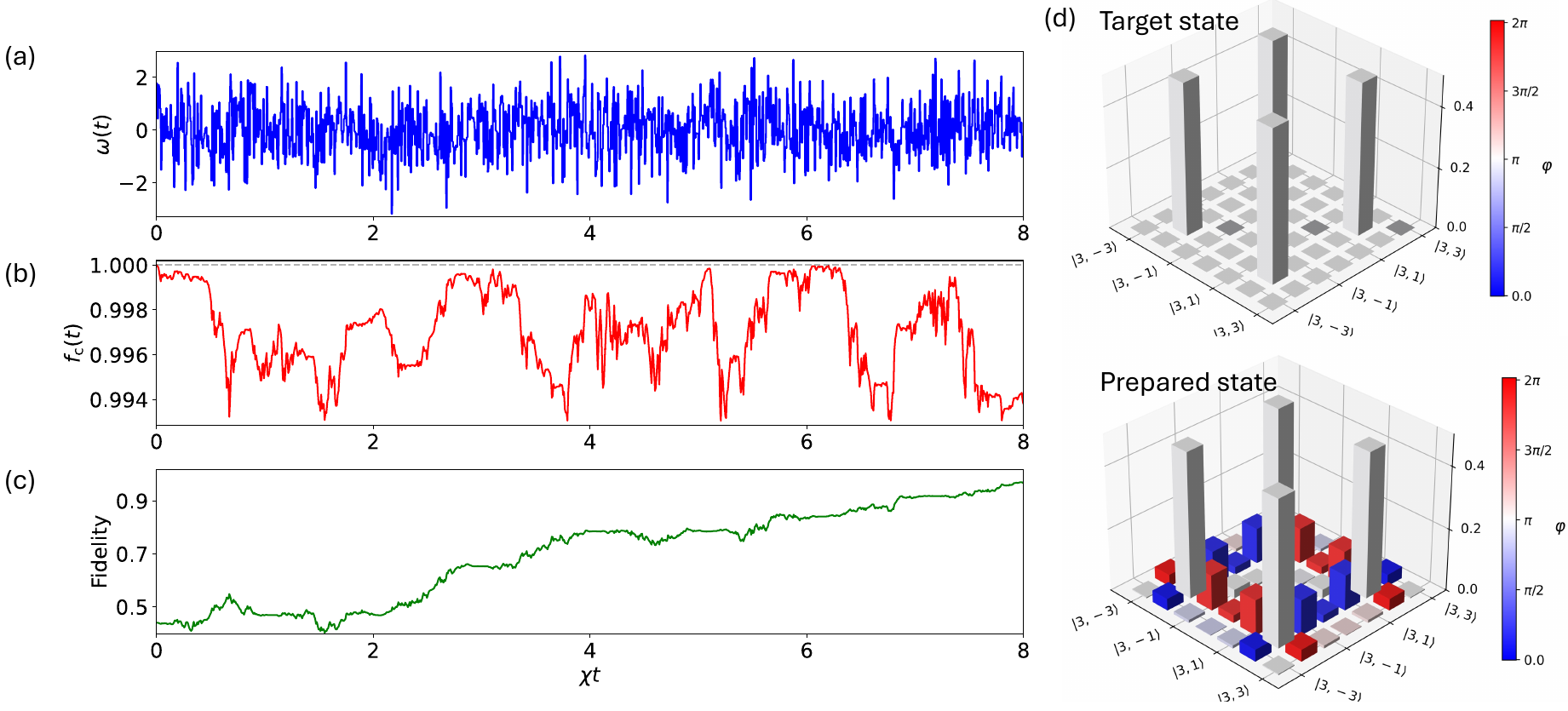}
\caption{Preparation of the nonlinear OFPS given in Eq.~(\ref{eq:non_optstate2}) with the one-axis twisting model. 
(a) Optimal control amplitude for the preparation process. (b) The evolved constraint value during the dynamics. 
(c) The evolved fidelity during the dynamics. (d) The tomography of the target and prepared states in the basis 
$\{\ket{j,m}\}$ . In the plots $\bar{n}=6$ and $N=5$. }
\label{fig:OFPSgene}
\end{figure*}

The specific preparation process of the OFPS is still an open-question and requires further investigations in the future. A possible 
direction in the linear case is the probabilistic superposition between the N00N state and $\ket{00}$ or $\ket{NN}$ state. In the nonlinear 
case, the preparation of the OFPS in Eq.~(\ref{eq:non_optstate2}) can be simulated with the one-axis twisting model with a transverse 
control field~\cite{Ma2011}. The Hamiltonian for this system is $\omega(t)J_x+\chi J_z^2$, where $J_{x}$ and $J_{z}$ are the angular 
momentum operators. $\omega(t)$ is the control and $\chi$ is a constant. Here we take $\chi=1$. In the eigenspace ($\{\ket{j,m}\}$) 
of $J_z$, the OFPS in Eq.~(\ref{eq:non_optstate2}) can be expressed by 
\begin{equation}
\frac{1}{\sqrt{2}}\left(\ket{\frac{\bar{n}}{2}, \frac{\bar{n}}{2}-N}+e^{i\theta}\ket{\frac{\bar{n}}{2}, N-\frac{\bar{n}}{2}}\right).
\label{eq:oat_target}
\end{equation}
Here $j$ is the total angular momentum and $m\in[-j,j]$ is the eigenvalue of $J_z$. In the following we take the state with $\theta=0$ 
as the target state for the preparation. To properly simulate the preparation of this OFPS, $m$ should be further constrained in the regime 
$[\frac{\bar{n}}{2}-N, N-\frac{\bar{n}}{2}]$. Next we use a simple case with $N=5$ and $\bar{n}=6$ to demonstrate the preparation process, 
as shown in Fig.~\ref{fig:OFPSgene}. The initial state is the renormalized state of $e^{i\frac{\pi}{2}J_y}\ket{\bar{n}/2, -\bar{n}/2}$ in the 
regime $m\in [-2, 2]$. The optimization is performed with the constrained optimization by linear approximation (COBYLA) 
algorithm~\cite{Powell1994, Powell1998, Powell2007}, which will be thoroughly introduced in our next paper~\cite{Qin2025}. In the optimization 
the constraint is set to be $f_{\mathrm{c}}=1$. Here $f_{\mathrm{c}}:=1-|\bra{3, -3}\ket{\psi(t)}|^2-|\bra{3, 3}\ket{\psi(t)}|^2$ with $\ket{\psi(t)}$ 
the evolved state. With the control amplitude given in Fig.~\ref{fig:OFPSgene}(a), the violation of the constraint is lower than $10^{-3}$ in the 
entire dynamics, as shown in Fig.~\ref{fig:OFPSgene}(b), indicating that the simulation basically fit the scenario discussed in this paper. 
Figure~\ref{fig:OFPSgene}(c) shows the fidelity between the evolved state and the target state in Eq.~(\ref{eq:oat_target}), and at the final time 
point the fidelity approaches to 96.8\%. The tomography of the target and prepared states in the basis $\{\ket{j,m}\}$ are given in 
Fig.~\ref{fig:OFPSgene}(d). We want to emphasize that this example only provides a preliminary preparation process of the OFPS. A more 
systematic study on the preparation is still ongoing and would be presented in another paper.  

\section{Comparison with entangled coherent state} 
\label{sec:comparison}

A more inspiring fact is that when the dimension of the state is large enough the given OFPS can provide better 
performance than the continuous-variable states with the same particle number. Lee \emph{et al.}~\cite{Lee2019} 
found that the single-mode optimal finite-dimensional state can overcome the squeezed vacuum state when the 
Fock dimension is large enough. In two-mode quantum interferometry, the entangled coherent state is well studied and 
outperforms the N00N state~\cite{Gerry2001,Gerry2002,Joo2011} in quantum parameter estimation. In the following 
we compare the performance between the OFPS and the entangled coherent state. 

The entangled coherent state is a very useful state in quantum metrology and can be expressed by~\cite{Gerry2001,Gerry2002,Joo2011}
\begin{equation}
C_{\alpha} \left(\ket{\alpha 0}+\ket{0 \alpha}\right),
\end{equation}
where $C_{\alpha}=1/\sqrt{2(1+e^{-|\alpha|^2})}$ is the normalization coefficient, and $\ket{\alpha}$ is 
the coherent state.

In the case of the linear phase shifts, the QFI for the entangled coherent state can be written as  
\begin{equation}
2|C_{\alpha}|^2|\alpha|^2\left(1+|\alpha|^2\right)
\end{equation}
due to the fact that $\expval{J^2_z}=|C_{\alpha}|^2 |\alpha|^2\left(1+|\alpha|^2\right)/2$ and $\expval{J_z}=0$. 
Here the average particle number $\bar{n}=2|C_{\alpha}|^2|\alpha|^2$. And for nonlinear phase shifts, the QFI can 
be written as 
\begin{equation}
2|C_{\alpha}|^2|\alpha|^2\left(|\alpha|^6+6|\alpha|^4+7|\alpha|^2+1\right),
\end{equation}
where $\expval{nJ_z}= 0$ has been applied. The QFIs for both cases can be  
rewritten into a function of $\bar{n}$ via the equation $\bar{n}=|\alpha|^2/(1+e^{-|\alpha|^2})$. 

\begin{figure}[tp]
\centering\includegraphics[width=8.7cm]{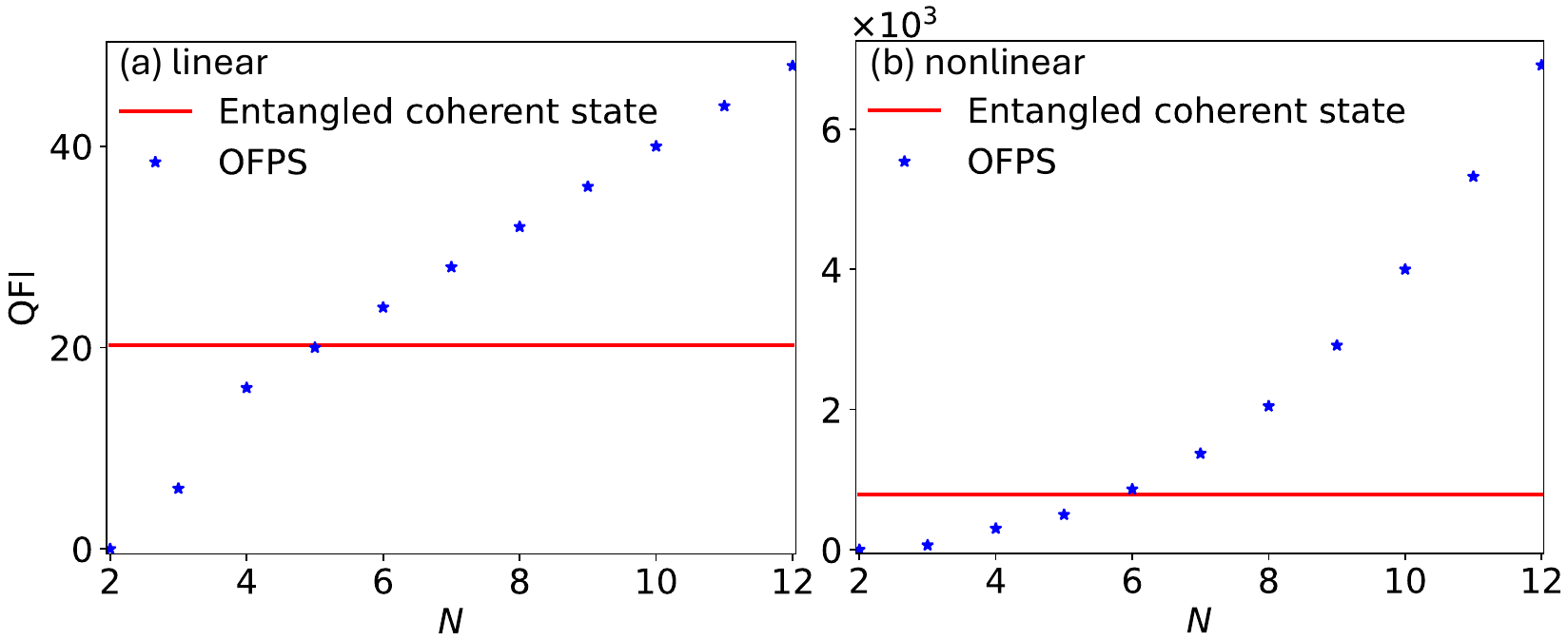}
\caption{Comparison of the QFI between the entangled coherent state (red line) and the OFPS (blue stars) for 
(a) linear phase shifts and (b) nonlinear phase shifts. The average particle number $\bar{n}=4$ in the plots. }
\label{fig:apx_cvcompare}
\end{figure}

The QFIs for the entangled coherent state and the OFPS are shown in Fig.~\ref{fig:apx_cvcompare}(a) for 
linear phase shifts and Fig.~\ref{fig:apx_cvcompare}(b) for nonlinear phase shifts in the case of $\bar{n}=4$. 
It can be seen that with the increase of $N$, the QFI of the OFPS would overcome that of the entangled 
coherent state, which could never be realized by the \noon state~\cite{Gerry2001,Gerry2002,Joo2011}. 

Currently, the continuous-variable states like the squeezed vacuum state has shown great power in various 
scenarios~\cite{Andersen2016,Zhang2021,LIGO2023} and it is quite possible that the preparation of the OFPS 
would be harder than the continuous-variable states under the current experimental quantum technologies. 
However, with the fast development of the finite-dimensional state~\cite{Deng2024}, it is possible that the 
preparation difficulty of the finite-dimensional state would be overcome and the theoretical advantage of the 
OFPS would be realized in practice. 

\section{Optimal measurements}
\label{sec:opt_measure}

\begin{figure*}[tp]
\includegraphics[width=18cm]{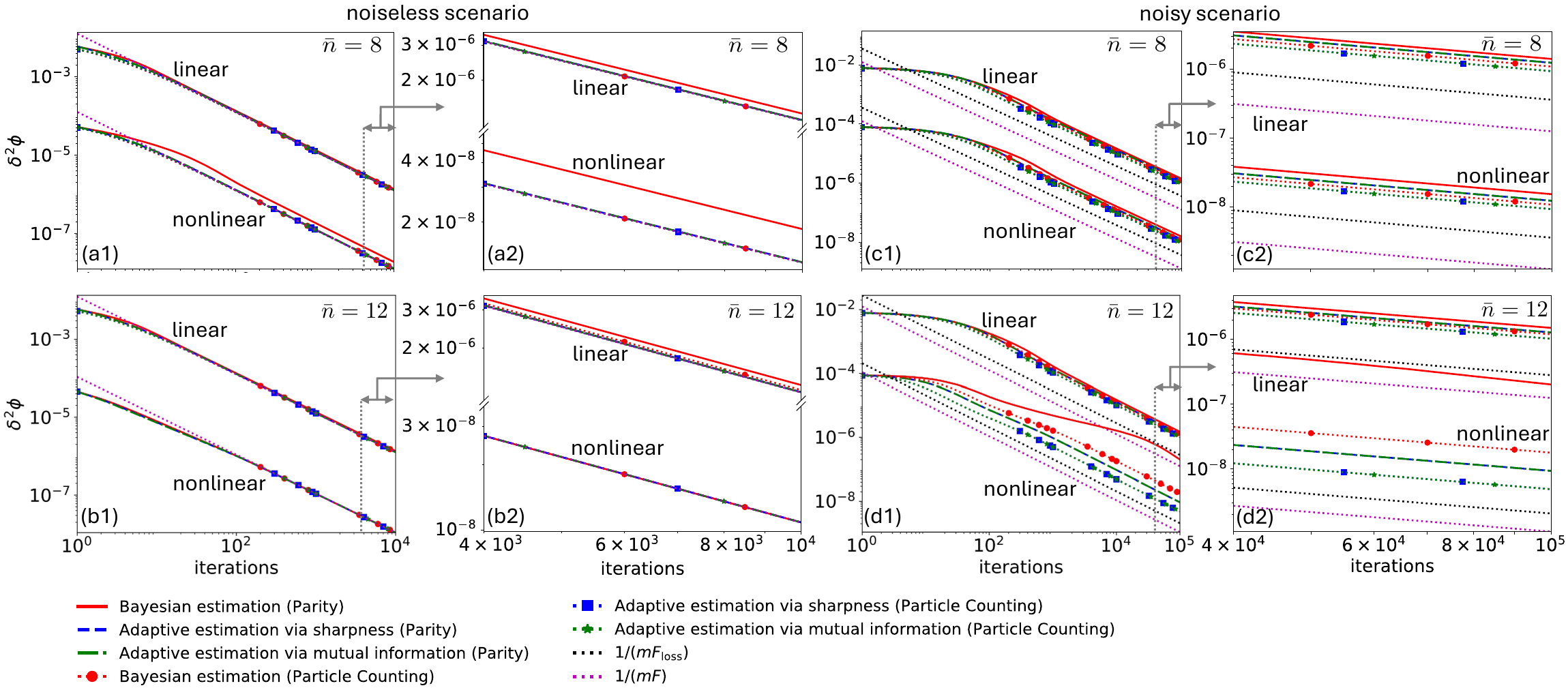}
\caption{Performance comparison between the adaptive schemes realized by the sharpness 
(dashed-blue line) and mutual information (dash-dotted-green line), and Bayesian estimations 
(solid-red line) in [(a1)-(a2), (b1)-(b2)] noiseless and [(c1)-(c2), (d1)-(d2)] noisy scenarios. 
2000 rounds of experiments are numerically simulated and all results in the plots are the average 
performance of them. The performance of all simulations are given in Appendix~\ref{sec:adapt}. 
In the figure $N=10$ and the true value of $\phi$ is taken as 0.2. In the noisy case 
the transmission rates $T_1=T_2=0.9$. }
\label{fig:adapt}
\end{figure*}

\begin{figure}[tp]
\includegraphics[width=8.cm]{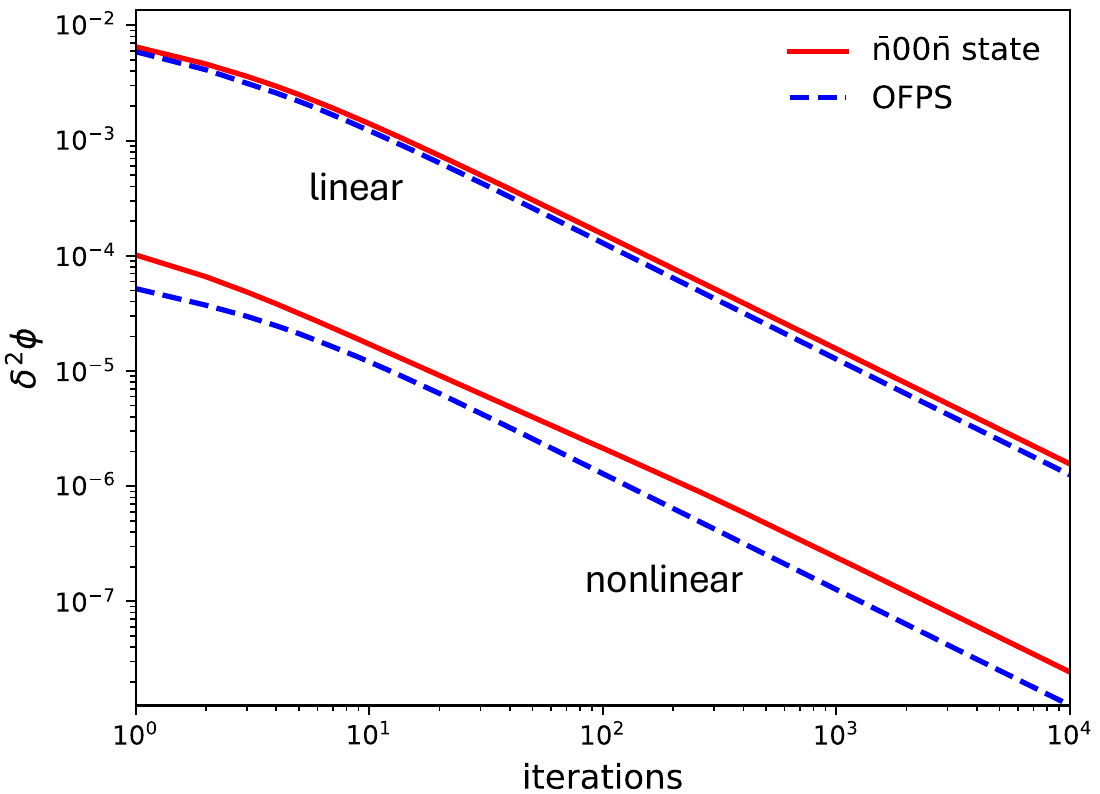}
\caption{Performance comparison between the OFPS and \noon state in the noiseless scenario with 
$\bar{n}=8$ and $N=10$. The solid-red and dashed-blue lines represent the Bayesian 
estimation for \noon state and adaptive estimation for the OFPS, respectively.}
\label{fig:compareNOON}
\end{figure}

A complete estimation scheme not only needs the optimal state, but also the optimal measurement to realize 
the predicted precision limit. Hence, the optimal measurement is always critical in quantum parameter 
estimation. In quantum optics, the parameterized state usually goes through a beam splitter first before the measurement is performed, 
such as in the Mach-Zehnder interferometer. Hence, here we follow this convention and use the one characterized 
by $\exp(i\pi J_x/2)$. 

In this work we consider both the parity and particle-counting measurements. In theory, the parity operator reads 
$\Pi_a = e^{i\pi a^{\dagger}a}=e^{i\frac{\pi}{2}n}e^{i\pi J_z}$, and the probability $P_{\pm}$ with respect to the 
result $\pm 1$ on mode $a$ is 
\begin{equation}
P_{\pm}=\frac{1}{2}\left(1\pm \expval{\Pi_a}\right).
\end{equation}
For the particle-counting measurement, the probability of detecting $m$ particles on mode $a$ is 
\begin{equation}
P_m =\sum_{j=0}^{2N}|\bra{mj}\ket{\psi}|^2
\end{equation}
with $\ket{\psi}$ a quantum state. As a matter of fact, both parity and particle-counting measurements can be the 
optimal measurements at the asymptotic limit, yet the optimality is only valid for some specific true values of $\phi$. 
For the linear phase shifts the parity and particle-counting measurements are only optimal when the true value of 
$\phi$ is $(\theta_1-\theta_2+2k\pi)/N-\pi/2$ with $k$ any integer, and for the nonlinear phase shifts they are 
optimal when the true value is $\left(\theta_1-\theta_2+2k\pi\right)/N^2-\pi/(2N)$ in the case that $\bar{n}\leq N$. 
The only case presenting the true-value independence of the optimality is that $\bar{n}$ is an integer in the regime 
$[N,\left\lfloor\frac{4N+1}{3}\right\rfloor]$. Detailed calculations for both parity and 
particle-counting measurements are given in Appendix~\ref{sec:measurement} and Appendix~\ref{sec:measurement1}.

In practice, the true value of $\phi$ is not tunable in most cases, which strongly limits the performance of 
parity and particle-counting measurements as the optimal measurements. To make sure these two measurements are 
always optimal for any true value, the adaptive measurement has to be involved~\cite{Holevo1984,Berry2000,Berry2001,
Hentschel2010,Huang2017,DiMario2020,Garcia2022,Zhang2022,Bargatin2005,Dobrzanski2017,Liu2022,Kurdzialek2023,
Miao2023,Tsang2012,Rubio2018}. In the adaptive scheme, a tunable phase is introduced in one arm, such as mode $a$. In the linear case, the 
operator for it is  $\exp(i\phi_{\mathrm{u}}a^{\dagger}a)$, and the operator for the total phase difference becomes 
$\exp(i(\phi+\phi_{\mathrm{u}})J_z)$. In the nonlinear case, the tunable phase can be introduced via the operator 
$\exp(i\phi_{\mathrm{u}}(a^{\dagger}a)^2)$ and the total phase difference then becomes $\exp(i(\phi+\phi_{\mathrm{u}})nJ_z)$. 
In this paper, both average sharpness function~\cite{Holevo1984,Berry2000,Berry2001,Hentschel2010,Huang2017,
DiMario2020,Garcia2022,Zhang2022} and average mutual information~\cite{DiMario2020,Garcia2022,Zhang2022,
Bargatin2005,Rzadkowski2017,Cover1991} are used as the objective functions for the update of $\phi_{\mathrm{u}}$. 

The conditional probabilities are periodic for both parity and particle-counting measurements (details see Appendix~\ref{sec:adapt}). 
In one period, two peaks exist and the Bayesian estimation cannot pick the right one, which will cause a wrong estimation. To avoid 
this problem, the prior distribution is taken as half of the period in this paper. For the sake of a fair performance comparison, 
the prior distribution in the adaptive measurement is taken as the same one as the Bayesian estimation. Specifically 
to say, the prior distribution in the demonstration is taken as a uniform distribution in the regime $[0,\frac{\pi}{10}]$ 
for all examples in the linear case. In the nonlinear case, the prior distribution is taken as a uniform distribution in the regime 
$\left[\frac{3\pi}{50},\frac{7\pi}{100}\right]$ for $\bar{n} = 8$, and $\left[\frac{\pi}{16},\frac{7\pi}{96}\right]$ for $\bar{n}=12$. 

The average performance of adaptive measurement for 2000 simulations of the experiment in the case of $N=10$, 
together with the Bayesian estimation, are illustrated in Figs.~\ref{fig:adapt}(a1) and \ref{fig:adapt}(b1) for the OFPSs 
in both regimes $\bar{n}< N$ ($\bar{n}=8$) and $\bar{n}>N$ ($\bar{n}=12$). It is not surprising that the performance 
with nonlinear phase shifts is better than that with linear phase shifts. The true value of $\phi$ is taken as 0.2, and both 
parity and particle-counting measurements at this point are not optimal. From the results of the last 6000 
rounds of iteration shown in Figs.~\ref{fig:adapt}(a2) and \ref{fig:adapt}(b2), it can be seen that the Bayesian estimation 
cannot reach the ultimate precision quantified by the QFI (dotted purple line), which is reasonable since the Bayesian 
estimation for both parity and particle-counting measurements can only reach the precision quantified by CFI, and in this 
case, the CFI differs from the QFI as these two measurements are not optimal for this specific true value. In the adaptive 
scheme, the sharpness and mutual information show consistent performance. More importantly, both parity and 
particle-counting measurements reach the precision quantified by the QFI in both linear and nonlinear cases, indicating 
that adaptive measurement can overcome the dependency of the measurement optimality on the true value. Hence, 
utilizing the adaptive scheme, the parity and particle-counting measurements are optimal to realize the ultimate precision 
quantified by the QFI, regardless of the true value. More details of the adaptive measurement can be found in 
Appendix~\ref{sec:adapt}. 

On the other hand, the performance of the adaptive measurement with the OFPS and Bayesian estimation with 
the \noon state is also compared in the noiseless scenario with $\bar{n}=8$ and $N=10$, as shown in 
Fig.~\ref{fig:compareNOON}. In this case, the parity measurement is optimal for the \noon state regardless 
of the true value, and thus adaptive measurement is unnecessary. It can be seen that the performance of the OFPS  
is slightly better than the \noon state in the case of linear phase shifts, and this advantage is enhanced when  
the nonlinear phase shifts are used. For the case of $\bar{n}=12$, the \noon state would outperform the optimal 
state since extra dimension resource is used. Once this resource of the OFPS also increases to the same amount, it 
comes back to the case of $\bar{n}\leq N$.  

\begin{figure*}[tp]
\centering
\includegraphics[width=16cm]{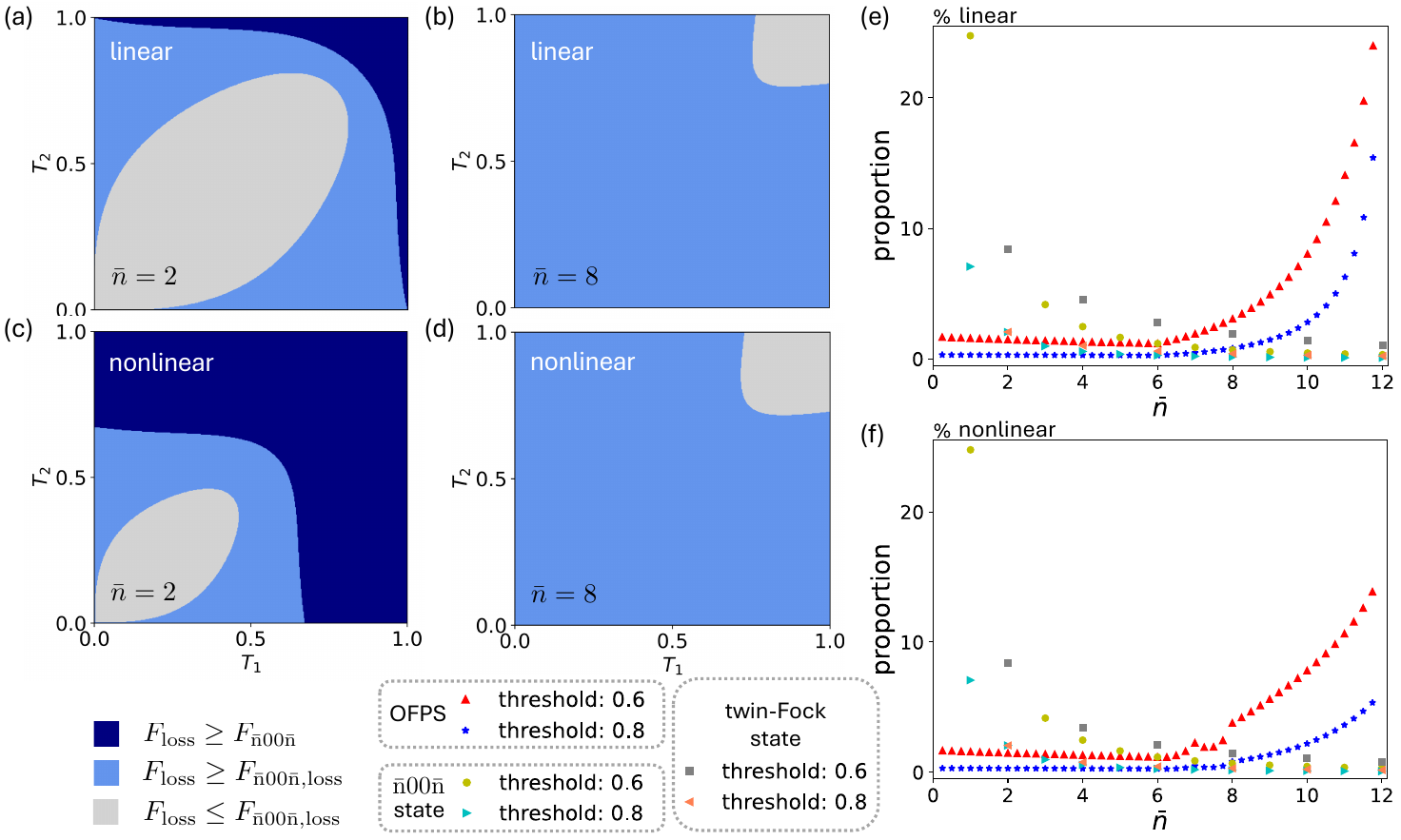}
\caption{[(a)-(d)] Performance comparison between the OFPS and \noon state in (a) linear case with 
$\bar{n}<N$ ($\bar{n}=2$), (b) linear case with $\bar{n}>N$ ($\bar{n}=8$), (c) nonlinear case with 
$\bar{n}<N$ ($\bar{n}=2$), and (d) nonlinear cases with $\bar{n}>N$ ($\bar{n}=8$). The variety of 
the proportion of the ratio $F_{\mathrm{loss}}/F$ that is larger than $0.6$ and $0.8$ with the change 
of average input particle numbers $\bar{n}$ for the OFPS, \noon state and twin-Fock state in both 
linear (e) and nonlinear (f) cases. $N=6$ in all figures.}
\label{fig:QFI_loss}
\end{figure*}

\section{Noisy performance}

The noise effect is essential to be considered in practice, and in phase estimation the particle loss is a major noise in general. 
In theory, the effect of particle loss can be modeled via a fictitious beam splitter on each 
arm~\cite{Barnett1998,Gardiner2004,Gardiner2004,Rubin2007,Huver2008,Dorner2009,Dobrzanski2009,Zhang2013,Liu2013,Knott2014}. 
The transmission rates $T_1$ and $T_2$ of these two fictitious beam splitters represent the remains of the input particles. 
When $T_{1}=1$ ($T_2=1$), no particle leaks from the arm of mode $a$ ($b$), and all particles leak out when $T_{1}=0$ 
($T_2=0$). The average performance of adaptive measurement under the noise of particle loss are shown in 
Figs.~\ref{fig:adapt}(c1) and \ref{fig:adapt}(d1) for $\bar{n}< N$ ($\bar{n}=8$) and $\bar{n}> N$ ($\bar{n}=12$), 
respectively. Here $\bar{n}$ is the average particle number of the input state. When the particle loss exists, the convergence 
of $\delta^2 \phi$ becomes slow, and we have to extend the iteration number in one experiment to $10^{5}$. Bayesian 
estimation requires more iterations to converge in the nonlinear case for parity measurement with $\bar{n}=12$, and its 
performance up to $10^{6}$ iterations is given in Appendix~\ref{sec:Calnoise}. From the last $6\times 10^{4}$ iterations given 
in Figs.~\ref{fig:adapt}(c2) and \ref{fig:adapt}(d2), it can be seen that both parity and particle-counting measurements cannot 
reach the precision quantified by the QFI, however, they can still overcome the precision given by their own CFI attained by 
the Bayesian estimation, and reach the maximum CFI with respect to all true values. This phenomenon immediately leads to 
the fact that the performance of particle-counting measurement is better than that of parity measurement under the particle 
loss since the maximum CFI is larger for the particle-counting measurement. The specific expressions of the maximum CFIs 
can be found in Appendix~\ref{sec:Calnoise}. 

Compared to the \noon state, i.e., $(\ket{\bar{n}0}+e^{i\theta}\ket{0\bar{n}})/\sqrt{2}$, the OFPSs  
not only present better performance in the lossless case, but also show the advantage under the particle loss for a large 
regime of $T_1$ and $T_2$, as illustrated in Figs.~\ref{fig:QFI_loss}(a) to \ref{fig:QFI_loss}(d) in the case of $N=6$. The blue 
regions (including both lightblue and darkblue regions) represent the regimes where the QFI of the OFPS ($F_{\mathrm{loss}}$) 
is larger than that of the \noon state ($F_{\mathrm{\bar{n}00\bar{n},loss}}$) under particle loss. It can be seen that the 
OFPS presents a significant advantage for small leakage or large yet unbalanced leakage when $\bar{n}<N$. More importantly, 
in both linear and nonlinear cases the lossy performance of the OFPS can even overcome the lossless performance of the 
\noon state ($F_{\mathrm{\bar{n}00\bar{n}}}$ represents the corresponding QFI) for not very large leakage when 
$\bar{n}<N$ [darkblue regimes in Figs.~\ref{fig:QFI_loss}(a) and \ref{fig:QFI_loss}(c)]. This advantage is remarkably significant in the 
nonlinear case. Hence, this result indicates that the OFPS is a better choice than the \noon  state when the average particle number 
is limited. In the case that $\bar{n}>N$, the \noon state outperforms the OFPS when $T_1$ and $T_2$ are large, as shown 
in Figs.~\ref{fig:QFI_loss}(b) and \ref{fig:QFI_loss}(d). However, in this case the dimension of the \noon state, which is $\bar{n}+1$, is larger 
than that of the OFPS, namely, $N+1$. This means more metrological resources are actually involved in the \noon state. Even though 
the used resources are less, the OFPS still presents a better performance with the increase of the leakage. This phenomenon indicates 
that the OFPS is a better choice for a large particle leakage when the average particle number is large or unlimited. 

\begin{figure}[tp]
\includegraphics[width=8.5cm]{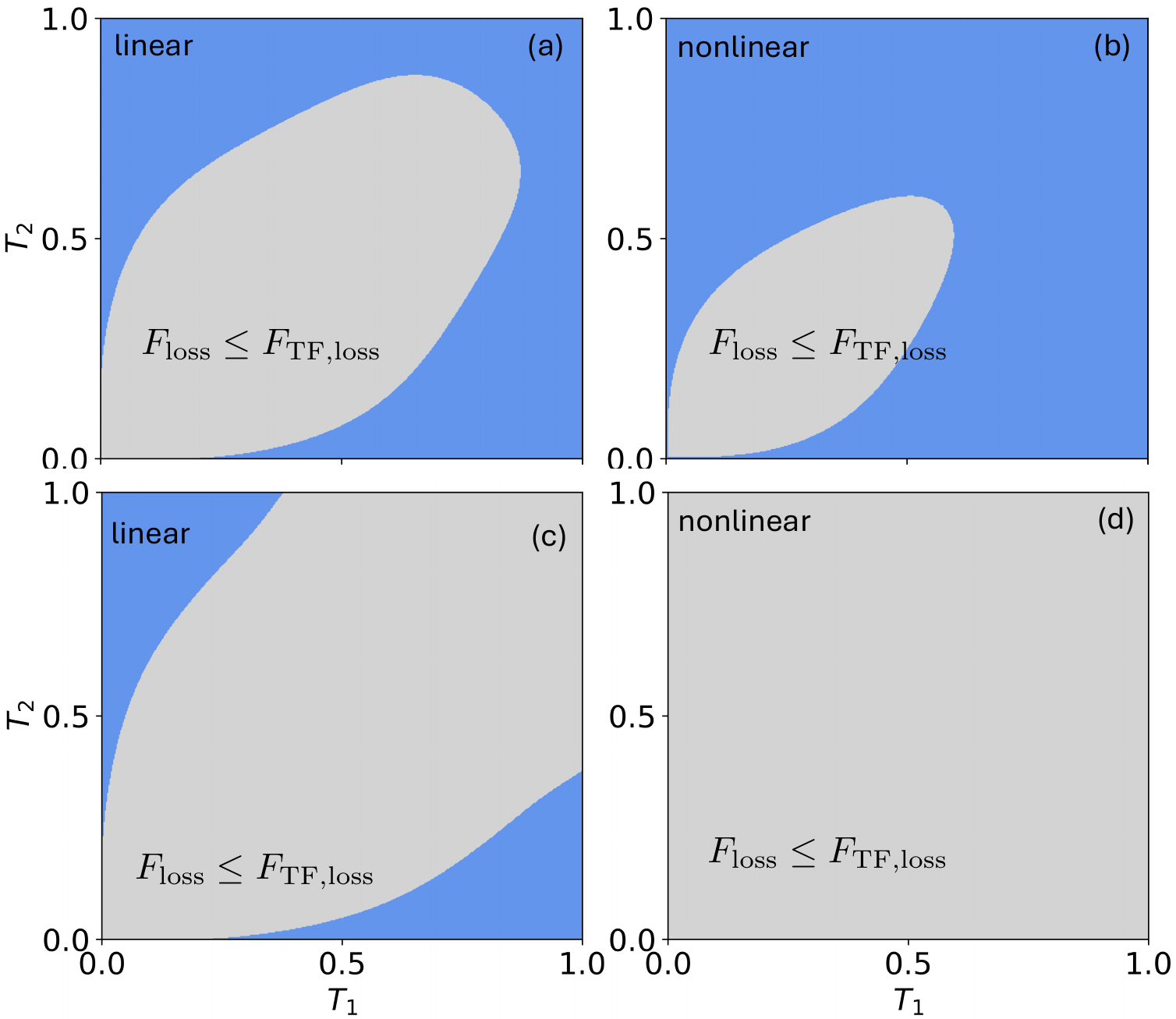}
\caption{Performance comparison between the OFPS and twin-Fock state in (a) linear and 
(b) nonlinear cases with $\bar{n}<N$ ($\bar{n}=4$), and (c) linear and (d) nonlinear cases with $\bar{n}>N$ 
($\bar{n}=8$).  The blue (gray) regions represent the regimes where the performance of the OFPS is better 
(worse) than that of the twin-Fock state. $N=6$ in all plots. }
\label{fig:twinFock}
\end{figure}

The twin-Fock state~\cite{Holland1993} $\ket{mm}$ ($\ket{m}$ is a Fock state) is another useful quantum state in quantum 
metrology, which shows close performance to the N00N state in ideal situations~\cite{Lang2014,Datta2011} and yet much 
more noise-resilient~\cite{Datta2011}. When a twin-Fock state directly connects to the phase shifts, the phase difference 
cannot be encoded into the state, which means the phase estimation cannot be executed. Hence, a beam splitter should 
be used before the phase shifts. Here we choose a 50:50 beam splitter described by the operator $\exp(-i \pi J_x/2)$. 
In the case of $\bar{n}=2$, the twin-Fock state $\ket{\bar{n}/2,\bar{n}/2}$ becomes the \noon state after going 
through the beam splitter $\exp(-i\pi J_x/2)$ and the corresponding performance have already been discussed in 
Figs.~\ref{fig:QFI_loss}(a) and \ref{fig:QFI_loss}(c). To further compare it with the OFPS, here we consider the case 
of $\bar{n}=4$ and $\bar{n}=8$. As shown in Fig.~\ref{fig:twinFock}, in the blue (gray) region the QFI of the OFPS  
($F_{\mathrm{loss}}$) is larger (smaller) than that of the twin-Fock state ($F_{\mathrm{TF, loss}}$). In the case of 
$\bar{n}=4$, similar to the case of $\bar{n}=2$, the OFPS presents better performance when either $T_1$ or $T_2$ is large 
[Fig.~\ref{fig:twinFock}(a)], and this advantage can be enhanced when the nonlinear phase shifts are applied 
[Fig.~\ref{fig:twinFock}(b)]. 

In the case of $\bar{n}=8$ ($\bar{n}>N$), the twin-Fock state presents better performance for most values of $T_1$ and $T_2$ 
[Fig.~\ref{fig:twinFock}(c)] and this advantage covers all values of $T_1$ and $T_2$ when the nonlinear phase shifts are used 
[Fig.~\ref{fig:twinFock}(d)]. This phenomenon is quite different from that with the \noon state, where the OFPS can still 
show significant advantage even the \noon state uses more resource of dimension. Hence, the twin-Fock state would be a better 
choice under noise in the case of $\bar{n}>N$.  However, one should notice that the OFPS ceases to be true optimal under noise, 
and the performance comparison between the twin-Fock state and the true optimal finite-dimensional states under noise would 
be given in our next paper. 

The robustness of performance is another important indicator in quantum metrology. Here we use the proportion of the ratio 
$F_{\mathrm{loss}}/F$ ($F$ is the lossless QFI) that is higher than a given threshold with respect to all values of $T_1$ 
and $T_2$ as the indicator of the robustness. The variety of robustness is illustrated in the case of $N=6$ with two values 
of threshold ($0.6$ and $0.8$) for both linear and nonlinear phase shifts, as shown in Figs.~\ref{fig:QFI_loss}(e) and \ref{fig:QFI_loss}(f). 
The basic behaviors with respect to these two values of threshold coincide with each other, indicating that the performance is not 
affected by the choice of the value of threshold. It can be seen that for a fixed Fock dimension the lowest robustness occurs 
around the point $\bar{n}=N$, which indicates that the \noon state presents a low robustness among all the OFPSs. When 
$\bar{n}\leq N$ the robustness does not show a significant change for both linear and nonlinear cases, however, when 
$\bar{n}\geq N$ it presents a remarkable improvement with the increase of $\bar{n}$. Interestingly, both the \noon and twin-Fock 
states show a completely contrary behavior. When the average particle number is small, their robustness is higher than that of the OFPS, 
however, with the increase of $\bar{n}$ the robustness of these two states reduce significantly. One should notice that the price 
for the improvement of robustness with the OFPS is the reduction of the QFI, since when $\bar{n}>N$ its QFI reduces yet that of the 
\noon state increases, as shown in Fig.~\ref{fig:OFPSvsN00N}. In the meantime, the robustness of the twin-Fock state is higher than the 
OFPS (\noon state) in the case that $\bar{n}=N$. 

\section{Conclusion}

In conclusion, the OFPS, together with the optimal measurement, has been provided for 
both linear and nonlinear quantum phase estimations. The given OFPSs reveal an important phenomenon that the dimension of the 
state could be treated as a metrological resource. Utilizing this feature, our schemes would be particularly useful in scenarios where 
weak light is required or the power of the probe is restricted, such as the biological detection or quantum measurement in the satellite 
and space station. When the particle number is fixed the measurement precision in our schemes can still be improved by preparing 
the OFPS with a higher dimension. In the meantime, the given schemes are applicable to both optical and condensed systems due to 
the extensive physical realizations of the operators of phase shifts and beam splitters, and could be widely applied in many mainstream 
quantum platforms in the near future. 

The OFPS provides a brand-new perspective for phase estimation in the quantum interferometry, and there are still several unsolved 
problems in this field that require further attentions. For instance, when the noise exists the OFPS may cease to be optimal 
mathematically and what is the true OFPS under noise is then an important problem. Besides, the preparation process of the OFPS in 
various quantum systems and the OFPS for nonlinear phase shifts with arbitrary nonlinearities are also worth to be further investigated 
in the future. As a matter of fact, the true OFPS under noise would be thoroughly discussed in our next paper~\cite{Qin2025}. 

\begin{acknowledgments}
The authors would like to thank Prof. Ling-Na Wu and two anonymous referees for their insightful suggestions. This work was supported by the 
National Natural Science Foundation of China (Grants No.\,12175075 and \,12575013). 

J.F.Q. and Y.X. contributed equally to this work.
\end{acknowledgments}

\section{DATA AVAILABILITY}

The data that support the findings of this article are openly available~\cite{zenodo2025}.

\appendix

\section{Proof of Theorem 1}
\label{sec:theorem1}

In this section we provide thorough proof of Theorem 1. The $(N+1)$-dimensional probe state can be 
expressed by
\begin{equation}
\ket{\psi_{\mathrm{in}}}=\sum^N_{i,j=0}c_{ij}\ket{ij},    
\end{equation}
where the coefficient $c_{ij}$ satisfies the normalization condition $\sum^N_{i,j=0}|c_{ij}|^{2}=1$. It is easy to see that the average 
particle number is 
\begin{equation}
\bar{n}=\bra{\psi_{\mathrm{in}}} a^{\dagger}a+b^{\dagger}b\ket{\psi_{\mathrm{in}}}
=\sum^N_{i,j=0}|c_{ij}|^{2}(i+j).
\end{equation}
In the following we denote $n:=a^{\dagger}a+b^{\dagger}b$ as the the operator for total particle number. 

We first consider the case of the linear phase shifts. In this case, the operator for the phase shift is 
\begin{equation}
e^{i(\phi_a a^{\dagger}a+\phi_b b^{\dagger}b)}=e^{i\frac{1}{2}\phi_{\mathrm{tot}}n}e^{i\phi J_z},
\end{equation}
where $\phi_{\mathrm{tot}}$ is the total phase and $\phi=\phi_a-\phi_b$ is the phase difference between two arms. Here 
\begin{equation}
J_z=\frac{1}{2}\left(a^{\dag}a-b^{\dag}b\right)    
\end{equation}
is a Schwinger operator. The other two Schwinger operators are 
\begin{eqnarray}
J_x &=& \frac{1}{2}\left(a^{\dag}b+ab^{\dag}\right), \\
J_y &=& \frac{1}{2i}\left(a^{\dag}b-ab^{\dag}\right). 
\end{eqnarray}

The QFI with respect to the phase difference for a pure parameterized state $\ket{\psi}$ 
can be written as 
\begin{equation}
F=4(\bra{\partial_{\phi}\psi}\ket{\partial_{\phi}\psi}-|\bra{\partial_{\phi}\psi}\ket{\psi}|^2).    
\end{equation}
In this case, since $\ket{\psi}=e^{i\frac{1}{2}\phi_{\mathrm{tot}}n}e^{i\phi J_z}\ket{\psi_{\mathrm{in}}}$, the QFI reads 
\begin{align}
F =& 4\left(\bra{\psi_{\mathrm{in}}}J^2_z\ket{\psi_{\mathrm{in}}}
-\bra{\psi_{\mathrm{in}}}J_z\ket{\psi_{\mathrm{in}}}^2\right) \nonumber \\
= & \sum^N_{i,j=0}P_{ij}\left(i\!-\!j\right)^{2}\!-\!\sum^N_{i,j,k,l=0}
P_{ij}P_{kl}\left(i\!-\! j\right)\left(k\!-\!l\right)\!,
\label{eq:apx_QFI_linear}
\end{align}
where $P_{ij}:=|c_{ij}|^2$. 

Utilizing the expression above, the problem of state optimization can be expressed by 
\begin{eqnarray}
\underset{P_{ij}}{\mathrm{max}} & &\sum^N_{i,j=0} P_{ij} (i-j)^2-\left[\sum^N_{i,j=0}P_{ij}(i-j)\right]^2, \nonumber \\
\text{s.t.} & & \begin{cases}
P_{ij}\in [0,1], \forall i,j, \\
\sum^N_{i,j=0} P_{ij}=1,  \\
\sum^N_{i,j=0} P_{ij}(i+j)=\bar{n},
\end{cases} 
\end{eqnarray}
where "s.t." is short for "subject to". To better solve this problem, we rewrite the subscripts of $P$ with $s=i+j$ 
and $d=(i-j)/2$. Here $s\in[0,2N]$ and 
\begin{equation}
\begin{cases}
d\in\left[-\frac{1}{2}s,\frac{1}{2}s\right], & s\in[0,N], \\
d\in\left[\frac{1}{2}s-N,N-\frac{1}{2}s\right], & s\in[N,2N].
\end{cases}    
\end{equation}
In the following we denote $x_s:=s/2$ when $s\in[0,N]$ and $x_s:=N-s/2$ when $s\in[N,2N]$, which gives a uniform expression 
of the regime for $d$, i.e., $d\in[-x_s,x_s]$. Then the optimization problem above can be rewritten into 
\begin{eqnarray}
\underset{P_{s,2d}}{\mathrm{max}} & &~4\left[\sum^{2N}_{s=0} \sum^{x_s}_{d=-x_s} d^2P_{s,2d}
-\left(\sum^{2N}_{s=0} \sum^{x_s}_{d=-x_s} d P_{s,2d} \right)^2\right], \nonumber \\
\text{s.t.} & & \begin{cases}
\sum^{x_s}_{d=-x_s}P_{s,2d}\in[0,1], \forall s, \\
\sum^{2N}_{s=0}\sum^{x_s}_{d=-x_s}P_{s,2d} =1,  \\
\sum^{2N}_{s=0} \sum^{x_s}_{d=-x_s} sP_{s,2d}=\bar{n}.
\end{cases} 
\end{eqnarray}
Notice that 
\begin{align}
& \sum^{2N}_{s=0} \sum^{x_s}_{d=-x_s} d^2P_{s,2d}-\left(\sum^{2N}_{s=0} \sum^{x_s}_{d=-x_s} d P_{s,2d} \right)^2 \nonumber \\  
\leq & \sum^{2N}_{s=0} \sum^{x_s}_{d=-x_s} d^2P_{s,2d},
\end{align}
and the equality can be attained when $\sum^{x_s}_{d=-x_s} d P_{s,2d}$ is zero. In the meantime, utilizing the condition 
$\sum^{x_s}_{d=-x_s} d P_{s,2d}=0$, 
\begin{align}
\sum^{x_s}_{d=-x_s} d^2P_{s,2d}=\sum^{x_s}_{d=-x_s} d^2P_{s,2d}-\left(\sum^{x_s}_{d=-x_s} dP_{s,2d}\right)^2,
\label{eq:apx_vartp}
\end{align}
which is nothing but the variance of $d$ with respect to the probability distribution $\{P_{s,2d}\}^{x_s}_{d=-x_s}$. According 
to the Popoviciu's inequality on variances~\cite{Popviciu1935}, the maximum value of Eq.~(\ref{eq:apx_vartp}) can only be attained 
when the distribution $\{P_{s,2d}\}^{x_s}_{d=-x_s}$ is a uniform bimodal one with peaks distributed at the boundaries, namely, 
\begin{align}
P_{s,2d} &=0,~\mathrm{for}~d\neq -x_s,x_s, \\
P_{s,-2x_s} &=P_{s,2x_s}.
\end{align}
The second condition is equivalent to 
\begin{equation}
\begin{cases}
|c_{0s}|^2= |c_{s0}|^2, & s\in[0,N],\\
|c_{s-N,N}|^2= |c_{N,s-N}|^2, & s\in[N,2N].
\end{cases}
\label{eq:apx_optcond1}
\end{equation}
Combining these two conditions, the optimization problem can be further rewritten into 
\begin{align}
&\underset{P_{ss},P_{s,2N-s}}{\mathrm{max}}  2\left[\sum^{N}_{s=0}s^2 P_{ss}
+\sum^{2N}_{s=N+1}(2N-s)^2 P_{s,2N-s}\right] \nonumber \\
&\text{s.t.}  \begin{cases}
P_{ss},P_{s,2N-s}\in\left[0,\frac{1}{2}\right], \forall s\neq 0,2N, \\ 
P_{00},P_{2N,0}\in [0,1], \\
\sum^{N}_{s=0}P_{ss}\!+\!\sum^{2N}_{s=N+1}P_{s,2N-s}=\frac{1}{2}(1\!+\!P_{00}\!+\!P_{2N,0}),  \\
\sum^{N}_{s=0}s P_{ss}\!+\!\sum^{2N}_{s=N+1}s P_{s,2N-s}=\frac{\bar{n}}{2}+N P_{2N,0}.
\end{cases} 
\end{align}
An equivalent writing way of the problem above is 
\begin{align}
&\underset{P_{ss},P_{s,2N-s}}{\mathrm{min}}  -2\left[\sum^{N}_{s=0}s^2 P_{ss}
+\sum^{2N}_{s=N+1}(2N-s)^2 P_{s,2N-s}\right] \nonumber \\
&\text{s.t.} \begin{cases}
P_{ss},P_{s,2N-s}\in\left[0,\frac{1}{2}\right], \forall s\neq 0,2N, \\ 
P_{00},P_{2N,0}\in [0,1], \\
\sum^{N}_{s=0}P_{ss}\!+\!\sum^{2N}_{s=N+1}P_{s,2N-s}=\frac{1}{2}(1\!+\!P_{00}\!+\!P_{2N,0}),  \\
\sum^{N}_{s=0}s P_{ss}\!+\!\sum^{2N}_{s=N+1}s P_{s,2N-s}= \frac{\bar{n}}{2}+N P_{2N,0}.
\end{cases} 
\end{align}

In the following we will use the Karush-Kuhn-Tucker (KKT) conditions~\cite{Boyd2004} to solve this optimization problem. 
For the sake of a better reading experience, we first introduce the KKT condition first. Consider the optimization problem 
\begin{eqnarray}
\underset{\bold{x}}{\mathrm{min}}& &f(\bold{x}), \\
\text{s.t.}& & \begin{cases} 
g_i(\bold{x}) = 0, & i=0, \cdots,p, \\
 h_i(\bold{x}) \leq 0, & i=0,\cdots,q,
\end{cases} 
\end{eqnarray}
where $f(\bold{x})$ is the objective function with the real variables $\bold{x}$ and $g_i(\bold{x}), i=0,\cdots,p$ 
[$h_i(\bold{x}),i=0,\cdots,q$] is the $i$th equality (inequality) constraint. The Lagrangian function $\mathcal{L}$ for 
this problem is 
\begin{equation}
\mathcal{L} = f(\bold{x})+\sum^p_{i=0}\lambda_i g_i(\bold{x})+\sum^q_{i=0}\nu_i h_i(\bold{x})
\end{equation}
with $\lambda_i\,(\nu_i)$ the Lagrange multiplier of $i$th equality (inequality) constraint. In this case, the optimal 
values (denoted by $\bold{x}^{*}$, 
$\lambda_i^*$, $\nu_i^*$) must satisfy the following conditions
\begin{equation}	
\begin{cases}
\nabla f(\bold{x}^*)+\sum^p_{i=0}\lambda^*_i\nabla g_i(\bold{x}^*) +\sum^q_{i=0}\nu^*_i\nabla h_i(\bold{x}^*) = 0, 
\nonumber\\
g_i(\bold{x}^*) = 0, \, i = 0,\cdots,p, \nonumber \\
h_i(\bold{x}^*) \leq 0,\, i = 0,\cdots,q, \nonumber \\
\nu^*_i\geq 0,\,i = 0,\cdots,q, \nonumber \\
\nu^*_ih_i(\bold{x}^*)=0,\,i = 0,\cdots,q.
\end{cases}
\end{equation}
In the first equation $\nabla$ represents the gradient. The last two equations are the dual feasibility condition and 
the complementary slackness condition. These conditions are usually called the KKT conditions. More details on the KKT 
conditions can be found in Ref.~\cite{Boyd2004}. 

Next, we will use the KKT conditions to find the optimal values of $P_{ss}$ and $P_{s,2N-s}$ (denoted by $P^*_{ss}$ and 
$P^*_{s,2N-s}$). In our problem, the Lagrangian function reads 
\begin{align}
\mathcal{L}= &-2\sum^N_{s=0} s^2 P_{ss}-2\sum^{2N}_{s=N+1} (2 N-s)^2 P_{s,2N-s} \nonumber  \\ 
&\!-\!2\sum^{N}_{s=1} {\nu}_s P_{ss}\!-\!2\sum^{2N-1}_{s=N+1} {\nu}_s P_{s,2N-s}\!-\!\nu_0 P_{00}\!-\!\nu_{2N} P_{2N,0}  \nonumber\\ 
&+\lambda_0\left(\!P_{00}+2\sum^{N}_{s=1} \!P_{ss}+2\!\!\!\!\sum^{2N-1}_{s=N+1}P_{s,2N-s}+P_{2N,0}-1\!\!\right) \nonumber\\ 
&+\lambda_1\left(2\sum^{N}_{s=0} s P_{ss}\!+\!2\!\sum^{2N-1}_{s=N+1} s P_{s,2N-s}\!+\!2N P_{2N,0}\!-\!\bar{n}\right),  
\end{align}
which indicates that the corresponding KKT conditions with respect to $P^*_{ss}$, $P^*_{s,2N-s}$, $\lambda^{*}_{0,1}$, 
and $\nu^{*}_{s}$ are of the form 
\begin{equation*}
\begin{cases}
s^2-\lambda^*_1 s-\lambda^*_0+\nu^*_s=0, s\in\mathbb{Z}_{[0,N]}, \\
(2N\!-\!s)^2-\lambda^*_1 s-\lambda^*_0+\nu^*_s=0, s\in\mathbb{Z}_{[N,2N]}, \\
\sum^{N}_{s=0} \!P^*_{ss}+\sum^{2N}_{s=N+1}\! P^*_{s,2N-s}=\frac{1}{2}\left(1\!+\!P_{00}\!+\!P_{2N,0}\right), \\
\sum^{N}_{s=0} s P^*_{ss}+\sum^{2N}_{s=N+1} s P^*_{s,2N-s}-\frac{\bar{n}}{2}\!-\! N P_{2N,0} = 0, \\
-P^*_{ss} \leq 0, s\in\mathbb{Z}_{[0,N]}, \\
-P^*_{s,2N-s} \leq 0, s\in\mathbb{Z}_{[N,2N]}, \\
\nu^*_s\geq 0, \forall s, \\
\nu^*_s P^*_{ss} = 0, s\in\mathbb{Z}_{[0,N]}, \\
\nu^*_s P^*_{s,2N-s} =0, s\in\mathbb{Z}_{[N,2N]}.
\end{cases}
\end{equation*}
Here $\mathbb{Z}_{[0,N]}$ ($\mathbb{Z}_{[N,2N]}$) is the set of integers from 0 ($N$) to $N$ ($2N$). As a matter of fact, 
the first two conditions are equivalent when $s=N$, so does $P^*_{ss}$ and $P^*_{s,2N-s}$.  

Now we apply these conditions to find the optimal values of $P^*_{ss}$ and $P^*_{s,2N-s}$. The conditions 
\begin{equation*}
\begin{cases}
s^2-\lambda^*_1 s-\lambda^*_0+\nu^*_s=0,  \\
\nu^*_s\geq 0  
\end{cases}
\end{equation*}
for $s\in\mathbb{Z}_{[0,N]}$ imply that in this case
\begin{equation}
f_0(s) := s^2-\lambda^*_1 s-\lambda^*_0\leq 0    
\end{equation}
Similarly, in the case that $s\in\mathbb{Z}_{[N,2N]}$, we can also obtain 
\begin{equation}
f_1(s) := s^2-\left(4N+\lambda^*_1\right)s-\lambda^*_0+4N^2\leq 0  
\end{equation}
via the conditions 
\begin{equation*}
\begin{cases}
(2N-s)^2-\lambda^*_1 s-\lambda^*_0+\nu^*_s=0,  \\
\nu^*_s\geq 0. 
\end{cases}
\end{equation*}
To simplify the discussion, in the following we take $f_0(s)$ and $f_1(s)$ as two continuous functions in the regime 
$s\in[0,N]$ and $s\in [N,2N]$. Notice that when $f_0(s)$ or $f_1(s)$ is less than zero, the corresponding $\nu^{*}_s$ has 
to be larger than zero since $f_{0,1}(s)+\nu^{*}_s=0$. In the meantime, in the KKT conditions $\nu^{*}_sP^{*}_{ss}=0$ 
($s\in\mathbb{Z}_{[0,N]}$) and $\nu^{*}_sP^{*}_{s,2N-s}=0$ ($s\in\mathbb{Z}_{[N,2N]}$), and when $\nu^{*}_s>0$, the only 
possible values of $P^{*}_{ss}$ and $P^{*}_{s,2N-s}$ are zero. Hence, the nonzero $P^{*}_{ss}$ and $P^{*}_{s,2N-s}$ must 
correspond to a vanishing $f_{0,1}(s)$. Notice that if no zero value exists for both $f_0(s)$ in the regime $s\in [0,N]$ 
and $f_1(s)$ in the regime $s\in [N,2N]$, then the optimal solution $P^*_{ss}$ and $P^{*}_{s,2N-s}$ are always zero, which 
is a trivial solution and is not considered in the following discussion. 

\begin{figure*}[tp]
\centering\includegraphics[width=16cm]{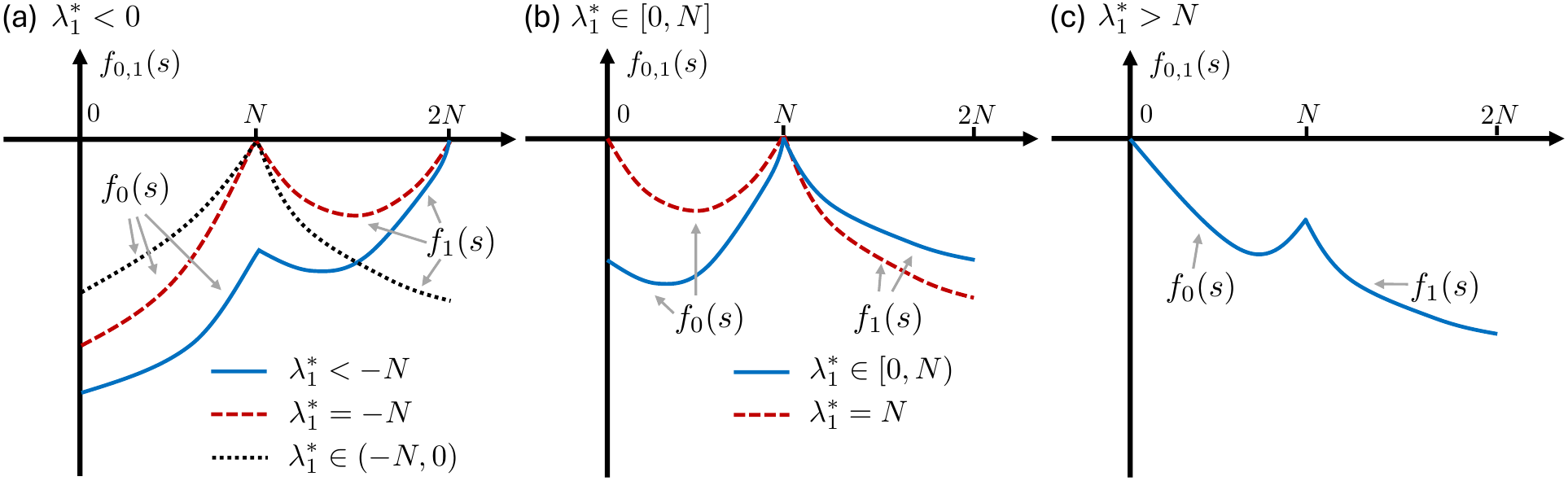}
\caption{Behaviors of $f_{0}(s)$ and $f_1(s)$ for (a) $\lambda^*_1<0$, 
(b) $\lambda^*_1\in [0,N]$, and (c) $\lambda^*_1>N$. }
\label{fig:apx_lambda}
\end{figure*}

Since both $f_0(s)$ and $f_1(s)$ are quadratic functions, the value of $f_{0,1}(s)$ can only be zero at the boundaries, of 
which the positions rely on the positions of the symmetric axes. It is easy to see that the symmetric axes for $f_0(s)$ and 
$f_1(s)$ are $s=\lambda^*_1/2$ and $s=2N+\lambda^*_1/2$, which means their positions are fully determined by the value of 
$\lambda^*_1$. Hence, the discussion below is divided into three parts according to the value of $\lambda^*_1$, i.e., 
$\lambda^*_1<0$, $\lambda^*_1\in\left[0,N\right]$ and $\lambda^*_1>N$, as illustrated in Fig.~\ref{fig:apx_lambda}.

In case that $\lambda^*_1<0$, the axis $s=\lambda^*_1/2$ is at the left side of $y$ axis, indicating that $f_0(s)$ can 
only be zero at the right boundary $s=N$. And when it happens [dotted black and dashed red lines in Fig.~\ref{fig:apx_lambda}(a)], 
noticing that $f_0(N)$ is always equivalent to $f_1(N)$, one can see that the symmetric axis $s=2N+\lambda^*_1/2$ cannot be at the 
left side of $s=3N/2$ since $f_1(s)$ has to be nonpositive in the regime $s\in [N,2N]$. When the symmetric axis is $s=3N/2$, 
i.e., $\lambda^*_1=-N$, $f_1(s)$ also reaches the value of zero at the right boundary $s=2N$. In this case, both $P^*_{NN}$ and 
$P^*_{2N,0}$ are nonzero, which means $c_{N0}$ and $c_{NN}$ is not zero. Together with the condition in Eq.~(\ref{eq:apx_optcond1}), 
one can immediately obtain the form of the optimal probe state in this case 
\begin{equation}
|c_{N0}|(e^{i\theta_1}\ket{0N}+e^{i\theta_2}\ket{N0})+|c_{NN}|\ket{NN}
\end{equation}
with $\theta_1,\theta_2\in[0,2\pi)$ two relative phases. Further utilizing the condition of normalization and the average particle 
number, $|c_{N0}|$ and $|c_{NN}|$ satisfy the equations
\begin{align}
2|c_{N0}|^2+|c_{NN}|^2 &=1, \\
2N\left(|c_{N0}|^2+|c_{NN}|^2\right) &=\bar{n}. 
\end{align}
The corresponding solutions are  
\begin{equation}
|c_{N0}|=\sqrt{\frac{2N-\bar{n}}{2N}},~~|c_{NN}|=\sqrt{\frac{\bar{n}-N}{N}}.    
\end{equation}
These solutions indicate that they are only physical when $\bar{n}\geq N$. Hence, when $\bar{n}\geq N$, one optimal probe state is 
of the form 
\begin{equation}
\sqrt{\frac{2N-\bar{n}}{2N}}\left(e^{i\theta_1}\ket{0N}+e^{i\theta_2}\ket{N0}\right)
+\sqrt{\frac{\bar{n}-N}{N}}\ket{NN}. 
\label{eq:apx_sol1}
\end{equation}

When the axis $s=2N+\lambda^*_1/2$ is at the right side of $s=3N/2$, $f_1(s)$ cannot be zero at the right boundary, indicating that 
the only nonzero $P^*_{ss}$ is just $P^*_{NN}$, i.e., $c_{N0}$. Therefore, the optimal probe state in 
this case is of the form 
\begin{equation}
|c_{N0}|\left(\ket{0N}+e^{i\theta}\ket{N0}\right)    
\end{equation}
with $\theta\in[0,2\pi)$ a relative phase. Utilizing the normalization condition, it can be expressed by 
\begin{equation}
\frac{1}{\sqrt{2}}\left(\ket{0N}+e^{i\theta}\ket{N0}\right). 
\label{eq:apx_sol2}
\end{equation}
One should notice that in this case the average particle number is $N$. Hence, this solution is only legitimate when $\bar{n}=N$. 
As a matter of fact, the solution in Eq.~(\ref{eq:apx_sol1}) reduces to Eq.~(\ref{eq:apx_sol2}) when $\bar{n}=N$. Therefore, these 
two solutions can be unified in Eq.~(\ref{eq:apx_sol1}). 

If $f_0(N)$ is not zero [solid blue line in Fig.~\ref{fig:apx_lambda}(a)], the only possible zero value for $f_1(s)$ is $f_1(2N)$. 
Hence, only $P^*_{2N,0}$ can be nonzero in this case, which means $c_{NN}$ is nonzero. However, 
one can see that the corresponding form of probe state is $c_{NN}\ket{NN}$, and the information of $\phi$ cannot be encoded into it 
due to the fact that $e^{i\phi J_z}\ket{NN}=\ket{NN}$. Hence, the optimal solution given in this case is unphysical. 

In the case that $\lambda^*_1\in [0,N]$, the symmetric axis $s=2N+\lambda^*_1/2\geq 2N$, indicating that the only possible zero value 
for $f_1(s)$ is its left boundary $s=N$, as illustrated in Fig.~\ref{fig:apx_lambda}(b). In this case, the left boundary of $f_0(s)$ 
can either be zero [dashed red line in Fig.~\ref{fig:apx_lambda}(b)] or not [solid blue line in Fig.~\ref{fig:apx_lambda}(b)], 
corresponding to $\lambda^*_1=N$ and $\lambda^*_1\in[0,N)$, respectively. Hence, when $\lambda^*_1=N$, $P^*_{00}$ and $P^*_{NN}$ 
are nonzero, i.e., $c_{00}$ and $c_{N0}$ are nonzero. Together with the condition in Eq.~(\ref{eq:apx_optcond1}), the corresponding 
optimal probe state reads 
\begin{equation}
|c_{00}|\ket{00}+|c_{N0}|(e^{i\theta_1}\ket{0N}+e^{i\theta_2}\ket{N0}).    
\end{equation}
Utilizing the normalization and average particle number conditions, the state above can be expressed by 
\begin{equation}
\sqrt{\frac{N-\bar{n}}{N}}\ket{0 0}+\sqrt{\frac{\bar{n}}{2N}}
\left(e^{i\theta_1}\ket{0 N} +e^{i\theta_2}\ket{N 0}\right),
\label{eq:apx_sol3}
\end{equation}
which is only legitimate when $\bar{n}\leq N$. In the case that $\lambda^*_1\in[0,N)$, the only zero point for both $f_0(s)$ and 
$f_1(s)$ is at $s=N$, indicating that only $P^*_{NN}$ can be nonzero. In this case the optimal state is also in the form of 
Eq.~(\ref{eq:apx_sol2}), and can also be covered by Eq.~(\ref{eq:apx_sol3}) by taking $\bar{n}=N$. 

In the case that $\lambda^*_1>N$, the symmetric axis $s=\lambda^*/2$ is at the right side of $s=N/2$, as illustrated in 
Fig.~\ref{fig:apx_lambda}(c), indicating that only the left boundary is possible to be zero for $f_0(s)$. In the meantime, 
the symmetric axis for $f_1(s)$ is still larger than $2N$, and hence $f_1(s)$ cannot be zero in the regime $s\in [N,2N]$. 
Thus, in this case only $P^*_{00}$ can be zero, which corresponds to the state $c_{00}\ket{00}$. It is easy to see that 
as in $\ket{NN}$, the phase difference $\phi$ cannot be encoded in the state $\ket{00}$, and this solution is unphysical. 

With the aforementioned discussions, the optimal probe states are solved without fully solving the KKT conditions. In summary, 
when $\bar{n}\in(0,N]$, the optimal probe state reads 
\begin{equation}
\sqrt{\frac{N-\bar{n}}{N}}\ket{0 0}+\sqrt{\frac{\bar{n}}{2N}}
\left(e^{i\theta_1}\ket{0 N} +e^{i\theta_2}\ket{N 0}\right)
\label{eq:apx_linOptStat}
\end{equation}
and when $\bar{n}\in[N,2N)$, the optimal probe state is
\begin{equation}
\sqrt{\frac{2N-\bar{n}}{2N}}\left(e^{i\theta_1}\ket{0 N}+e^{i\theta_2}\ket{N 0}\right)
+\sqrt{\frac{\bar{n}-N}{N}}\ket{N N}. 
\label{eq:apx_linOptStat1}
\end{equation}
The theorem is then proved. \hfill $\blacksquare$

Utilizing Eq.~(\ref{eq:apx_QFI_linear}), the QFI for the state (\ref{eq:apx_linOptStat}) is in the form 
\begin{equation}
F = \bar{n}N,
\end{equation} 
and for the state (\ref{eq:apx_linOptStat1}) it is 
\begin{equation}
F = N(2N-\bar{n}). 
\end{equation}

\section{Proofs of Theorems 2-4 and corresponding corollaries}
\label{sec:theorem2}

In this section we provide the thorough proof of the theorems with the nonlinear phase shifts. For two nonlinear phase shifts, 
the operator for the phase shift reads 
\begin{align}
& e^{i[\phi_a(a^{\dagger}a)^2+\phi_b(b^{\dagger}b)^2]} \nonumber \\
=& e^{i\frac{1}{2}\phi_{\mathrm{tot}}[(a^{\dagger}a)^2+(b^{\dagger}b)^2]}
e^{i\frac{1}{2}\phi[(a^{\dagger}a)^2-(b^{\dagger}b)^2]} \nonumber \\ 
=& e^{i\frac{1}{2}\phi_{\mathrm{tot}}[(a^{\dagger}a)^2+(b^{\dagger}b)^2]}
e^{i\phi n J_z},
\end{align}
where $\phi_{\mathrm{tot}}=\phi_a+\phi_b$ and $\phi=\phi_a-\phi_b$. Hence, the parameterized state is 
\begin{equation}
\ket{\psi} = e^{i\frac{1}{2}\phi_{\mathrm{tot}}[(a^{\dagger}a)^2+(b^{\dagger}b)^2]}
e^{i\phi n J_z}\ket{\psi_{\mathrm{in}}}.    
\end{equation}
The corresponding QFI then reads 
\begin{align}
F =& 4\left(\bra{\psi_{\mathrm{in}}}n^2 J^2_z\ket{\psi_{\mathrm{in}}}
-|\bra{\psi_{\mathrm{in}}}n J_z \ket{\psi_{\mathrm{in}}}|^2\right) \nonumber \\
=& \sum^N_{i,j=0}P_{ij}(i^2\!-\!j^2)^{2}\!-\!\sum^N_{i,j,k,l=0}\!\!P_{ij}P_{kl}(i^2\!-\!j^2)(k^2\!-\!l^2),
\label{eq:apx_QFI_nonlinear}
\end{align}
where $P_{ij}:=|c_{ij}|^2$.

As in the linear case, here we rewrite $P_{ij}$ to $P_{s,2d}$ with $s=i+j$ and $d=(i-j)/2$, and the 
optimization problem can then be expressed by 
\begin{eqnarray}
\underset{P_{s,2d}}{\mathrm{max}} & &~4\left[\sum^{2N}_{s=0} \sum^{x_s}_{d=-x_s} s^2d^2P_{s,2d}
-\left(\sum^{2N}_{s=0} s \sum^{x_s}_{d=-x_s} d P_{s,2d} \right)^2\right], \nonumber \\
\text{s.t.} & & \begin{cases}
\sum^{x_s}_{d=-x_s}P_{s,2d}\in[0,1], \forall s, \\
\sum^{2N}_{s=0}\sum^{x_s}_{d=-x_s}P_{s,2d} =1,  \\
\sum^{2N}_{s=0} \sum^{x_s}_{d=-x_s} sP_{s,2d}=\bar{n}, 
\end{cases} 
\end{eqnarray}
where $x_s$ is defined the same as that in the previous section, i.e., $x_s:=s/2$ for $s\in\mathbb{Z}_{[0,N]}$ and 
$x_s:=N-s/2$ for $s\in\mathbb{Z}_{[N,2N]}$. Notice that 
\begin{align}
& \sum^{2N}_{s=0} \sum^{x_s}_{d=-x_s} s^2d^2P_{s,2d}
-\left(\sum^{2N}_{s=0} s \sum^{x_s}_{d=-x_s} d P_{s,2d}\right)^2 \nonumber \\  
\leq & \sum^{2N}_{s=0} \sum^{x_s}_{d=-x_s} s^2d^2P_{s,2d},
\end{align}
and the equality is attained when $\sum^{x_s}_{d=-x_s}d P_{s,2d}=0$. With the condition $\sum^{x_s}_{d=-x_s} 
d P_{s,2d}=0$, one can further have 
\begin{align}
\sum^{x_s}_{d=-x_s} d^2P_{s,2d}=\sum^{x_s}_{d=-x_s} d^2P_{s,2d}-\left(\sum^{x_s}_{d=-x_s} dP_{s,2d}\right)^2
\label{eq:apx_nonvartp}
\end{align}
which is just the variance of $d$ with respect to the probability distribution $\{P_{s,2d}\}^{x_s}_{d=-
x_s}$, similarly to the linear case. Hence, according to the Popoviciu’s inequality on variances~\cite{Popviciu1935}, 
the maximum value of Eq.~(\ref{eq:apx_nonvartp}) can only be attained when 
\begin{align}
P_{s,2d} &=0,~\mathrm{for}~d\neq -x_s,x_s, \\
P_{s,-2x_s} &=P_{s,2x_s}.
\end{align}
Same as in the linear case, the second condition is equivalent to 
\begin{equation}
\begin{cases}
|c_{0s}|^2= |c_{s0}|^2, & s\in\mathbb{Z}_{[0,N]},\\
|c_{s-N,N}|^2= |c_{N,s-N}|^2, & s\in\mathbb{Z}_{[N,2N]}.
\label{eq:apx_optcond2}
\end{cases}
\end{equation}
Combining these two conditions, the optimization problem can be further rewritten into 
\begin{align*}
& \underset{P_{ss},P_{s,2N-s}}{\mathrm{max}} 2\left[\sum^{N}_{s=0}s^4 P_{ss}
+\sum^{2N}_{s=N+1}s^2(2N-s)^2 P_{s,2N-s}\right] \nonumber \\
& \text{s.t.} \begin{cases}
P_{ss},P_{s,2N-s}\in\left[0,\frac{1}{2}\right], \forall s\neq 0, 2N, \\
P_{00},P_{2N,0}\in [0,1], \\
\sum^{N}_{s=0} P_{ss}\!+\!\sum^{2N}_{s=N+1} P_{s,2N-s}=\frac{1}{2}\left(1\!+\!P_{00}\!+\!P_{2N,0}\right)\!, \\
\sum^{N}_{s=0}s P_{ss}\!+\!\sum^{2N}_{s=N+1}s P_{s,2N-s}= \frac{\bar{n}}{2}+NP_{2N,0}, 
\end{cases} 
\end{align*}
where the maximization problem is equivalent to the minimization problem as follows: 
\begin{equation*}
\underset{P_{ss},P_{s,2N-s}}{\mathrm{min}}~-2\left[\sum^{N}_{s=0}s^4 P_{ss}
+\sum^{2N}_{s=N+1}s^2(2N-s)^2 P_{s,2N-s}\right]\!.  
\end{equation*}

The Lagrangian function for the expression above reads 
\begin{align}
\mathcal{L}=&-2\sum^N_{s=0} s^4 P_{ss}-2\sum^{2N}_{s=N+1} s^2(2 N-s)^2 P_{s,2N-s} \nonumber \\
&-\!2\!\sum^{N}_{s=1} {\nu}_s P_{ss}\!-\!2\!\!\sum^{2N-1}_{s=N+1} {\nu}_s P_{s,2N-s}\!-\!\nu_0 P_{00}\!-\!\nu_{2N} P_{2N,0} \nonumber \\
&\!+\!\lambda_0\!\left(\!\!P_{00}+2\sum^{N}_{s=1} P_{ss}+2\!\!\!\sum^{2N-1}_{s=N+1}\!\!\!\! P_{s,2N-s}+P_{2N,0}-1\!\!\right) \nonumber \\
&\!+\!\lambda_1\!\left(2\sum^{N}_{s=0} s P_{ss}+2\!\sum^{2N-1}_{s=N+1} s P_{s,2N-s}\!+\!2N P_{2N,0}\!-\!\bar{n}\right)\!,  
\end{align}
and the corresponding KKT conditions are 
\begin{equation}
\begin{cases}
s^4-\lambda^*_1 s-\lambda^*_0+\nu^*_s=0, s\in\mathbb{Z}_{[0,N]}, \\
s^2(2 N-s)^2-\lambda^*_1 s-\lambda^*_0+\nu^*_s=0, s\in\mathbb{Z}_{[N,2N]},\\
\sum^{N}_{s=0} P^*_{ss}+\!\sum^{2N}_{s=N+1} \!P^*_{s,2N-s}=\!\frac{1}{2}\left(1\!+\!P_{00}\!+\!P_{2N,0}\right), \\
\sum^{N}_{s=0} s P^*_{ss}+\sum^{2N}_{s=N+1} s P^*_{s,2N-s}-\frac{\bar{n}}{2}\!-\!N P_{2N,0} = 0, \\
-P^*_{ss} \leq 0, s\in\mathbb{Z}_{[0,N]}, \\
-P^*_{s,2N-s} \leq 0, s\in\mathbb{Z}_{[N,2N]}, \\
\nu^*_s\geq 0, \forall s, \\
\nu^*_s P^*_{ss}=0, s\in\mathbb{Z}_{[0,N]}, \\
\nu^*_s P^*_{s,2N-s}=0, s\in\mathbb{Z}_{[N,2N]}.
\end{cases}
\end{equation}
Now define two continuous functions 
\begin{equation}
g_0(s):=s^4-\lambda^*_1 s-\lambda^*_0  
\end{equation}
for $s\in [0,N]$ and 
\begin{equation}
g_1(s):=s^2(2 N-s)^2-\lambda^*_1 s-\lambda^*_0
\end{equation}
for $s\in [N,2N]$. $g_0(s)=g_1(s)$ when $s=N$. As in the linear case, $P^*_{ss}$ is only possible to be nonzero when 
$g_0(s)=0$ due to the fact that $g_0(s)+\nu^*_s=0$, $\nu^*_s\geq 0$, and $\nu^*_s P^*_{ss}=0$ for $s\in\mathbb{Z}_{[0,N]}$. 
Same relation exists between $P^*_{s,2N-s}$ and $g_1(s)$ for $s\in\mathbb{Z}_{[N,2N]}$. 

\begin{figure}[tp]
\centering\includegraphics[width=8.cm]{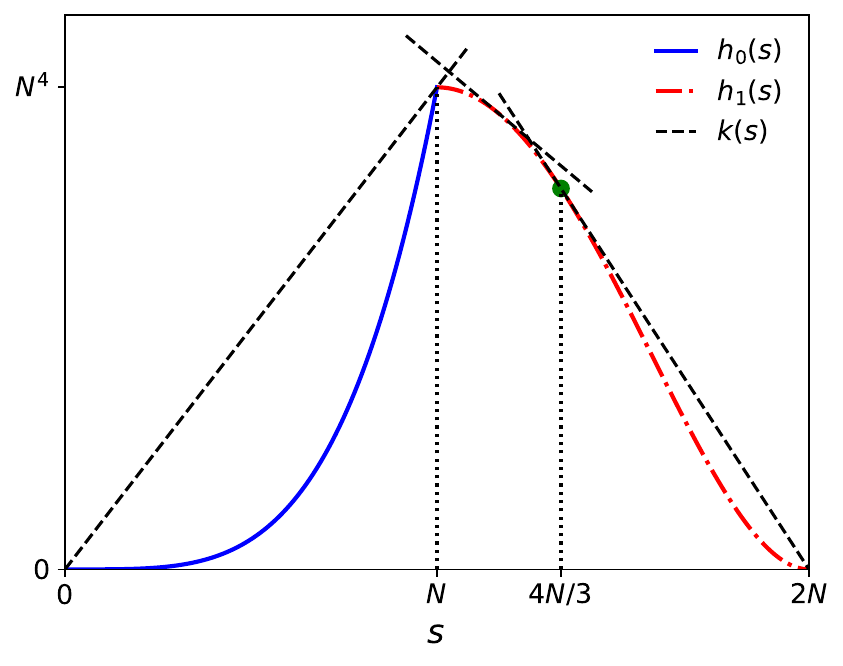}
\caption{Schematic of locating the zero points for $g_0(s)$ and $g_1(s)$. The solid 
blue line, dash-dotted red line, and dashed black represent the functions $h_0(s)$, 
$h_1(s)$ and, $k(s)$, respectively.}
\label{fig:apx_non_KKT}
\end{figure}

Different from the linear case, here both $g_0(s)$ and $g_1(s)$ are proportional to $s^4$, indicating that it is 
not easy to solve their zero points analytically. To find the zero points, we further denote continuous functions 
$h_0(s):=s^4$ for $s\in [0,N]$, $h_1(s):=s^2(2N-s)^2$ for $s\in [N,2N]$, and $k(s):=\lambda^*_1s+\lambda^*_0$ for 
all values $s$, i.e., $s\in [0,2N]$. Utilizing these functions, the zero points of $g_0(s)$ and $g_1(s)$ can be found 
from the geometric perspective given in Fig.~\ref{fig:apx_non_KKT}. The zero points of $g_0(s)$ [$g_1(s)$] is nothing 
but the intersection between $h_0(s)$ [$h_1(s)$] and $k(s)$. Due to the fact that both $h_0(s)$ and $h_1(s)$ are 
no larger than $k(s)$, i.e., the line of $k(s)$ (dashed black line) has to be always on top of the lines of $h_0(s)$ 
(solid blue line) and $h_1(s)$ (dash-dotted red line), the only possible intersections between $k(s)$ and $h_0(s)$ 
are the original point and the point of $h_0(N)$, as shown in the figure. Therefore, the corresponding nonzero 
$P^*_{ss}$ in this case are $P^{*}_{00}$ and $P^*_{NN}$, i.e., $|c_{00}|$ and $|c_{N0}|$, which means the optimal 
probe state can be expressed by 
\begin{equation}
|c_{00}|\ket{00}+|c_{N0}|(e^{i\theta_1}\ket{0N}+e^{i\theta_2}\ket{N0})    
\end{equation}
with $\theta_1,\theta_2\in[0,2\pi)$ two relative phases. Utilizing the normalization and average particle 
number conditions, $|c_{00}|$ and $|c_{N0}|$ are fully determined, the specific form of the optimal probe 
state reads 
\begin{equation}
\sqrt{\frac{N-\bar{n}}{N}}\ket{0 0} 
+\sqrt{\frac{\bar{n}}{2N}} \left(e^{i\theta_1}\ket{0 N}
+e^{i\theta_2}\ket{N 0}\right),
\end{equation}
where $\bar{n}\leq N$. Notice that it is possible that only one intersection, either $h_0(0)$ or $h_0(N)$, exists in 
this case. However, the state corresponding to the nonzero $P^{*}_{00}$ is $\ket{00}$, which cannot encode the 
phases. In the meantime, the state corresponding to the nonzero $P^*_{NN}$ is contained by the expression above by 
taking $\bar{n}=N$. Theorem 2 is then proved. \hfill $\blacksquare$

Regarding Theorems 3 and 4, the situation between $h_1(s)$ and $k(s)$ is similar. As a matter of fact, $h_1(s)$ is first concave 
and then convex from $N$ to $2N$. On the concave part, the legitimate intersection between $h_1(s)$ and $k(s)$ only exists 
when $k(s)$ is the tangent line of $h_1(s)$ due to the fact that $h_1(s)\leq k(s)$. However, this legality stops when the intersection 
between the tangent line and $s$ axis reaches $2N$, as shown in Fig.~\ref{fig:apx_non_KKT}. When it happens, the value of 
$s$ for the intersection between $h_1(s)$ and $k(s)$ (green dot in the figure) is $4N/3$. In the meantime, similarly to 
$h_0(s)$, in the regime $s\in [4N/3,2N]$, the intersections between $h_1(s)$ and $k(s)$ can only the point of $h_1(4N/3)$ 
and $h_1(2N)$. Hence, the nonzero $P^*_{s,2N-s}$ could be those $P^*_{s,2N-s}$ for $s\in [N,4N/3]$, and $P^*_{4N/3,2N/3}$ 
and $P^*_{2N,0}$ for $s\in [4N/3,2N]$. In the case that $s\in [N,4N/3]$, $P^*_{s,2N-s}$ corresponds to the coefficient 
$|c_{N,s-N}|$, which means the form of optimal probe state in this case reads 
\begin{equation}
|c_{N,s-N}|\left(\ket{s-N,N}+e^{i\theta}\ket{N,s-N}\right).  
\end{equation}
Here $\theta\in [0,2\pi)$ is a relative phase and we assumed that the Fock states are continuous states. In the case that 
$s\in[4N/3,2N]$, $P^*_{4N/3,2N/3}$ and $P^*_{2N,0}$ correspond to $|c_{N,s-N}|$ and $|c_{NN}|$, and the optimal probe 
state can be expressed by 
\begin{equation}
\left|c_{N,\frac{1}{3}N}\!\right|\left(\!e^{i\theta_1}\ket{\frac{1}{3}N,N}
+e^{i\theta_2}\ket{N,\frac{1}{3}N}\!\right)\!+\!|c_{NN}|\ket{NN}  
\end{equation}
with $\theta_1,\theta_2$ two relative phases. Utilizing the normalization and average particle number conditions,  
these two states can be specifically written as 
\begin{equation}
\frac{1}{\sqrt{2}}\left(\ket{\bar{n}-N, N}+e^{i\theta}\ket{N,\bar{n}-N}\right)    
\end{equation}
for $\bar{n}\in [N,4N/3]$ and 
\begin{align}
&\sqrt{\frac{3(2N-\bar{n})}{4N}}\left(e^{i\theta_1} \ket{\frac{1}{3}N, N}
+e^{i\theta_2}\ket{N, \frac{1}{3}N}\right) \nonumber \\
& + \sqrt{\frac{3\bar{n}-4N}{2N}} \ket{NN} 
\end{align} 
for $\bar{n}\in [4N/3,2N]$. Similarly to the discussion of $h_0(s)$, it is possible that only one point between $P^{*}_{4N/3,2N/3}$ 
and $P^*_{2N,0}$ is nonzero for $s\in[4N/3,2N]$, however, $P^*_{2N,0}$ corresponds to $\ket{NN}$, which cannot encode the phases, 
and the state corresponding to $P^{*}_{4N/3,2N/3}$ is already contained in the expression above. 

In summary, taking into account the continuous Fock states assumption, the optimal probe states for nonlinear phase shifts read 
\begin{widetext}
\begin{equation}
\begin{cases}
\sqrt{\frac{N-\bar{n}}{N}}\ket{0 0}+\sqrt{\frac{\bar{n}}{2N}}\left(e^{i\theta_1}\ket{0 N}+e^{i\theta_2}
\ket{N 0}\right), & \bar{n}\in(0,N],\\
\frac{1}{\sqrt{2}}\left(\ket{\bar{n}-N, N} 
+e^{i\theta}\ket{N, \bar{n}-N}\right), & \bar{n}\in\left [N,\frac{4N}{3}\right], \\
\sqrt{\frac{3(2N-\bar{n})}{4N}}\left(e^{i\theta_1} 
\ket{\frac{1}{3}N, N} +e^{i\theta_2}\ket{N,\frac{1}{3}N}\right)
+\sqrt{\frac{3\bar{n}-4N}{2N}} \ket{NN}, & \bar{n}\in\left[\frac{4N}{3},2N\right).
\end{cases}
\label{eq:apx_NonlinOptstat}
\end{equation}
\end{widetext}
Utilizing Eq.~(\ref{eq:apx_QFI_nonlinear}), the QFIs for above states are   
\begin{equation}
F=\begin{cases}
\bar{n}N^3, & \bar{n}\in(0,N],\\
\bar{n}^2 (2N-\bar{n})^2, &  \bar{n}\in[N,\frac{4N}{3}], \\
\frac{32}{27} (2N-\bar{n})N^3, & \bar{n}\in[\frac{4N}{3},2N).
\end{cases}
\end{equation}

As we constantly emphasized, the assumption of continuous Fock states are used in the expressions above, namely, 
it is assumed that $\ket{\bar{n}-N}$ and $\ket{\frac{1}{3}N}$ are valid Fock states. However, they are actually not 
when $\bar{n}-N$ and $N/3$ are not integers. Hence, for the most general case that $\bar{n}$ and $N/3$ are not 
integers, the true OFPS have to be further discussed. In the following we provide thorough discussions on the 
true solutions of OFPS when $\bar{n}$ is not an integer. 

Due to the previous discussions, the types of intersections between $h_1(s)$ and $k(s)$ are 
different in the regimes $s\in [N,4N/3]$ and $s\in [4N/3,2N]$, as shown in Fig.~\ref{fig:apx_non_KKT}. When the 
condition that $s\in\mathbb{Z}$ ($\mathbb{Z}$ is the set of integers) is involved, the tangent line of $h_1(s)$ 
for a continuous $s$ may not be accessible. Since $4N/3$ may not be an integer, we rewrite these two regimes into 
$[N, \lfloor 4N/3\rfloor]$ and $[\lfloor 4N/3\rfloor+1,2N]$. Here $\lfloor \cdot\rfloor$ is the floor function. 

\begin{figure}[tp]
\centering\includegraphics[width=8.cm]{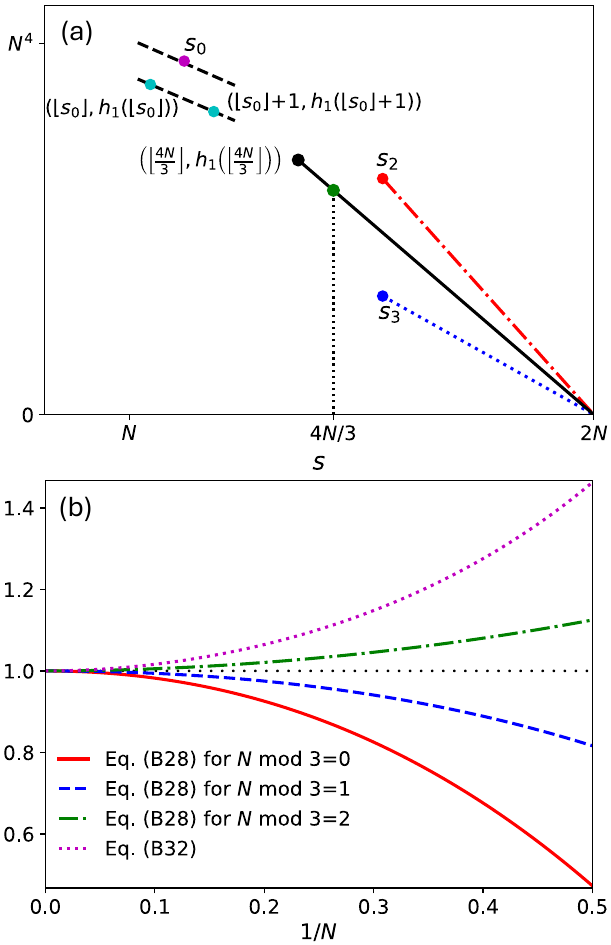}
\caption{(a) Schematic of locating the legitimate intersections between $h_1(s)$ and $k(s)$.
(b) Monotonicity performance of Eq.~(\ref{eq:apx_4N3}) for $N\bmod 3=0,1,2$, and that of 
Eq.~(\ref{eq:apx_lb4N3}) for $\lfloor s_0\rfloor= 4N/3-1$.}
\label{fig:apx_non_PhyAnaly}
\end{figure}

We first discuss the regime $s\in [N, \lfloor 4N/3\rfloor]$. In this regime, all points could be the intersection when 
the integer condition is not involved. Now let us denote $s_0$ as the intersection between $h_1(s)$ and its tangent 
line, then when the integer condition is considered, the possible intersections are actually $(\lfloor s_0\rfloor,
h_1(\lfloor s_0\rfloor))$ and $(\lfloor s_0\rfloor+1,h_1(\lfloor s_0\rfloor+1))$, as shown in 
Fig.~\ref{fig:apx_non_PhyAnaly}(a). Three cases exist here: either of these two points is the intersection or both 
of them are. Now let us first check whether both of them can be the intersections simultaneously. If this case is a 
legitimate one, the intersection between the line through these two points (dashed black line) and the $s$ axis has 
to be on the right side of the point $(2N,0)$. As a matter of fact, this line can be expressed by 
\begin{align}
& \left[h_1(\lfloor s_0\rfloor+1)-h_1(\lfloor s_0\rfloor)\right] s 
-\lfloor s_0\rfloor h_1(\lfloor s_0\rfloor+1) \nonumber \\
& +(\lfloor s_0\rfloor+1)h_1(\lfloor s_0\rfloor), 
\end{align}
where $h_1(\lfloor s_0\rfloor)=\lfloor s_0\rfloor^2(2N-\lfloor s_0\rfloor)^2$ and 
$h_1(\lfloor s_0\rfloor+1)=(\lfloor s_0\rfloor+1)^2(2N-\lfloor s_0\rfloor-1)^2$. 
It is easy to see that the value of $s$ for the intersection between the line above and the $s$ axis is 
\begin{equation}
\lfloor s_0\rfloor+\frac{h_1(\lfloor s_0\rfloor)}{h_1(\lfloor s_0\rfloor)-h_1(\lfloor s_0\rfloor+1)}.   
\label{eq:apx_xcross}
\end{equation}
If the value of Eq.~(\ref{eq:apx_xcross}) is no less than $2N$, the inequality 
\begin{equation}
\frac{h_1(\lfloor s_0\rfloor)}{h_1(\lfloor s_0\rfloor)-h_1(\lfloor s_0\rfloor+1)}\geq 2N-\lfloor s_0\rfloor
\end{equation}
must hold. Due to the fact that $h_1(s)$ is a monotonic decreasing function, $h_1(\lfloor s_0\rfloor)\geq 
h_1(\lfloor s_0\rfloor+1)$, which means the inequality above can be further rewritten into 
\begin{equation}
\frac{h_1(\lfloor s_0\rfloor+1)}{h_1(\lfloor s_0\rfloor)}
\geq \frac{2N-\lfloor s_0\rfloor-1}{2N-\lfloor s_0\rfloor}.
\end{equation}
It can be seen that $2N-\lfloor s_0\rfloor-1\geq 2N/3-1$ since $\lfloor s_0\rfloor\leq \lfloor 4N/3\rfloor\leq 4N/3$, 
which means $2N-\lfloor s_0\rfloor-1\geq 0$ for $N\geq 2$. When $N=1$, $\lfloor s_0\rfloor=1$ and  $2N-\lfloor s_0\rfloor-1=0$, 
the inequality above naturally holds since $h_1(s)$ is always nonnegative. Once it holds, the inequality above can further 
reduce to 
\begin{equation}
\frac{\left(\lfloor s_0\rfloor+1\right)^2(2N-\lfloor s_0\rfloor-1)}{\lfloor s_0\rfloor^2(2N-\lfloor s_0\rfloor)}\geq 1.   
\label{eq:apx_s0}
\end{equation}
The lefthand term can be written as 
\begin{equation}
\left(1+\frac{1}{\lfloor s_0\rfloor}\right)^2\left(1-\frac{1}{2N-\lfloor s_0\rfloor}\right),   
\label{eq:apx_s0tp1}
\end{equation}
which is obviously a monotonic decreasing function with respect to $\lfloor s_0\rfloor$. 

Recall that $s_0\in [N, \lfloor 4N/3\rfloor]$, the minimum value of the expression above must be attained at 
$\lfloor 4N/3\rfloor$. However, the fact is that for different values of $N$, the expression 
\begin{equation}
\left(1+\frac{1}{\lfloor 4N/3\rfloor}\right)^2\left(1-\frac{1}{N-\lfloor N/3\rfloor}\right)   
\label{eq:apx_4N3}
\end{equation}
is not always no less than 1, which means the inequality (\ref{eq:apx_s0}) does not always hold. When $N\bmod 3 =2$, i.e., 
the remainder of $N$ divided by $3$ is 2, $\lfloor 4N/3\rfloor=(4N-2)/3$ and the expression above reduces to 
\begin{equation}
\left(1+\frac{3/N}{4-2/N}\right)^2\left(1-\frac{3/N}{2+2/N}\right).    
\end{equation}
This expression is a monotonic increasing with respect to $1/N$ [dash-dotted green line in Fig.~\ref{fig:apx_non_PhyAnaly}(b)], 
and thus its minimum value is $1$, which can be attained when $1/N\rightarrow 0$. Hence, in this case the inequality 
(\ref{eq:apx_s0}) always holds for any value of $\lfloor s_0\rfloor$ satisfying $\lfloor s_0\rfloor\leq \lfloor 4N/3\rfloor$, 
indicating that both points $(\lfloor s_0\rfloor, h_1(\lfloor s_0\rfloor))$ and $(\lfloor s_0\rfloor+1,h_1(\lfloor s_0\rfloor+1))$ 
can be the intersections simultaneously. When $N\bmod 3=0$, the expression (\ref{eq:apx_4N3}) reduces to 
\begin{equation}
\left(1+\frac{3/N}{4}\right)^2\left(1-\frac{3/N}{2}\right),     
\end{equation}
and when $N\bmod 3=1$, it reduces to 
\begin{equation}
\left(1+\frac{3/N}{4-1/N}\right)^2\left(1-\frac{3/N}{2+1/N}\right).     
\end{equation}
These two expressions are monotonic decreasing functions with respect to $1/N$ [solid red and dashed blue lines in  
Fig.~\ref{fig:apx_non_PhyAnaly}(b)], and the minimum values are less than 1, indicating that the inequality (\ref{eq:apx_s0}) 
does not always hold. However, in these two cases, the inequality (\ref{eq:apx_s0}) always holds for $\lfloor s_0\rfloor \leq 
\lfloor 4N/3\rfloor-1$. This is due to the fact in this case $\lfloor s_0\rfloor \leq4N/3-1$ for any value of $N$, then  
the lower bound of the expression (\ref{eq:apx_s0tp1}) is 
\begin{equation}
\left(1+\frac{3/N}{4-3/N}\right)^2\left(1-\frac{3/N}{2+3/N}\right).      
\label{eq:apx_lb4N3}
\end{equation}
This expression is a monotonic increasing function with respect to $1/N$ [dotted purple line in Fig.~\ref{fig:apx_non_PhyAnaly}(b)]. 
Since its minimum value with respect to $1/N$ is $1$, this lower bound is no less than $1$, indicating that Eq.~(\ref{eq:apx_s0tp1}) 
is always no less than 1 for $\lfloor s_0\rfloor \leq\lfloor 4N/3\rfloor-1$. Hence, the inequality (\ref{eq:apx_s0}) always holds  
for $\lfloor s_0\rfloor \leq\lfloor 4N/3\rfloor-1$ regardless the value of $N$. 

Based on the analysis above, one can see that the inequality (\ref{eq:apx_s0}) always holds when $\lfloor s_0\rfloor \leq \lfloor 
4N/3\rfloor-1$, and when $\lfloor s_0\rfloor=\lfloor 4N/3\rfloor$, it holds for $N\bmod 3=2$ and does not hold for $N\bmod 3=0,1$. 
The fact that the inequality (\ref{eq:apx_s0}) always holds for $\lfloor s_0\rfloor \leq \lfloor 4N/3\rfloor-1$ means that in 
this regime $P^{*}_{\lfloor s_0\rfloor,2N-\lfloor s_0\rfloor}$ and $P^{*}_{\lfloor s_0\rfloor+1,2N-\lfloor s_0\rfloor-1}$ are 
nonzero, and the corresponding optimal state is of the form 
\begin{align*}
& |c_{N,\lfloor s_0\rfloor+1-N}|\left(\ket{\lfloor s_0\rfloor\!+\!1\!-\!N,N}
\!+\!e^{i\theta_1}\ket{N,\lfloor s_0\rfloor\!+\!1\!-\!N}\right) \\
& +|c_{N,\lfloor s_0\rfloor-N}|\left(e^{i\theta_2}\ket{\lfloor s_0\rfloor-N,N}
+e^{i\theta_3}\ket{N, \lfloor s_0\rfloor-N}\right)
\end{align*}
with $\theta_{1,2,3}\in [0,2\pi)$. Further utilizing the normalization condition and the average particle number condition, one 
can obtain that
\begin{eqnarray}
|c_{N,\lfloor s_0\rfloor+1-N}|^2 &=& \frac{\bar{n}-\lfloor s_0\rfloor}{2}, \\
|c_{N,\lfloor s_0\rfloor-N}|^2 &=& \frac{\lfloor s_0\rfloor+1-\bar{n}}{2}.
\end{eqnarray}
Due to the fact that both $|c_{N,\lfloor s_0\rfloor+1-N}|^2$, $|c_{N,\lfloor s_0\rfloor-N}|^2$ are nonnegative, it is easy to 
see that 
\begin{equation}
\lfloor s_0\rfloor\leq \bar{n} \leq \lfloor s_0\rfloor+1,    
\end{equation}
which indicates that $\lfloor s_0\rfloor=\lfloor\bar{n}\rfloor$  due to the fact that $\bar{n}$ is not an integer. Then the optimal 
probe state can be written as 
\begin{align}
& \sqrt{\frac{\bar{n}-\lfloor \bar{n}\rfloor}{2}}\left(\ket{\lfloor \bar{n}\rfloor\!+\!1\!-\!N, N}
\!+\!e^{i\theta_1}\ket{N, \lfloor\bar{n}\rfloor\!+\!1\!-\!N}\right)\nonumber\\
& +\sqrt{\frac{1-(\bar{n}-\lfloor\bar{n}\rfloor)}{2}}\left(e^{i\theta_2} \ket{\lfloor \bar{n}\rfloor\!-\!N, N}
\!+\!e^{i\theta_3}\ket{N, \lfloor\bar{n}\rfloor\!-\!N}\right),
\label{eq:apx_optstate_tp1}
\end{align}
where $\bar{n}$ satisfies $\lfloor\bar{n}\rfloor\leq \lfloor 4N/3\rfloor-1$. It coincides with the form in Eq.~(\ref{eq:apx_NonlinOptstat}) 
for an integer $\bar{n}$.

Notice that it is possible only one point between $(\lfloor s_0\rfloor, h_1(\lfloor s_0\rfloor))$ and 
$(\lfloor s_0\rfloor+1,h_1(\lfloor s_0\rfloor+1))$ is the intersection. If so, only $P^{*}_{\lfloor s_0\rfloor,2N-\lfloor s_0\rfloor}$ 
or $P^{*}_{\lfloor s_0\rfloor+1,2N-\lfloor s_0\rfloor-1}$ is nonzero. When $P^{*}_{\lfloor s_0\rfloor,2N-\lfloor s_0\rfloor}$ is 
nonzero, the formula of the optimal probe state is 
\begin{equation}
|c_{N,\lfloor s_0\rfloor-N}|\left(\ket{\lfloor s_0\rfloor-N,N}
+e^{i\theta}\ket{N, \lfloor s_0\rfloor-N}\right).    
\end{equation}
The normalization and average particle number conditions give 
\begin{equation}
|c_{N,\lfloor s_0\rfloor-N}|=\frac{1}{\sqrt{2}},~~\bar{n}=\lfloor s_0\rfloor.     
\end{equation}
This means it is only possible when $\bar{n}$ is an integer. The optimal probe state then reads 
\begin{equation}
\frac{1}{\sqrt{2}}\left(\ket{\bar{n}-N,N}+e^{i\theta}\ket{N, \bar{n}-N}\right), 
\label{apx_optstate_tp2}
\end{equation}
which is nothing but the optimal state given in Eq.~(\ref{eq:apx_NonlinOptstat}) for $\bar{n}\in[N,4N/3]$. This result is quite 
reasonable since the optimal state is legitimate in physics as long as $\bar{n}$ is an integer. In the meantime it indicates that 
$P^{*}_{\lfloor s_0\rfloor,2N-\lfloor s_0\rfloor}$ cannot be zero when $\bar{n}$ is not an integer. In the case that $P^{*}_{\lfloor 
s_0\rfloor+1,2N-\lfloor s_0\rfloor-1}$ is nonzero, the same result can be obtained via a similar analysis. Hence, in the regime 
$\lfloor\bar{n}\rfloor\leq \lfloor 4N/3\rfloor-1$, the physical legitimate optimal probe state is the one given in 
Eq.~(\ref{eq:apx_optstate_tp1}). 

In the case that $\lfloor s_0\rfloor=\lfloor 4N/3\rfloor$, the inequality (\ref{eq:apx_s0}) holds for $N\bmod 3=2$, which means 
Eq.~(\ref{eq:apx_optstate_tp1}) is still the optimal probe state. For $N\bmod 3=0,1$, the inequality (\ref{eq:apx_s0}) does not 
hold, indicating that $(\lfloor s_0\rfloor, h_1(\lfloor s_0\rfloor))$ and $(\lfloor s_0\rfloor+1,h_1(\lfloor s_0\rfloor+1))$ 
cannot be the intersections simultaneously. As a matter of fact, only $(\lfloor s_0\rfloor, h_1(\lfloor s_0\rfloor))$ can be 
the intersection in this case and the corresponding formula for the optimal probe state is also in the form of 
Eq.~(\ref{apx_optstate_tp2}), yet an extra requirement is that $\bar{n}$ has to be an integer, which means it cannot be the 
intersection when $\bar{n}$ is not an integer.  Combing this result with the one for $\lfloor s_0\rfloor\leq \lfloor 4N/3\rfloor-1$, 
it can be seen that the optimal probe state for $\lfloor s_0\rfloor\leq\lfloor 4N/3\rfloor$ is just in the form of 
Eq.~(\ref{eq:apx_optstate_tp1}), but $\bar{n}$ satisfies $\lfloor\bar{n}\rfloor\leq \lfloor 4N/3\rfloor$ for $N\bmod 3=2$ 
and $\bar{n}\leq \lfloor 4N/3\rfloor$ for $N\bmod 3=0,1$.

Next we discuss the regime of $s\in [\lfloor 4N/3\rfloor+1,2N)$. For $s\in [4N/3,2N)$ the intersections between $h_1(s)$ and $k(s)$ 
are $(4N/3,h_1(4N/3))$ and $(2N,0)$ when $s$ is continuous. In the case that $s$ is discrete, i.e., $s\in \mathbb{Z}$, 
$(4N/3,h_1(4N/3))$ may not be a legitimate point anymore. Then the position of $(\lfloor 4N/3\rfloor+1,h_1(\lfloor 4N/3\rfloor+1))$ 
becomes crucial. As shown in Fig.~\ref{fig:apx_non_PhyAnaly}(a), if this point is above the line through the points 
$(\lfloor 4N/3\rfloor, h_1(\lfloor 4N/3\rfloor))$ and $(2N,0)$ (solid black line), demonstrated by the point $s_2$ in the plot, 
then $(\lfloor 4N/3\rfloor+1,h_1(\lfloor 4N/3\rfloor+1))$ and $(2N,0)$ can be the intersections simultaneously since all points on 
$h_1(s)$ are under the line through these two points (dash-dotted red line). If $(\lfloor 4N/3\rfloor+1,h_1(\lfloor 4N/3\rfloor+1))$ 
is under the solid black line, demonstrated by the point $s_3$ in the plot, then this point and $(2N,0)$ cannot be the intersections 
simultaneously since the point $(\lfloor 4N/3\rfloor, h_1(\lfloor 4N/3\rfloor))$ is above the line through them (dotted blue line). 
Hence, in this case the legitimate intersections are $(\lfloor 4N/3\rfloor,h_1(\lfloor 4N/3\rfloor))$ and $(2N,0)$. Based on the 
discussions in the case of $\lfloor s_0\rfloor=\lfloor 4N/3\rfloor$, we already know that $(\lfloor 4N/3\rfloor+1,h_1(\lfloor 
4N/3\rfloor+1))$ is $s_2$ when $N\bmod 3=2$ and it is $s_3$ when $N\bmod 3=0,1$. Now we discuss them one by one. 

When $N\bmod 3=2$, $(\lfloor 4N/3\rfloor+1,h_1(\lfloor 4N/3\rfloor+1))$ and $(2N,0)$ can be the intersections simultaneously, 
indicating that $P^*_{\lfloor 4N/3\rfloor+1,2N-\lfloor 4N/3\rfloor-1}$ and $P^*_{2N,0}$ are nonzero. The corresponding form of 
the optimal probe state then reads 
\begin{align}
& \left|c_{N,\lfloor\frac{N}{3}\rfloor+1}\right|\left(e^{i\theta_1}
\ket{\left\lfloor\frac{N}{3}\right\rfloor+1,N}\right. 
\nonumber \\
& +\left.e^{i\theta_2}\ket{N,\left\lfloor\frac{N}{3}\right\rfloor\!+\!1}\right)+|c_{NN}|\ket{NN}.  
\end{align}
Here $\theta_1, \theta_2\in[0,2\pi)$ are two relative phases. Utilizing the normalization and average particle number conditions, 
it becomes 
\begin{align}
& \sqrt{\frac{2N-\bar{n}}{2(N-\left\lfloor\frac{N}{3}\right\rfloor-1)}}
\left(e^{i\theta_1}\ket{\left\lfloor\frac{N}{3}\right\rfloor\!+\!1,N}\right. 
\nonumber \\
& +\left.e^{i\theta_2}\ket{N,\left\lfloor\frac{N}{3}\right\rfloor\!+\!1}\right)
+\sqrt{\frac{\bar{n}-\left\lfloor\frac{4N}{3}\right\rfloor-1}
{N-\left\lfloor\frac{N}{3}\right\rfloor-1}}\ket{NN}, 
\label{eq:apx_Nm3_tp1}
\end{align}
where $\bar{n}$ satisfies $\bar{n}\geq \lfloor 4N/3\rfloor+1$. In the meantime, $P^*_{2N,0}$ cannot be the only nonzero point 
due to the previous discussion. When $P^*_{\lfloor 4N/3\rfloor+1,2N-\lfloor 4N/3\rfloor-1}$ is the only nonzero point, the 
formula of the optimal state is 
\begin{equation*}
\left|c_{N,\lfloor\frac{N}{3}\rfloor+1}\right|
\left(\ket{\left\lfloor\frac{N}{3}\right\rfloor+1,N}
+e^{i\theta}\ket{N,\left\lfloor\frac{N}{3}\right\rfloor+1}\right).
\end{equation*}
Here $\theta\in[0,2\pi)$ is a relative phase. According to the normalization and average particle number conditions, it becomes 
\begin{equation}
\frac{1}{\sqrt{2}}\left(\ket{\left\lfloor\frac{N}{3}\right\rfloor\!+\!1,N}+e^{i\theta}
\ket{N,\left\lfloor\frac{N}{3}\right\rfloor\!+\!1}\right),
\label{eq:apx_Nm3_tp3}
\end{equation}
where $\bar{n}=\lfloor 4N/3 \rfloor+1$. It can be seen that this state is already 
contained in Eq.~(\ref{eq:apx_Nm3_tp1}). And when $\bar{n}$ is not an integer, $P^*_{\lfloor 4N/3\rfloor+1,2N-\lfloor 4N/3\rfloor-1}$ 
cannot the only nonzero point. 

When $N\bmod 3=0,1$, the legitimate intersections are $(\lfloor 4N/3\rfloor,h_1(\lfloor 4N/3\rfloor))$ and $(2N,0)$,
which means that $P^*_{\lfloor 4N/3\rfloor,2N-\lfloor 4N/3\rfloor}$ and $P^*_{2N,0}$ are nonzero. The 
optimal state can then be written as
\begin{align}
& \left|c_{N,\lfloor\frac{N}{3}\rfloor}\right|\left(e^{i\theta_1}
\ket{\left\lfloor\frac{N}{3}\right\rfloor,N}+e^{i\theta_2}\ket{N,\left\lfloor\frac{N}{3}\right\rfloor}\right) \nonumber \\
& +|c_{NN}|\ket{NN}.  
\end{align}
Utilizing the normalization and average particle number conditions, the state above can be specifically written as  
\begin{align}
& \sqrt{\frac{2N-\bar{n}}{2(N-\left\lfloor\frac{N}{3}\right\rfloor)}}
\left(e^{i\theta_1}\ket{\left\lfloor\frac{N}{3}\right\rfloor,N}+e^{i\theta_2}\ket{N,\left\lfloor\frac{N}{3}\right\rfloor}\right) \nonumber \\
& +\sqrt{\frac{\bar{n}-\left\lfloor\frac{4N}{3}\right\rfloor}
{N-\left\lfloor\frac{N}{3}\right\rfloor}}\ket{NN}, 
\label{eq:apx_Nm3_tp2}
\end{align}
where $\bar{n}$ satisfies $\bar{n}\geq \lfloor 4N/3 \rfloor$. The state corresponding to the case that 
$P^*_{\lfloor 4N/3\rfloor,2N-\lfloor 4N/3\rfloor}$ is the only nonzero point is of the form 
\begin{equation}
\frac{1}{\sqrt{2}}\left(\ket{\left\lfloor\frac{N}{3}\right\rfloor,N}+e^{i\theta}
\ket{N,\left\lfloor\frac{N}{3}\right\rfloor}\right)
\end{equation}
with $\bar{n}=\lfloor 4N/3 \rfloor$, which is already contained in Eq.~(\ref{eq:apx_Nm3_tp2}). And when $\bar{n}$ is not 
an integer, $P^*_{\lfloor 4N/3\rfloor,2N-\lfloor 4N/3\rfloor}$ cannot the only nonzero point. 

In summary, for $N\bmod 3=2$, the optimal state is Eq.~(\ref{eq:apx_optstate_tp1}) for $\lfloor\bar{n}\rfloor\leq 
\lfloor 4N/3\rfloor$, which is equivalent to $\bar{n}<\lfloor 4N/3\rfloor+1$, and Eq.~(\ref{eq:apx_Nm3_tp1}) for 
$\bar{n}\geq \lfloor 4N/3\rfloor+1$. As a matter of fact, taking $\bar{n}= \lfloor 4N/3\rfloor+1$ in 
Eq.~(\ref{eq:apx_optstate_tp1}), it just reduces to the state in Eq.~(\ref{eq:apx_Nm3_tp3}). Hence, one can also 
state that the optimal state is Eq.~(\ref{eq:apx_optstate_tp1}) for $\bar{n}\leq \lfloor 4N/3\rfloor+1$. For 
$N\bmod 3=0,1$, the optimal state is Eq.~(\ref{eq:apx_optstate_tp1}) for $\bar{n}\leq \lfloor 4N/3\rfloor$ and 
Eq.~(\ref{eq:apx_Nm3_tp2}) for $\bar{n}\geq \lfloor 4N/3\rfloor$. Then the optimal states can be 
unified into the following expressions:
\begin{widetext}
\begin{equation}
\begin{cases}  
\sqrt{\frac{\bar{n}-\lfloor \bar{n}\rfloor}{2}}\left(\ket{\lfloor \bar{n}\rfloor\!+\!1\!-\!N, N}
\!+\!e^{i\theta_1}\ket{N, \lfloor\bar{n}\rfloor\!+\!1\!-\!N}\right) &  \\
+\sqrt{\frac{1-(\bar{n}-\lfloor\bar{n}\rfloor)}{2}}\left(e^{i\theta_2} \ket{\lfloor \bar{n}\rfloor\!-\!N, N}
\!+\!e^{i\theta_3}\ket{N, \lfloor\bar{n}\rfloor\!-\!N}\right), & \bar{n}\in\left[N, \left\lfloor\frac{4N+1}{3}\right\rfloor\right], \\
\\
\sqrt{\frac{2N-\bar{n}}{2\left(N-\left\lfloor\frac{N+1}{3}\right\rfloor\right)}}
\left(e^{i\theta_1}\ket{\left\lfloor\frac{N+1}{3}\right\rfloor,N}+e^{i\theta_2}\ket{N,\left\lfloor\frac{N+1}{3}\right\rfloor}\right) 
+\sqrt{\frac{\bar{n}-\left\lfloor\frac{4N+1}{3}\right\rfloor}{N-\left\lfloor\frac{N+1}{3}\right\rfloor}}\ket{NN}, 
& \bar{n}\in\left[\left\lfloor\frac{4N+1}{3}\right\rfloor,2N\right).
\end{cases}
\label{eq:apx_gener_phy_optstate}
\end{equation}
Theorems 3 and 4 and corresponding corollaries are then proved. \hfill $\blacksquare$
\end{widetext}

Utilizing Eq.~(\ref{eq:apx_QFI_nonlinear}), the expressions of QFI for the optimal states above are 
\begin{align}
F = &(\bar{n}-\lfloor\bar{n}\rfloor)(\lfloor\bar{n}\rfloor+1)^2(2N-\lfloor\bar{n}\rfloor-1)^2\nonumber\\
 & + (1+\lfloor\bar{n}\rfloor-\bar{n})\lfloor\bar{n}\rfloor^2(2N-\lfloor\bar{n}\rfloor)^2
\end{align}
for $\bar{n}\in\left[N, \left\lfloor\frac{4N+1}{3}\right\rfloor\right]$, and
\begin{align}
F = &\frac{2N-\bar{n}}{N-\left\lfloor\frac{N+1}{3}\right\rfloor}\left(\left\lfloor\frac{4N+1}{3}\right\rfloor\right)^2
\left(N-\left\lfloor\frac{N+1}{3}\right\rfloor\right)^2
\end{align}
for $\bar{n}\in\left[\left\lfloor\frac{4N+1}{3}\right\rfloor,2N\right)$.

\section{Optimal probe states in the Mach-Zehnder interferometer}
\label{sec:MZstate}

In the previous sections we provide the OFPSs for both linear and nonlinear phase shifts. In practice, 
the phase estimation is usually performed in the Mach-Zehnder interferometer (MZI), in which a beam splitter exists 
in front of the phase shifts. Here we use a 50:50 beam splitter represented by the operator $\exp(-i\frac{\pi}{2}J_x)$. 
Hence, the optimal probe state must take the form $\exp(i\frac{\pi}{2}J_x)\ket{\psi_{\mathrm{opt}}}$ with 
$\ket{\psi_{\mathrm{opt}}}$ the OFPS we previously gave.  

\subsection{Linear case}

For a two-mode Fock state $\ket{n_1 n_2}$, $\exp(i\frac{\pi}{2}J_x)\ket{n_1 n_2}$ can be calculated as 
\begin{align}
& e^{i\frac{\pi}{2}J_x}\ket{n_1 n_2} \nonumber \\
=&\left(\frac{1}{\sqrt{2}}\right)^{n_1+n_2}\sum^{n_1}_{k=0}\sum^{n_2}_{l=0}\binom{n_1}{k}\binom{n_2}{l}
i^{k+l}\frac{\sqrt{(n_1-k+l)!}}{\sqrt{n_1!}} \nonumber \\
& \times\frac{\sqrt{(n_2+k-l)!}}{\sqrt{n_2!}}\ket{n_1-k+l,n_2+k-l}, 
\label{eq:apx_Jxrotation}
\end{align}
where $\ket{n_{1}}=\frac{1}{\sqrt{n_{1}!}}(a^{\dag})^{n_{1}}\ket{0}$, 
$\ket{n_{2}}=\frac{1}{\sqrt{n_{2}!}}(b^{\dag})^{n_{2}}\ket{0}$, and 
\begin{eqnarray}
e^{i\frac{\pi}{2}J_{x}}a^{\dagger}e^{-i\frac{\pi}{2}J_{x}} 
&=& \frac{1}{\sqrt{2}}\left(a^{\dagger}+i b^{\dagger}\right), \\
e^{i\frac{\pi}{2}J_{x}}b^{\dagger}e^{-i\frac{\pi}{2}J_{x}}
&=& \frac{1}{\sqrt{2}}\left(b^{\dagger}+i a^{\dagger}\right) 
\end{eqnarray}
have been applied. 

In the case of $\bar{n}\leq N$, the OFPS without the beam splitter is given in Eq.~(\ref{eq:apx_linOptStat}). 
Therefore, with Eq.~(\ref{eq:apx_Jxrotation}) it can be seen that the OFPS in the MZI reads 
\begin{align}
& \sqrt{1-\frac{\bar{n}}{N}}\ket{00}+ 2^{-\frac{1}{2}(N+1)}\sqrt{\frac{\bar{n}}{N}}
\sum^N_{k=0} i^{k}\binom{N}{k}^{\frac{1}{2}}
\nonumber \\
& \times \left(e^{i\theta_1}\ket{k,N-k}+e^{i\theta_2}\ket{N-k,k}\right).
\label{eq:apx_linearMZIopt}
\end{align}
In the case of $\bar{n}\geq N$, the OFPS without the beam splitter is given in Eq.~(\ref{eq:apx_linOptStat1}). 
Hence, the OFPS in the MZI is of the form 
\begin{align}
& 2^{-\frac{1}{2}(N+1)}\!\!\!\sqrt{2-\frac{\bar{n}}{N}} 
\sum_{k=0}^{N}i^{k}\binom{N}{k}^{\frac{1}{2}}\left(e^{i\theta_1}\ket{k,N-k}\right.\nonumber\\
&\left.+e^{i\theta_2}\ket{N\!-\!k,k}\right)\!+\!2^{-N}\sqrt{\frac{\bar{n}}{N}-1}
\sum^N_{k,l=0}\binom{N}{k}\binom{N}{l}i^{k+l}\nonumber\\
& \times\frac{\sqrt{(N-k+l)!(N+k-l)!}}{N!}\ket{N\!-\!k+\!l,N\!+\!k\!-\!l}. 
\end{align}

\subsection{Nonlinear case}

Now we provide the optimal probe states in the MZI with nonlinear phase shifts. In the case that $\bar{n}\leq N$, 
the optimal probe state without the beam splitter is the same as that in the linear case. Hence, the optimal probe
state in the MZI also takes the form of Eq.~(\ref{eq:apx_linearMZIopt}).

When $\bar{n}\geq N$, the legitimate optimal probe states without the beam splitter are given in 
Eq.~(\ref{eq:apx_gener_phy_optstate}). Utilizing Eq.~(\ref{eq:apx_Jxrotation}), the OFPS in the MZI in 
the regime $\bar{n}\in\left[N,\left\lfloor\frac{4N+1}{3}\right\rfloor\right]$ can be expressed by 
\begin{widetext}
\begin{align}
& 2^{-(\frac{1}{2}\lfloor \bar{n}\rfloor+1)}\sqrt{\bar{n}-\lfloor \bar{n}\rfloor}
\sum^{\lfloor \bar{n}\rfloor\!+\!1\!-\!N}_{k=0}\sum^{N}_{l=0}\binom{\lfloor \bar{n}\rfloor\!+\!1\!-\!N}{k}
\binom{N}{l}i^{k+l}\sqrt{\frac{(\lfloor \bar{n}\rfloor+1-N-k+l)!(N+k-l)!}{N!(\lfloor \bar{n}\rfloor+1-N)!}}\nonumber\\
& \times \left(\ket{\lfloor \bar{n}\rfloor+1-N-k+l,N+k-l} 
+e^{i\theta_1}\ket{N+k-l,\lfloor \bar{n}\rfloor+1-N-k+l}\right)\nonumber\\
& + 2^{-\frac{1}{2}(\lfloor \bar{n}\rfloor+1)}\sqrt{1-(\bar{n}-\lfloor \bar{n}\rfloor)}
\sum^{\lfloor \bar{n}\rfloor-N}_{s=0}\sum^N_{t=0}\binom{\lfloor \bar{n}\rfloor-N}{s}\binom{N}{t}i^{s+t}
\sqrt{\frac{(\lfloor \bar{n}\rfloor-N-s+t)!(N+s-t)!}{(\lfloor \bar{n}\rfloor-N)!N!}}\nonumber\\
&\times\left(e^{i\theta_2}\ket{\lfloor \bar{n}\rfloor-N-s+t,N+s-t}+
e^{i\theta_3}\ket{N+s-t,\lfloor \bar{n}\rfloor-N-s+t}\right).
\end{align}
In the regime $\bar{n}\in\left[\left\lfloor\frac{4N+1}{3}\right\rfloor,2N\right)$, 
the optimal probe state in the MZI reads
\begin{align}
& 2^{-N}\sqrt{\frac{\bar{n}-\zeta-N}{N-\zeta}}\sum^N_{k,l=0}
\binom{N}{k}\binom{N}{l}i^{k+l}\frac{\sqrt{(N-k+l)!(N+k-l)!}}{N!}\ket{N-k+l,N+k-l}\nonumber\\
& + 2^{-\frac{1}{2}(N+\zeta+1)}\sqrt{\frac{2N-\bar{n}}{N-\zeta}}
\sum^{\zeta}_{s=0}\sum^N_{t=0}\binom{\zeta}{s}\binom{N}{t}i^{s+t}\sqrt{\frac{(\zeta-s+t)!(N+s-t)!}{\zeta!N!}} \nonumber\\
& \times \left(e^{i\theta_1}\ket{\zeta-s+t,N+s-t}+e^{i\theta_2}\ket{N+s-t,\zeta-s+t}\right),
\end{align}
where $\zeta:=\left\lfloor\frac{N+1}{3}\right\rfloor$. 
\end{widetext}

\section{Quantum Fisher information matrix for the OFPSs}
\label{sec:QFIM}

In this appendix we provide the calculation of the quantum Fisher information matrix (QFIM) for the phase difference $\phi$ and 
phase summation $\phi_{\mathrm{tot}}$. For a pure state $\ket{\psi}$, the  $jk$th entry of the QFIM can be calculated via the equation
\begin{equation}
F_{jk} = 4\mathrm{Re}\left(\expval{\partial_j \psi|\partial_k \psi}-\expval{\partial_j \psi|\psi}\expval{\psi|\partial_k \psi}\right),
\end{equation}
where $j,k=\phi,\phi_{\mathrm{tot}}$. 

For linear phase shifts, recall that the parameterization operator is $\mathrm{e}^{\frac{i}{2}\phi_{\mathrm{tot}}n} \mathrm{e}^{i\phi J_z}$, 
then the non-diagonal entry $F_{\phi_{\mathrm{tot}},\phi}$ of the QFIM with respect to a probe state $\ket{\psi_{\mathrm{in}}}$ reads 
\begin{equation*}
F_{\phi_{\mathrm{tot}},\phi} = 2\mathrm{Re}\left(\bra{\psi_{\mathrm{in}}}n J_z \ket{\psi_{\mathrm{in}}}
-\bra{\psi_{\mathrm{in}}}n \ket{\psi_{\mathrm{in}}}\bra{\psi_{\mathrm{in}}}J_z \ket{\psi_{\mathrm{in}}}\right).
\end{equation*}
In the regime $\bar{n}\in(0,N]$ the OFPS reads  
\begin{equation}
\sqrt{\frac{N-\bar{n}}{N}}\ket{0 0}+\sqrt{\frac{\bar{n}}{2N}}
\left(e^{i\theta_1}\ket{0 N} +e^{i\theta_2}\ket{N 0}\right),
\label{eq:apx_OPFS1}
\end{equation}
where $\theta_1,\theta_2\in[0,2\pi)$ are the relative phases. Utilizing this OFPS as the probe state, it can be seen that 
$\bra{\psi_{\mathrm{in}}}nJ_z \ket{\psi_{\mathrm{in}}}=0$ and $\bra{\psi_{\mathrm{in}}}J_z \ket{\psi_{\mathrm{in}}}=0$, 
indicating that $F_{\phi_{\mathrm{tot}},\phi}=0$. In the regime $\bar{n}\in[N,2N)$, the OFPS reads 
\begin{equation*}
\sqrt{\frac{2N-\bar{n}}{2N}}\left(e^{i\theta_1}\ket{0 N}\!+\!e^{i\theta_2}
\ket{N 0}\right)\!+\!\sqrt{\frac{\bar{n}-N}{N}}\ket{N N}. 
\end{equation*}
For this OFPS, one can also find that $\bra{\psi_{\mathrm{in}}}n J_z \ket{\psi_{\mathrm{in}}}=0$ and  
$\bra{\psi_{\mathrm{in}}} J_z \ket{\psi_{\mathrm{in}}}=0$, which means $F_{\phi_{\mathrm{tot}},\phi}=0$. 

For the nonlinear phase shifts, the parameterization operator is $e^{\frac{i}{2}\phi_{\mathrm{tot}}[(a^{\dagger}a)^2+(b^{\dagger}b)^2]}
e^{i\phi n J_z}$. Regarding a pure probe state $\ket{\psi_{\mathrm{in}}}$, the non-diagonal entry reads 
\begin{align*}
F_{\phi_{\mathrm{tot}},\phi}=&2\mathrm{Re}\left(\bra{\psi_{\mathrm{in}}}
\big[(a^{\dagger}a)^2+(b^{\dagger}b)^2\big]nJ_z\ket{\psi_{\mathrm{in}}}\right. \\
& - \left.\bra{\psi_{\mathrm{in}}}(a^{\dagger}a)^2+(b^{\dagger}b)^2\ket{\psi_{\mathrm{in}}}
\bra{\psi_{\mathrm{in}}}nJ_z\ket{\psi_{\mathrm{in}}} \right).
\end{align*}
In the regime $\bar{n}\in(0,N]$, the OFPS is still the state given in Eq.~(\ref{eq:apx_OPFS1}). For this state, it is easy to see that 
$\bra{\psi_{\mathrm{in}}}[(a^{\dagger}a)^2+(b^{\dagger}b)^2]nJ_z\ket{\psi_{\mathrm{in}}}=0$ and 
$\bra{\psi_{\mathrm{in}}}nJ_z\ket{\psi_{\mathrm{in}}}=0$, indicating that $F_{\phi_{\mathrm{tot}},\phi}=0$. 
In the regime $\bar{n}\in\left[N, \left\lfloor\frac{4N+1}{3}\right\rfloor\right]$, the OFPS is 
\begin{align}
& \sqrt{\frac{\bar{n}-\lfloor \bar{n}\rfloor}{2}}
\left(\ket{\lfloor \bar{n}\rfloor\!+\!1\!-\!N, N}
\!+\!e^{i\theta_1}\ket{N, \lfloor\bar{n}\rfloor\!+\!1\!-\!N}\right)\notag\\
&+\sqrt{\frac{1-(\bar{n}-\lfloor\bar{n}\rfloor)}{2}}
\left(e^{i\theta_2} \ket{\lfloor \bar{n}\rfloor\!-\!N, N}
\!+\!e^{i\theta_3}\ket{N, \lfloor\bar{n}\rfloor\!-\!N}\right), 
\end{align}
where $\theta_1,\theta_2$, and $\theta_3$ are the relative phases. In this case, notice that 
\begin{align*}
& \bra{\lfloor \bar{n}\rfloor\!+\!1\!-\!N, N}[(a^{\dagger}a)^2\!+\!(b^{\dagger}b)^2]nJ_z\ket{\lfloor \bar{n}\rfloor\!+\!1\!-\!N, N}  \\
& \!+\!\bra{N, \lfloor \bar{n}\rfloor\!+\!1\!-\!N}[(a^{\dagger}a)^2\!+\!(b^{\dagger}b)^2]nJ_z\ket{N, \lfloor \bar{n}\rfloor\!+\!1\!-\!N}=0,
\end{align*}
and 
\begin{align*}
& \bra{\lfloor \bar{n}\rfloor-\!N, N}[(a^{\dagger}a)^2\!+\!(b^{\dagger}b)^2]nJ_z\ket{\lfloor \bar{n}\rfloor-\!N, N}  \\
& \!+\!\bra{N, \lfloor \bar{n}\rfloor-\!N}[(a^{\dagger}a)^2\!+\!(b^{\dagger}b)^2]nJ_z\ket{N, \lfloor \bar{n}\rfloor-\!N}=0,
\end{align*}
then $\bra{\psi_{\mathrm{in}}}[(a^{\dagger}a)^2+(b^{\dagger}b)^2]nJ_z\ket{\psi_{\mathrm{in}}}=0$. Similarly, 
\begin{align*}
& \bra{\lfloor \bar{n}\rfloor\!+\!1\!-\!N, N}nJ_z\ket{\lfloor \bar{n}\rfloor\!+\!1\!-\!N, N} \\
& +\bra{N, \lfloor \bar{n}\rfloor\!+\!1\!-\!N}nJ_z\ket{N, \lfloor \bar{n}\rfloor\!+\!1\!-\!N}=0,
\end{align*}
and 
\begin{align*}
& \bra{\lfloor \bar{n}\rfloor-\!N, N}nJ_z\ket{\lfloor \bar{n}\rfloor-\!N, N}  \\
& +\bra{N, \lfloor \bar{n}\rfloor-\!N}nJ_z\ket{N, \lfloor \bar{n}\rfloor-\!N}=0.
\end{align*}
This fact means $\bra{\psi_{\mathrm{in}}}nJ_z\ket{\psi_{\mathrm{in}}}=0$. Hence, $F_{\phi_{\mathrm{tot}},\phi}=0$ in this case. 

Furthermore, in the regime $\bar{n}\!\in\!\left[\left\lfloor\frac{4N+1}{3}\right\rfloor,2N\right)$, the OFPS is 
\begin{equation}
\sqrt{\frac{2N-\bar{n}}{2\left(N-\zeta\right)}}\Big(e^{i\theta_1}\ket{\zeta N}
+e^{i\theta_2} \ket{N \zeta}\Big)+\sqrt{\frac{\bar{n}-N-\zeta}{N-\zeta}}\ket{NN}.
\end{equation}
Here $\zeta := \left\lfloor\frac{N+1}{3}\right\rfloor$, and $\theta_1,\theta_2$ are the relative phases. In this case,
\begin{align*}
& \bra{\zeta N}[(a^{\dagger}a)^2\!+\!(b^{\dagger}b)^2]nJ_z\ket{\zeta N}  \\
& \!+\!\bra{N \zeta }[(a^{\dagger}a)^2\!+\!(b^{\dagger}b)^2]nJ_z\ket{N \zeta}=0,
\end{align*}
which means $\bra{\psi_{\mathrm{in}}}[(a^{\dagger}a)^2+(b^{\dagger}b)^2]nJ_z\ket{\psi_{\mathrm{in}}}=0$. In the mean time, 
\begin{equation}
\bra{\zeta N}nJ_z\ket{\zeta N} +\bra{N \zeta }nJ_z\ket{N \zeta}=0,
\end{equation}
and it means $\bra{\psi_{\mathrm{in}}}nJ_z\ket{\psi_{\mathrm{in}}}=0$. Hence, $F_{\phi_{\mathrm{tot}},\phi}=0$ in this case. 

Hence, in both linear and nonlinear cases, the QFIM for the aforementioned OFPSs are diagonal. This fact indicates that the 
measurement of phase difference does not require the information of the phase summation in both cases. 

\section{Parity measurement}
\label{sec:measurement}

\subsection{Linear case}

The parity operator for the $a$th mode is 
\begin{equation}
\Pi_a = e^{i\pi a^{\dagger}a}=e^{i\frac{\pi}{2}n}e^{i\pi J_z}, 
\end{equation}
where $n=a^{\dagger}a+b^{\dagger}b$ is the operator for the total particle number and commutes with all $J_x$, $J_y$, 
and $J_z$. Recall that the state before the measurement is $e^{i\frac{\pi}{2}J_x}e^{i\frac{1}{2}\phi_{\mathrm{tot}}n}
e^{i\phi J_z}\ket{\psi_{\mathrm{in}}}$. Then the expected value of the parity operator reads 
\begin{align}
\expval{\Pi_a}
& =\bra{\psi_{\mathrm{in}}}e^{-i\phi J_z} e^{-i\frac{\pi}{2}J_x}e^{i\frac{\pi}{2}n}e^{i\pi J_z} 
e^{i\frac{\pi}{2}J_x}e^{i\phi J_z}\ket{\psi_{\mathrm{in}}}\nonumber\\
& = \bra{\psi_{\mathrm{in}}}e^{i\frac{\pi}{2}n}e^{-i\phi J_z}e^{-i\pi J_y}e^{i\phi J_z}\ket{\psi_{\mathrm{in}}}, 
\label{eq:apx_parity1}
\end{align}
where the equality $e^{-i\frac{\pi}{2}J_x}e^{i\pi J_z}e^{i\frac{\pi}{2}J_x}=e^{-i\pi J_y}$ has been applied. 

In the case that $\bar{n}\leq N$, the OFPS reads 
\begin{equation}
\sqrt{\frac{N-\bar{n}}{N}}\ket{0 0}
+\sqrt{\frac{\bar{n}}{2N}}\left(e^{i\theta_1}\ket{0 N}+e^{i\theta_2}\ket{N 0}\right).
\label{eq:apx_psi}
\end{equation}
Substituting it into Eq.~(\ref{eq:apx_parity1}), and further utilizing 
\begin{equation}
e^{i\phi J_z}\ket{n_1 n_2} = e^{i\frac{\phi}{2}(n_1-n_2)}\ket{n_1 n_2},    
\label{eq:apx_expJz}
\end{equation}
where $\ket{n_{1(2)}}$ is a Fock state with respect to mode $a$ ($b$), and 
\begin{align}
&  e^{-i\pi J_y}\ket{n_1 n_2} \nonumber \\
&= \frac{\left(e^{-i\pi J_y}a^{\dagger}e^{i\pi J_y}\right)^{n_1}}{\sqrt{n_1!}}
\frac{\left(e^{-i\pi J_y}b^{\dagger}e^{i\pi J_y}\right)^{n_2}}{\sqrt{n_2!}}\ket{00} \nonumber \\
&= \frac{\left(-a^{\dagger}\right)^{n_2}}{\sqrt{n_2!}}\frac{\left(b^{\dagger}\right)^{n_1}}{\sqrt{n_1!}}\ket{00} \nonumber \\
&= e^{i\pi n_2}\ket{n_2 n_1},
\label{eq:apx_expJy}
\end{align}
where $e^{-i\pi J_y}a^{\dagger}e^{i\pi J_y}=b^{\dagger}$ and $e^{-i\pi J_y}b^{\dagger}e^{i\pi J_y}=-a^{\dagger}$ 
have been applied, one can obtain the expression  
\begin{equation}
\expval{\Pi_a}=1-\frac{\bar{n}}{N}\left(1-\cos\beta_1\right),  
\end{equation}
where 
\begin{equation}
\beta_{1}:=\theta_2-\theta_1+\frac{\pi}{2}N+\phi N.     
\label{eq:apx_beta1}
\end{equation}

The variance $\delta^2\phi$ of measuring $\phi$ via $\expval{\Pi_a}$ can be evaluated through the 
error propagation relation
\begin{equation}
\delta^2\phi=\frac{\expval{\Pi^2_a}-\expval{\Pi_a}^2}
{\left|\partial_{\phi}\expval{\Pi_a}\right|^2}.     
\end{equation}
As a matter of fact, here $\expval{\Pi^2_a}=1$ due to the fact that $\Pi^2_a=\openone$ with $\openone$ the 
identity operator. Applying the expression of $\expval{\Pi_a}$, $\delta^2\phi$ can be expressed by 
\begin{equation}
\delta^2\phi=\frac{1}{\bar{n}N}\frac{2(1-\cos\beta_{1})}{\sin^2 \beta_{1}}
-\frac{1}{N^2}\frac{(1-\cos\beta_{1})^2}{\sin^2 \beta_{1}}.      
\end{equation}
One may notice that $\delta^2\phi$ depends on $\phi$, indicating that the true value of $\phi$ could affect the performance 
of parity measurement. When the value of $\beta_{1}$ is very close to $2k\pi$ ($k$ is any integer), i.e., 
$\beta_{1}=2k\pi+\delta\beta_{1}$ with $\delta\beta_{1}$ a small quantity, $\delta^2\phi$ reduces to 
\begin{equation}
\delta^2\phi=\frac{1}{\bar{n}N}-\frac{1}{4N^2}\delta^2\beta_{1},   
\end{equation}
which means that 
\begin{equation}
\lim_{\delta\beta_{1}\rightarrow 0}\delta^2\phi=\frac{1}{\bar{n}N}.    
\end{equation}
Noticing that the QFI in this case is $\bar{n}N$, the parity measurement is optimal when the value of $\beta_{1}$ equals 
to $2k\pi$, which means the true value of $\phi$ ($\phi_{\mathrm{true}}$) has to be in the form  
\begin{equation}
\phi_{\mathrm{true}}=\frac{1}{N}\left(\theta_1-\theta_2+2k\pi\right)-\frac{\pi}{2},~k\in\mathbb{Z},  
\label{eq:apx_optphi}
\end{equation}
where $\mathbb{Z}$ is the set of integers. 

Now we discuss the performance of parity measurement from the perspective of the classical Fisher information (CFI), 
which is 
\begin{equation}
I=\frac{\left(\partial_{\phi} P_{+}\right)^2}{P_{+}}+\frac{\left(\partial_{\phi} P_{-}\right)^2}{P_{-}},    
\end{equation}
where $P_{\pm}$ is the probability of obtaining the result $\pm 1$ by measuring $\expval{\Pi_a}$. It can be seen that 
\begin{align}
P_{+} & = 1-\frac{\bar{n}}{2N}\left(1-\cos \beta_1 \right), \label{eq:apx_parity_LinProleq} \\
P_{-} & = \frac{\bar{n}}{2N}\left(1-\cos \beta_1 \right), \label{eq:apx_parity_LinProleq1}
\end{align}
which can be obtained via the equations $\expval{\Pi_a}=P_{+}-P_{-}$ and $P_{+}+P_{-}=1$. With these expressions, 
the CFI can be calculated as 
\begin{align}
I = \frac{\bar{n}N^2\sin^2\beta_{1}}{(1-\cos\beta_{1})\left[2N-\bar{n}(1-\cos\beta_{1})\right]},
\end{align}
which directly gives 
\begin{equation}
\lim_{\beta_{1}\rightarrow 2k\pi}I=\bar{n}N. 
\end{equation}
Therefore, this equation means that the CFI can reach the QFI when the true value of $\phi$ satisfies Eq.~(\ref{eq:apx_optphi}). 

In the case that $\bar{n}\geq N$, the OFPS reads 
\begin{equation*}
\sqrt{\frac{2N-\bar{n}}{2N}}\left(e^{i \theta_1}\ket{0N}+e^{i\theta_2} \ket{N0}\right)
+\sqrt{\frac{\bar{n}-N}{N}}\ket{NN}.    
\end{equation*}
The value of $\expval{\Pi_a}$ can then be calculated as 
\begin{equation}
\expval{\Pi_a} =\frac{\bar{n}-N}{N}+\frac{2N-\bar{n}}{N}\cos\beta_1. 
\end{equation}
Utilizing the error propagation relation, $\delta^2 \phi$ can be expressed by 
\begin{equation}
\delta^2\phi=\frac{2(1-\cos\beta_{1})}{N(2N-\bar{n})\sin^2\beta_{1}}
-\frac{\left(1-\cos\beta_{1}\right)^2}{N^2\sin^2\beta_{1}},
\end{equation}
and its limit is 
\begin{equation}
\lim_{\beta_{1}\rightarrow 2k\pi}\delta^2\phi=\frac{1}{N(2N-\bar{n})}.
\end{equation}
In this case, the QFI is just $N(2N-\bar{n})$, indicating that the parity measurement is optimal when 
\begin{equation}
\phi_{\mathrm{true}}=\frac{1}{N}\left(\theta_1-\theta_2+2k\pi\right)-\frac{\pi}{2}. 
\end{equation}

From the perspective of CFI, the conditional probability $P_{\pm}$ in this case reads 
\begin{align}
P_{+} & = 1-\frac{2N-\bar{n}}{2N}\left(1-\cos\beta_1\right), \label{eq:apx_parity_LinProgeq} \\
P_{-} & = \frac{2N-\bar{n}}{2N}\left(1-\cos\beta_1\right). \label{eq:apx_parity_LinProgeq1}
\end{align}
The CFI is 
\begin{equation}
I = \frac{(2N-\bar{n})N^2\sin^2\beta_{1}}{(1-\cos\beta_{1})
\left[2N-(2N-\bar{n})(1-\cos\beta_{1})\right]},
\end{equation}
and $\lim_{\beta_{1}\rightarrow 2k\pi}I=(2N-\bar{n})N$.

\subsection{Nonlinear case} 

In the nonlinear case, the state before the measurement is $e^{i\frac{\pi}{2}J_x}e^{i\frac{1}{2}\phi_{\mathrm{tot}}
[(a^{\dagger}a)^2+(b^{\dagger}b)^2]} e^{i \phi n J_z}\ket{\psi_\mathrm{in}}$. Then the expectation of the parity operator is 
\begin{align}
\expval{\Pi_a} = 
& \bra{\psi_{\mathrm{in}}}e^{-i\phi n J_z}
e^{-i\frac{1}{2}\phi_{\mathrm{tot}}[(a^{\dagger}a)^2+(b^{\dagger}b)^2]}e^{-i\frac{\pi}{2}J_x} \nonumber\\
&  \times e^{i\frac{\pi}{2}n}e^{i\pi J_z} e^{i\frac{\pi}{2}J_x}e^{i\frac{1}{2}\phi_{\mathrm{tot}}
[(a^{\dagger}a)^2+(b^{\dagger}b)^2]}e^{i \phi n J_z}\ket{\psi_\mathrm{in}} \nonumber \\
= & \bra{\psi_\mathrm{in}}e^{i\frac{\pi}{2}n}e^{-i\phi n J_z}e^{-i\frac{1}{2}
\phi_{\mathrm{tot}}[(a^{\dagger}a)^2+(b^{\dagger}b)^2]}e^{-i\pi J_y} \nonumber \\
&  \times e^{i\frac{1}{2}\phi_{\mathrm{tot}}[(a^{\dagger}a)^2+(b^{\dagger}b)^2]}
e^{i \phi n J_z}\ket{\psi_\mathrm{in}}, 
\end{align} 
where the equality $e^{-i\frac{\pi}{2}J_x}e^{i\pi J_z}e^{i\frac{\pi}{2}J_x}=e^{-i\pi J_y}$ has been applied. 

In the case of $\bar{n}\leq N$, the OFPS reads 
\begin{equation}
\sqrt{\frac{N-\bar{n}}{N}}\ket{0 0}+\sqrt{\frac{\bar{n}}{2N}}\left(e^{i\theta_1}\ket{0 N}+e^{i\theta_2}
\ket{N 0}\right).    
\end{equation}
Utilizing Eq.~(\ref{eq:apx_expJy}) and the equality $e^{i\phi n J_z}\ket{n_1 n_2} = e^{i\frac{1}{2}\left(n_1^2-n_2^2\right) \phi}\ket{n_1 n_2}$, 
$\expval{\Pi_a}$ can be expressed by 
\begin{equation}
\expval{\Pi_a}=1-\frac{\bar{n}}{N}\left(1-\cos\beta_{2}\right), 
\end{equation}
where 
\begin{equation}
\beta_{2}:=\theta_2-\theta_1+\frac{\pi}{2}N+\phi N^2.   
\label{eq:apx_beta2}
\end{equation}
The variance $\delta^2\phi$ obtained from the error propagation relation can be written as 
\begin{equation}
\delta^2\phi=\frac{1}{\bar{n}N^3}\frac{2(1-\cos\beta_{2})}{\sin^2\beta_{2}}
-\frac{1}{N^4}\frac{(1-\cos\beta_{2})^2}{\sin^2\beta_{2}}.   
\end{equation}
Its limit for $\beta_{2}\rightarrow 2k\pi$ is 
\begin{equation}
\lim_{\beta_{2}\rightarrow 2k\pi}\delta^2\phi=\frac{1}{\bar{n}N^3}.   
\end{equation}
In this case, the QFI reads $\bar{n}N^3$, therefore, same with the linear case, the parity measurement is optimal when 
the value of $\beta_{2}$ approaches to $2k\pi$, which means the true value of $\phi$ ($\phi_{\mathrm{true}}$) needs to be 
\begin{equation}
\phi_{\mathrm{true}}=\frac{1}{N^2}\left(\theta_1-\theta_2+2k\pi\right)-\frac{\pi}{2N},~k\in\mathbb{Z}.   
\end{equation}

From the perspective of CFI, the probabilities $P_{+}$ and $P_{-}$ read 
\begin{align}
P_{+} & = 1-\frac{\bar{n}}{2N}\left(1-\cos\beta_2\right), \label{eq:apx_parity_NonProleq} \\
P_{-} & = \frac{\bar{n}}{2N}\left(1-\cos\beta_2\right), \label{eq:apx_parity_NonProleq1}
\end{align}
and the CFI can then be expressed by 
\begin{equation}
I=\frac{\bar{n}N^4\sin^2\beta_{2}}{(1-\cos\beta_{2})\left[2N-\bar{n}(1-\cos\beta_{2})\right]}.
\end{equation}
It can be further found that 
\begin{equation}
\lim_{\beta_2\rightarrow 2k\pi}I=\bar{n}N^3.
\end{equation}

In the case of $\bar{n}\geq N$, we demonstrate a simple case that $\bar{n}\in[N,\left\lfloor\frac{4N+1}{3}\right\rfloor]$ 
is an integer. In this case, the OFPS is 
\begin{equation}
\frac{1}{\sqrt{2}}\left(\ket{\bar{n}-N,N}+e^{i\theta}\ket{N,\bar{n}-N}\right). 
\end{equation}
The value of $\expval{\Pi_a}$ is given by
\begin{equation}
\expval{\Pi_a}=\cos\gamma 
\end{equation}
with 
\begin{equation}
\gamma := \theta+\frac{\pi}{2}(2N-\bar{n})+\phi \bar{n}(2N-\bar{n}).     
\label{eq:apx_gamma}
\end{equation}
Then $\delta^2\phi$ can be calculated as 
\begin{equation}
\delta^2\phi=\frac{1}{\bar{n}^2(2N-\bar{n})^2}, 
\end{equation}
which is independent of the true value of $\phi$. Notice that here the QFI is $\bar{n}^2(2N-\bar{n})^2$, and thus the parity 
measurement is optimal for all possible true values of $\phi$. From the perspective of CFI, $P_{\pm}$ is in the form 
\begin{equation}
P_{+}=\frac{1}{2}\left(1+\cos\gamma\right),\,\,\, P_{-}=\frac{1}{2}\left(1-\cos\gamma\right).
\label{eq:apx_parity_NonProgeq}
\end{equation}
The CFI can then be expressed by 
\begin{align}
I=\bar{n}^2(2N-\bar{n})^2.
\end{align}

\section{Particle-counting measurement}
\label{sec:measurement1}

\subsection{Linear case}

For the particle-counting measurement, the probability of detecting $m$ particles on mode $a$ is 
\begin{equation}
P_m =\sum_{j=0}^{2N}|\bra{mj}\ket{\psi}|^2
\end{equation}
with $\ket{\psi}$ a quantum state. Recall that the state before the measurement in the linear case is 
\begin{equation}
e^{i\frac{\pi}{2}J_x}e^{i\frac{1}{2}\phi_{\mathrm{tot}}n}e^{i\phi J_z}\ket{\psi_{\mathrm{in}}}.
\end{equation}
The probability $P_m$ for this state is
\begin{align}
P_m &= \sum_{j=0}^{2N}\left|\bra{mj}e^{i\frac{\pi}{2}J_x}e^{i\frac{1}{2}\phi_{\mathrm{tot}}n}
e^{i\phi J_z}\ket{\psi_{\mathrm{in}}}\right|^2 \nonumber \\
&= \sum_{j=0}^{2N}\left|\bra{mj}e^{i\frac{\pi}{2}J_x}e^{i\phi J_z}\ket{\psi_{\mathrm{in}}}\right|^2.
\end{align}

In the case that $\bar{n}\leq N$, the OFPS is given in Eq.~(\ref{eq:apx_linOptStat}), and 
$P_m$ can be calculated as 
\begin{align}
P_m =& \frac{N-\bar{n}}{N}\delta_{0m}+h(m-N)2^{-N}\frac{\bar{n}}{N} \nonumber \\
& \times \binom{N}{m}\left[1+(-1)^m\cos\beta_1\right],
\label{eq:apx_phocount_LinProleq}
\end{align}
where $\beta_1$ is defined in Eq.~(\ref{eq:apx_beta1}) and $h(m-N)$ is the step function defined by 
\begin{equation}
h(m-N):=
\begin{cases}
1, & m-N\leq 0, \\
0, & m-N>0.
\end{cases}
\label{eq:apx_step}
\end{equation}
Its derivative with respect to $\phi$ is 
\begin{equation}
\partial_{\phi} P_m = h(m-N)(-1)^{m+1} 2^{-N}\bar{n}\binom{N}{m} \sin\beta_1.
\end{equation}
The fact that the probability $P_m$ has no contribution to the CFI when $m>N$ means that the CFI reads 
$I=\sum^{N}_{m=0}(\partial_{\phi}P_m)^2/P_m$. 

The general expression of the CFI is tedious. However, when $\beta_1=2k\pi$, i.e., $\phi_{\mathrm{true}}=\frac{1}{N}
\left(\theta_1-\theta_2+2k\pi\right)-\frac{\pi}{2}$, $\partial_{\phi} P_m$ is zero, and only the terms 
$(\partial_{\phi}P_m)^2/P_m$ with a vanishing $P_m$ would contribute to the CFI. From 
Eq.~(\ref{eq:apx_phocount_LinProleq}), it can be seen that this only happens when $m$ is odd. Hence, utilizing 
Bernoulli's rule, the CFI becomes $\sum^{\tau_N}_{j=0}2\partial^2_{\phi}P_{2j+1}$, where $\tau_N=(N-1)/2$ for an odd 
$N$ and $\tau_N=(N-2)/2$ for an even $N$. Substituting the expression of $\partial_{\phi}P_m$ into this expression, 
it can be further calculated as
\begin{equation}
I=\bar{n}N2^{-N+1}\sum^{\tau_N}_{j=0}\binom{N}{2j+1}=\bar{n} N,
\end{equation}
where the equality $\sum^{\tau_N}_{j=0}\binom{N}{2j+1}=2^{N-1}$ has been applied. This result indicates 
that when $\phi_{\mathrm{true}}=\frac{1}{N}\left(\theta_1-\theta_2+2k\pi\right)-\frac{\pi}{2}$, the CFI in this 
case reaches the QFI, and the particle-counting measurement is optimal. As a matter of fact, this calculation process 
also shows the reason why the parity and particle-counting measurements are optimal simultaneously when $\phi_{\mathrm{true}}
=\frac{1}{N}\left(\theta_1-\theta_2+2k\pi\right)-\frac{\pi}{2}$. At this point, $P_m$ vanishes when $m$ is odd, which 
means $P_+$ is one and $P_-$ is zero. This is just the case that parity measurement is optimal.  

In the case that $\bar{n}\geq N$, utilizing OFPS given in Eq.~(\ref{eq:apx_linOptStat1}), $P_m$ reads 
\begin{align}
P_m =& h(m\!-\!N)2^{-N}\left(2-\frac{\bar{n}}{N}\right) \binom{N}{m}\left[1+(-1)^m\cos\beta_1\right] \nonumber \\
& +2^{-2N}\left(\frac{\bar{n}}{N}-1\right)\frac{m!(2N-m)!}{\left(N!\right)^2}\chi^2_1, 
\label{eq:apx_phocount_LinProgeq}
\end{align}
where $\chi_1$ is defined by 
\begin{equation}
\chi_1:=\sum^{\min\{N,m\}}_{k=\max\{0,m-N\}}(-1)^k\binom{N}{k}\binom{N}{m-k}.    
\end{equation}
And $\partial_{\phi}P_m$ reads 
\begin{equation}
\partial_{\phi}P_m = h(m\!-\!N)(-1)^{m+1} 2^{-N} (2N-\bar{n})\binom{N}{m}\sin\beta_1. 
\end{equation}
As in the case that $\bar{n}\leq N$, the general expression of CFI here is tedious. However, when 
$\phi_{\mathrm{true}}=\frac{1}{N}\left(\theta_1-\theta_2+2k\pi\right)-\frac{\pi}{2}$, only the terms 
$(\partial_{\phi}P_m)^2/P_m$ with an odd $m$ satisfying $m\leq N$ would contribute to the CFI due to the fact 
that 
\begin{align}
& \sum^m_{k=0}(-1)^k\binom{N}{k}\binom{N}{m-k} \nonumber \\
=& \sum^{\frac{1}{2}(m-1)}_{l=0}\left[(-1)^l+(-1)^{m-l}\right]\binom{N}{l}\binom{N}{m-l}=0.   
\end{align}
Hence, the CFI can be calculated as 
\begin{equation}
I=N(2N-\bar{n})2^{-N+1}\!\sum^{\tau_N}_{j=0}\binom{N}{2j+1}=N(2N-\bar{n}),    
\end{equation}
which means that the CFI reaches the QFI at this point and the particle-counting measurement is thus optimal. 

\subsection{Nonlinear case}

For nonlinear phase shifts, when $\bar{n}\leq N$, the OFPS is the same as the linear 
case, as given in Eq.~(\ref{eq:apx_NonlinOptstat}). Then $P_m$ can be expressed by 
\begin{align}
P_m =& \frac{N-\bar{n}}{N}\delta_{0m}+h(m-N)2^{-N}\frac{\bar{n}}{N} \nonumber \\
& \times \binom{N}{m}\left[1+(-1)^m\cos\beta_2\right],
\label{eq:apx_phocount_NonProleq}
\end{align}
and its derivative with respect to $\phi$ is
\begin{equation}
\partial_{\phi}P_m=h(m-N)(-1)^{m+1}2^{-N}\bar{n}N \binom{N}{m}\sin\beta_2.
\end{equation}
respectively. In the case that $\beta_2=2k\pi$, i.e., $\phi_{\mathrm{true}}=\frac{1}{N^2}\left(\theta_1-\theta_2+2k\pi\right)
-\frac{\pi}{2N}$, utilizing the same calculation procedure in the linear case, the CFI can be calculated as $\bar{n}N^3$, 
which indicates that the CFI at this point reaches the QFI and the particle-counting measurement is optimal. 

When $\bar{n}\geq N$, we only consider the case that $\bar{n}\in [N, \left\lfloor\frac{4N+1}{3}\right\rfloor]$ 
is an integer, which means the OFPS is 
\begin{equation}
\frac{1}{\sqrt{2}}\left(\ket{\bar{n}-N,N}+e^{i\theta}\ket{N,\bar{n}-N}\right).     
\end{equation}
With this state, $P_m$ reads 
\begin{equation}
P_m = 2^{-\bar{n}}\frac{m!(\bar{n}-m)!}{(\bar{n}-N)!N!}\left[1+(-1)^{m}\cos\gamma\right]\chi^2_2
\label{eq:apx_phocount_NonProgeq}
\end{equation}
for $m\leq \bar{n}$, and $P_m=0$ for $m>\bar{n}$. Here $\gamma$ is defined in Eq.~(\ref{eq:apx_gamma}), and 
$\chi_2$ is defined by 
\begin{equation}
\chi_2:=\sum^{\min\{N,m\}}_{k=\max\{0,N+m-\bar{n}\}}(-1)^{k}\binom{N}{k}\binom{\bar{n}-N}{m-k}.
\end{equation}
In the meantime, 
$\partial_{\phi} P_m$ is 
\begin{equation}
\partial_{\phi} P_m = 2^{-\bar{n}}\bar{n}(2N-\bar{n})\sin\gamma\frac{m!(\bar{n}-m)!}{(\bar{n}-N)!N!}
(-1)^{m+1}\chi^2_2
\end{equation}
for $m\leq \bar{n}$ and zero for $m>\bar{n}$. Utilizing the expressions of $P_m$ and $\partial_{\phi}P_m$, the CFI 
can be written as 
\begin{equation*}
I=\bar{n}^2(2N-\bar{n})^2\sum^{\bar{n}}_{m=0}\frac{2^{-\bar{n}}m!(\bar{n}-m)!}{(\bar{n}-N)!N!}
\frac{\sin^2\gamma}{1+(-1)^m\cos\gamma}\chi^2_2.    
\end{equation*}
Noticing that 
\begin{equation}
\frac{\sin^2\gamma}{1+(-1)^m\cos\gamma}=2-[1+(-1)^m\cos\gamma],    
\end{equation}
the CFI reduces to 
\begin{align*}
I=& \bar{n}^2(2N-\bar{n})^2\sum^{\bar{n}}_{m=0}\left(\frac{2^{-\bar{n}}m!(\bar{n}-m)!}{(\bar{n}-N)!N!}
2\chi^2_2-P_m\right) \\
=& \bar{n}^2(2N-\bar{n})^2\left(-1+2\sum^{\bar{n}}_{m=0}\frac{2^{-\bar{n}}m!(\bar{n}-m)!}{(\bar{n}-N)!N!}
\chi^2_2\right),
\end{align*}
where the normalization relation $\sum^{\bar{n}}_{m=0}P_m=1$ is applied. Further notice that the normalization 
relation is independent of the value of $\gamma$, and when $\cos\gamma=0$, the normalization relation reduces to 
\begin{equation}
\sum^{\bar{n}}_{m=0}\frac{2^{-\bar{n}}m!(\bar{n}-m)!}{(\bar{n}-N)!N!}\chi^2_2=1.    
\end{equation}
With this equation, the CFI further reduces to 
\begin{equation}
I=\bar{n}^2(2N-\bar{n})^2,    
\end{equation}
which is nothing but the QFI in this case. Hence, the particle-counting measurement is optimal in this case,  
regardless of the true values. 

\section{Adaptive measurement}
\label{sec:adapt}

\begin{figure*}[tp]
\centering\includegraphics[width=13.cm]{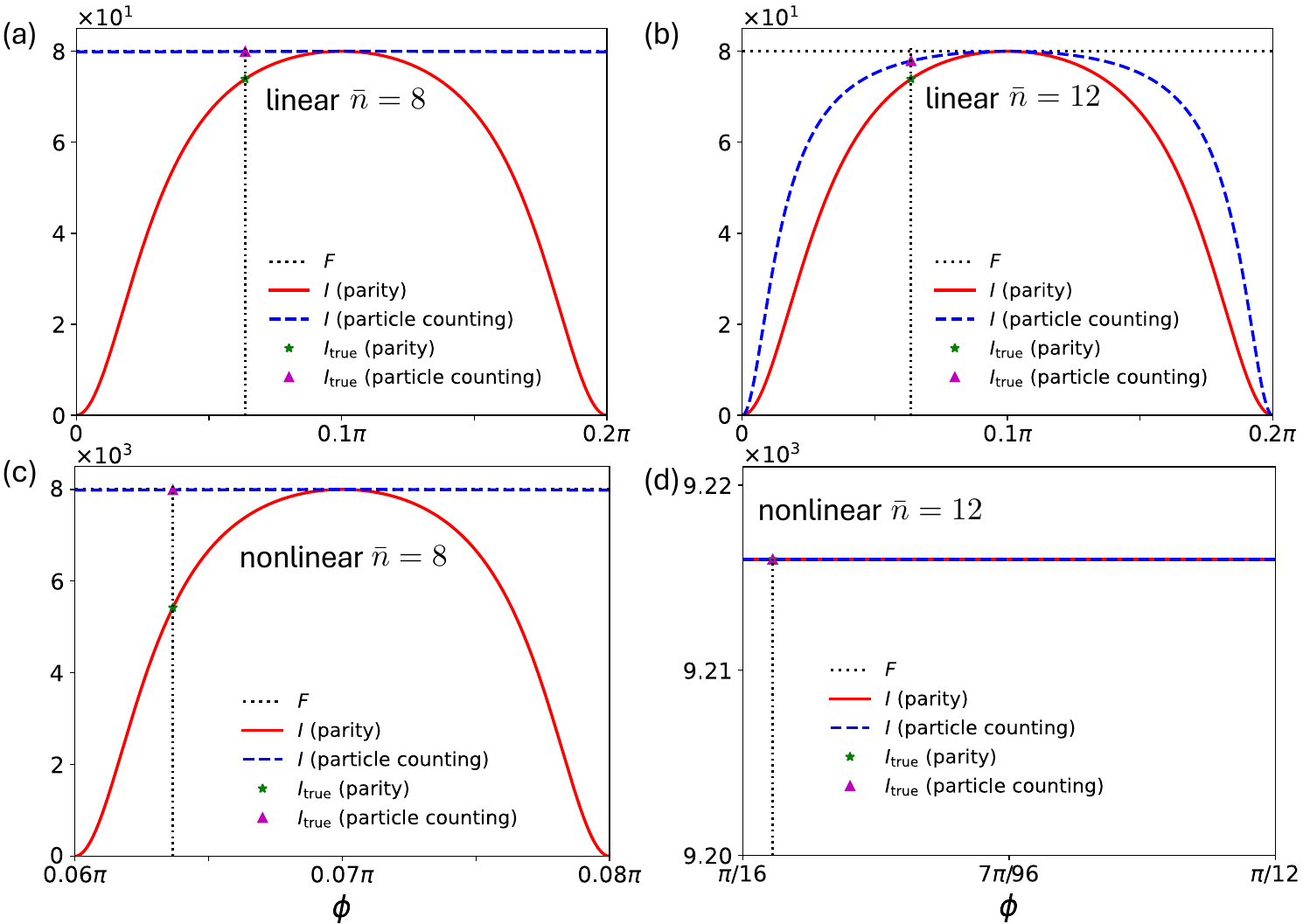}
\caption{CFI and QFI for parity and particle-counting measurements in the case of both linear 
and nonlinear cases with different values of average particle number. (a) and (b) show the 
results of the linear case with $\bar{n}=8$ and $\bar{n}=12$, respectively. (c) and (d) 
show the results of the nonlinear case with $\bar{n}=8$ and $\bar{n}=12$, respectively.
The dotted black line, solid red line, and dashed blue line represent the QFI, the CFI for 
parity measurement, and the CFI for particle-counting measurement, respectively. In the figure 
$N=10$.}
\label{fig:apx_CQFI}
\end{figure*}

The optimality of the parity and particle-counting measurement usually relies on the true value of $\phi$. 
As shown in Fig.~\ref{fig:apx_CQFI}, in the linear case with $\bar{n}=8,12$, the CFI with respect to the 
parity (solid red line) and particle-counting measurement (dashed blue line) can only reach the QFI (dotted black 
line) at some specific value of $\phi$. A similar phenomenon occurs in the nonlinear case with $\bar{n}=8$. In 
the nonlinear case with $\bar{n}=12$, both parity and particle-counting measurements are optimal for all values 
of $\phi$. 

To overcome the dependence of optimality on the true value, adaptive measurement has to be involved. In the 
adaptive measurement, a tunable phase $\phi_{\mathrm{u}}$ is included on mode $a$, and the total phase difference 
now becomes $\phi+\phi_{\mathrm{u}}$. In each round of the measurement, parity or particle-counting measurements are 
performed and a new value of $\phi_u$ is calculated and used in the next round. The specific process of adaptive 
measurement and corresponding thorough calculations can be found in a recent review~\cite{Liu2022}. 

\begin{figure*}[tp]
\centering\includegraphics[width=16.8cm]{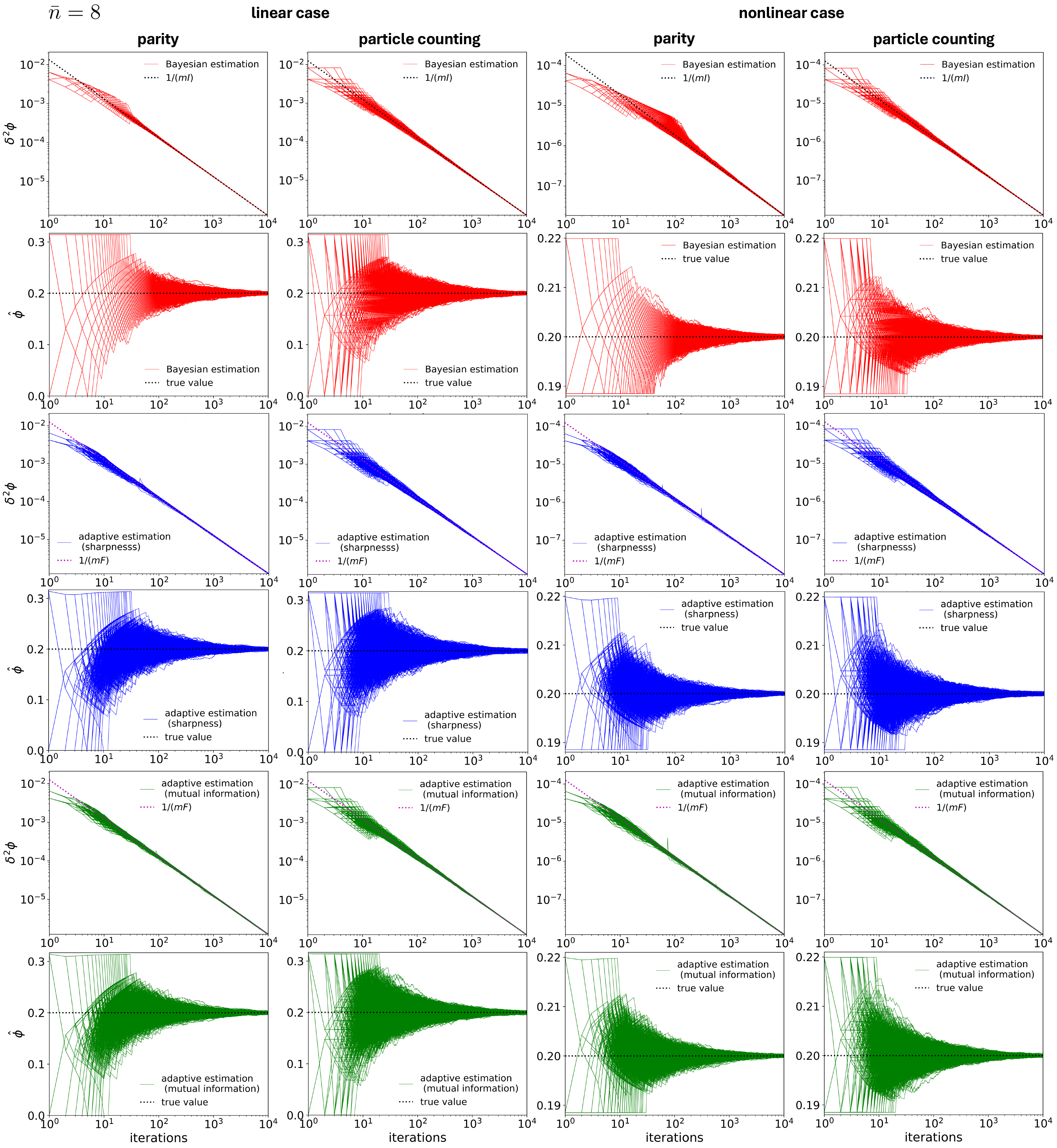}
\caption{Performance of $\hat{\phi}$ and $\delta^2 \phi$ of 2000 rounds simulations 
for the adaptive measurement in the case of $\bar{n}=8$. In the figure the true value of $\phi$ is 
taken as 0.2 and $N=10$.}
\label{fig:apx_n8}
\end{figure*}

\begin{figure*}[tp]
\centering\includegraphics[width=16.8cm]{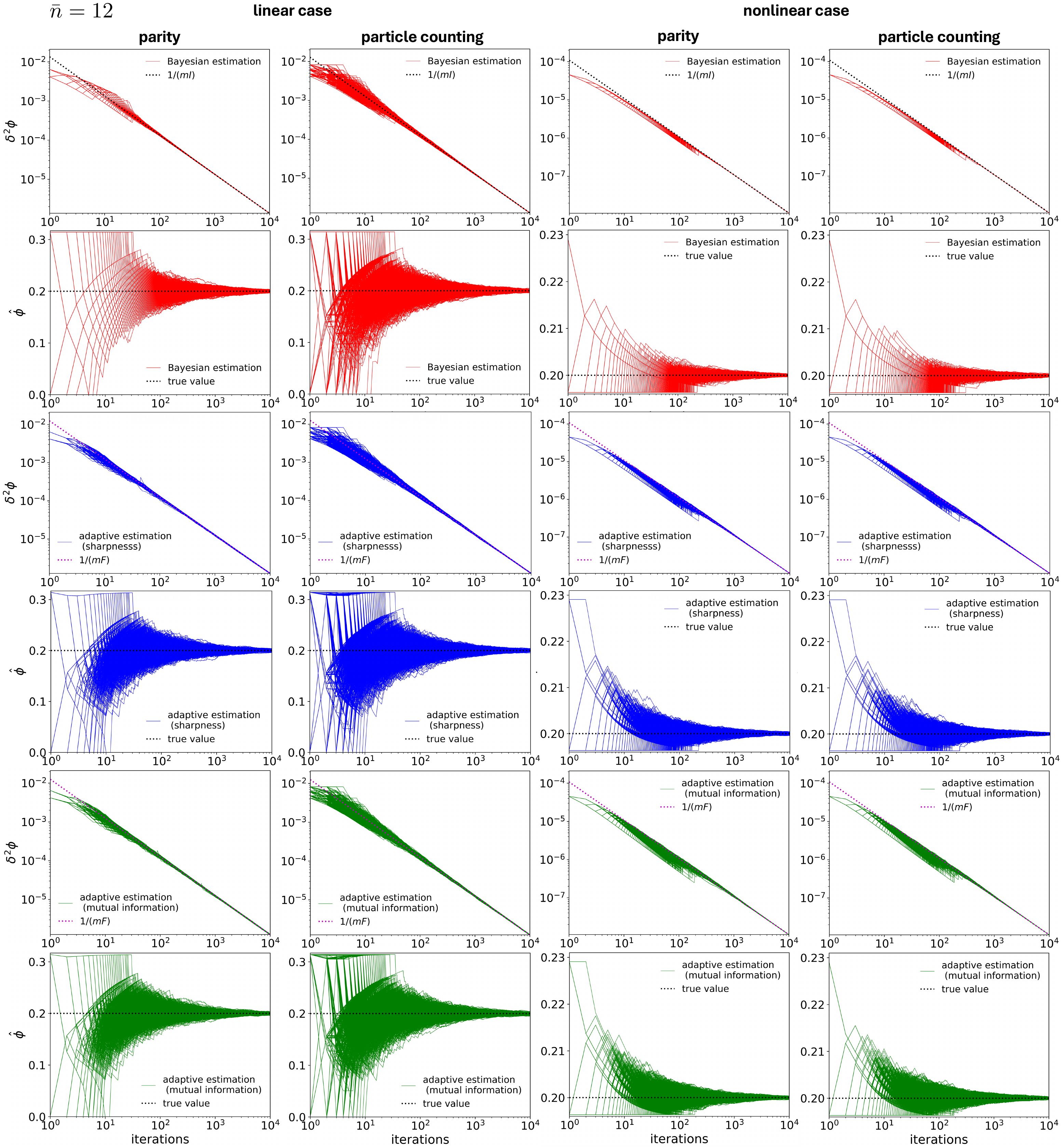}
\caption{Performance of $\hat{\phi}$ and $\delta^2 \phi$ of 2000 rounds simulations 
for the adaptive measurement in the case of $\bar{n}=12$. In the figure the true value of $\phi$ is 
taken as 0.2 and $N=10$.}
\label{fig:apx_n12}
\end{figure*}

In this paper, we use the average sharpness functions~\cite{Holevo1984,Berry2000,Berry2001,Hentschel2010,Huang2017,
DiMario2020,Garcia2022,Zhang2022} and mutual information~\cite{DiMario2020,Garcia2022,Zhang2022,Bargatin2005,
Rzadkowski2017,Cover1991} as the objective function to update $\phi_{\mathrm{u}}$. The sharpness function in the $(k+1)$th round 
of iteration can be expressed by~\cite{Holevo1984,Berry2000}
\begin{equation}
S_{k+1}(\phi_{\mathrm{u}})= \frac{\left|\int^{2\pi}_{0} P(y|\phi,\phi_{\mathrm{u}})P_{k+1}(\phi)e^{i\phi} d\phi \right|}
{\int^{2\pi}_0 P(y|\phi,\phi_{\mathrm{u}})P_{k+1}(\phi) d\phi},
\label{eq:apx_sharpness}
\end{equation}
where $P_{k+1}(\phi)$ is the prior probability in $(k+1)$th round. It is updated via the Bayes' rule, namely, it is 
taken as the posterior distribution $P_{k}(\phi|y,\phi_{{\mathrm{u}},k-1})$ obtained in $k$th round. According to 
the Bayes' theorem, the posterior distribution can be expressed by 
\begin{equation}
P_{k}(\phi|y,\phi_{{\mathrm{u}},k-1}) = \frac{P(y|\phi,\phi_{{\mathrm{u}},k-1})P_{k}(\phi)}
{\int^{2\pi}_0 P(y|\phi,\phi_{{\mathrm{u}},k-1})P_{k}(\phi)d\phi}, 
\end{equation}
where $\phi_{{\mathrm{u}},k-1}$ is the value of $\phi_{\mathrm{u}}$ obtained in the $(k-1)$th round and used in the $k$th 
round. $P_{k}(\phi)$ is the prior distribution in the $k$th round. $P(y|\phi,\phi_{{\mathrm{u}},k-1})$ is the conditional 
probability for the result $y$. For parity measurement, in the linear case $P(y|\phi,\phi_{{\mathrm{u}},k-1})$ is in the 
forms of Eqs.~(\ref{eq:apx_parity_LinProleq}) and (\ref{eq:apx_parity_LinProleq1}) when $\bar{n}\leq N$, and in the forms 
of Eqs.~(\ref{eq:apx_parity_LinProgeq}) and (\ref{eq:apx_parity_LinProgeq1}) when $\bar{n}\geq N$. In the nonlinear case, 
it takes the form of Eqs.~(\ref{eq:apx_parity_NonProleq}) and (\ref{eq:apx_parity_NonProleq1}) when $\bar{n}\leq N$, and 
Eq.~(\ref{eq:apx_parity_NonProgeq}) when $\bar{n}\geq N$. For the measurement of particle counting, it takes the form of 
Eqs.~(\ref{eq:apx_phocount_LinProleq}) and (\ref{eq:apx_phocount_LinProgeq}) in the linear case, and 
Eqs.~(\ref{eq:apx_phocount_NonProleq}) and (\ref{eq:apx_phocount_NonProgeq}) in the nonlinear case. For the formulas 
of conditional probability mentioned above, $\phi$ in the formulas should be replaced with $\phi+\phi_{\mathrm{u}}$.  

An alternative choice of sharpness is replacing $\exp(i\phi)$ in Eq.~(\ref{eq:apx_sharpness}) with $\exp(i2\pi\phi/T)$, 
as done in Refs.~\cite{Berry2000,Berry2001,Huang2017}. Here $T$ is the period of the conditional probability. However, 
the performance of the adaptive measurement has no significant difference for these two formulas according to our test. 
Hence, in this paper we use Eq.~(\ref{eq:apx_sharpness}) as the objective function. 

In the $(k+1)$th round, the value of $\phi_{\mathrm{u}}$ (denoted by $\phi_{{\mathrm{u}},k+1}$) is taken as the argument 
that can maximize the average sharpness, 
\begin{equation}
\phi_{{\mathrm{u}},k+1}\!=\!\mathrm{argmax}\,\sum_{y}\left|\int^{2\pi}_0 e^{i\phi}
P(y|\phi,\phi_{{\mathrm{u}}})P_{k+1}(\phi)d\phi \right|.      
\end{equation}

Apart from the sharpness function, the mutual information can also be used as the objective function for the update of 
$\phi_{\mathrm{u}}$. In our case, the average mutual information in the $(k+1)$th round of iteration can be expressed 
by~\cite{DiMario2020,Cover1991}
\begin{align}
\mathcal{I}_{k+1}(\phi_{\mathrm{u}}) =& \sum_y \int^{2\pi}_0 d\phi\,P(y|\phi,\phi_{{\mathrm{u}}})P_{k+1}(\phi) 
\nonumber \\ 
& \times\log_2 \left[\frac{P(y|\phi,\phi_{{\mathrm{u}}})}{\int^{2\pi}_0 P(y|\phi,\phi_{{\mathrm{u}}})
P_{k+1}(\phi)d\phi}\right].   
\end{align} 
The value of $\phi_{\mathrm{u}}$ in the $(k+1)$th round is taken as 
\begin{equation}
\phi_{{\mathrm{u}},k+1}=\mathrm{argmax}~\mathcal{I}_{k+1}(\phi_{\mathrm{u}}).
\end{equation}

In this paper, the experimental results are simulated via a random number $s\in [0,1]$. The regime $[0,1]$ is 
separated into $m$ parts according to the distribution of the conditional probability. Here $m$ is the number 
of measurement results. The width of the $k$th ($k=1,2,\dots,m$) regime is equivalent to the value of the conditional 
probability for the $k$th result. In one round of the simulation, a random value of $s$ is generated, and if this 
value is located in the $k$th regime, then the $k$th result is then taken as the simulated experimental result. 

The classical estimation in this paper is finished by the maximum a posterior method, namely, the estimated value 
$\hat{\phi}$ in the $k$th round is obtained via the following equation
\begin{equation}
\hat{\phi}_k=\mathrm{argmax}~P_k(\phi|y,\phi_{{\mathrm{u}},k-1}).    
\end{equation}
The variance $\delta^2 \phi$ in the $k$th round can be calculated by 
\begin{align}
\delta^2 \phi=& \int\phi^2 P_k(\phi|y,\phi_{{\mathrm{u}},k-1}) d\phi \nonumber \\
& -\left(\int\phi P_k(\phi|y,\phi_{{\mathrm{u}},k-1}) d\phi\right)^2.    
\end{align}

In the adaptive measurement, the true value of $\phi$ in all examples is taken as $0.2$. The corresponding values 
of CFI are illustrated in Fig.~\ref{fig:apx_CQFI}. $2000$ rounds of experiments are simulated and the corresponding 
performance of $\hat{\phi}$ and $\delta^2\phi$ are shown in Fig.~\ref{fig:apx_n8} for $\bar{n}=8$ and 
Fig.~\ref{fig:apx_n12} for $\bar{n}=12$. The average performance of $2000$ rounds is given in the main text. The 
true values of $\phi$ in these figures are taken as $0.2$.

\section{Calculations under the noise of particle loss}
\label{sec:Calnoise}

\subsection{Expressions of the reduced density matrices}

The particle loss in the MZI can be modeled by the fictitious beam splitters~\cite{Barnett1998,Gardiner2004,
Gardiner2004,Rubin2007,Huver2008,Dorner2009,Dobrzanski2009,Zhang2013,Liu2013,Knott2014}, which can be expressed by 
\begin{eqnarray}
B_{ac} &=& e^{i\frac{\eta_{1}}{2}\left(a^{\dag}c+ac^{\dag}\right)}, \\
B_{bd} &=& e^{i\frac{\eta_{2}}{2}\left(b^{\dag}d+bd^{\dag}\right)},
\end{eqnarray}
where $c$ and $d$ are two fictitious modes representing the particle loss. The transmission coefficients for these 
two beam splitters are $T_1=\cos^2(\eta_1/2)$ and $T_2=\cos^2(\eta_2/2)$. When $T_1=1$ ($T_2=1$), no particle leaks 
from $a$ ($b$) mode, and when $T_1=0$ ($T_2$=0), all particles leak from $a$ ($b$) mode. As a matter of fact, these 
two fictitious beam splitters can be placed either in front of or behind the phase shifts, which would not cause 
different results~\cite{Huver2008,Dorner2009}.

Taking into account the fictitious modes $c$ and $d$, the total probe state can be written as 
\begin{equation}
\ket{\psi_{\mathrm{tot}}}=\ket{\psi_{\mathrm{opt}}}\ket{0}_{c}\ket{0}_{d}.     
\end{equation}
After going through the fictitious beam splitters, the state becomes mixed and the corresponding density matrix can 
be expressed by 
\begin{equation}
\rho=\mathrm{Tr}_{cd}\left(B_{bd}B_{ac}\ket{\psi_{\mathrm{tot}}}
\bra{\psi_{\mathrm{tot}}}{B^{\dag}_{ac}}{B^{\dag}_{bd}}\right),
\end{equation}
where $\mathrm{Tr}_{cd}(\cdot)$ is the partial trace on the modes $c$ and $d$. Notice that $\ket{\psi_{\mathrm{opt}}}$ 
already includes the influence of the first beam splitter, if there is one. The state above is actually the state before 
going through the phase shifts. 

Now let us first consider the OFPS for $\bar{n}\leq N$ in the linear case, which is 
\begin{equation}
\sqrt{\frac{N-\bar{n}}{N}}\ket{0 0} +\sqrt{\frac{\bar{n}}{2N}}
\left(e^{i\theta_1}\ket{0 N}+e^{i\theta_2}\ket{N 0}\right). 
\end{equation}
Utilizing the equations
\begin{align}
& e^{i \frac{\eta_{1}}{2}(a^{\dag}c+ac^{\dag})}\ket{N 0}\ket{0}_{c}\ket{0}_d \nonumber \\
= & \sum_{k=0}^N\binom{N}{k}^{\frac{1}{2}} i^k T_1^{\frac{1}{2}(N-k)}R_1^{\frac{k}{2}}\ket{N-k,0}\ket{k}_c\ket{0}_d,
\end{align}
and 
\begin{align}
& e^{i \frac{\eta_{2}}{2}(b^{\dag}d+bd^{\dag})}\ket{0 N}\ket{0}_{c}\ket{0}_d \nonumber \\
 = & \sum_{k=0}^N\binom{N}{k}^{\frac{1}{2}}\!\!i^k T_2^{\frac{1}{2}(N-k)} R_2^{\frac{k}{2}}\ket{0,N-k}\ket{0}_c\ket{k}_d,
\end{align}
where $R_{1(2)}=1-T_{1(2)}$, the reduced density matrix can be expressed by 
\begin{align}
\rho = & \frac{N-\bar{n}}{N} \ket{0 0}\bra{0 0}+\sqrt{\frac{\bar{n}(N-\bar{n})}{2 N^2}}\rho_1\nonumber\\
&+\frac{\bar{n}}{2 N}\sum_{k=0}^N \binom{N}{k}\rho_{2,k}+\frac{\bar{n}}{2 N} \rho_3,
\label{eq:apx_linear_rho_lossleq}
\end{align}
where 
\begin{eqnarray}
\rho_1 &=& T_1^{\frac{N}{2}}\left(e^{-i\theta_2}\ket{0 0}\bra{N 0}+e^{i\theta_2}\ket{N 0}\bra{0 0}\right) \nonumber \\
& & +T_2^{\frac{N}{2}}\left(e^{-i\theta_1}\ket{0 0}\bra{0 N}+e^{i\theta_1}\ket{0 N}\bra{0 0}\right),
\end{eqnarray}
and 
\begin{eqnarray}
\rho_{2,k} &=& T_1^{N-k}R_1^k\ket{N-k,0}\bra{N-k,0} \nonumber \\
& & +T_2^{N-k}R_2^k\ket{0,N-k}\bra{0,N-k},    
\end{eqnarray}
and 
\begin{equation}
\rho_3=\left(T_1 T_2\right)^{\frac{N}{2}}\left[e^{i(\theta_1-\theta_2)}\ket{0 N}\bra{N 0}
+e^{i(\theta_2-\theta_1)}\ket{N 0}\bra{0 N}\right].   
\end{equation}

In the linear case with $\bar{n}\geq N$, the OFPS reads 
\begin{equation}
\sqrt{\frac{2N-\bar{n}}{2N}}\left(e^{i\theta_1}\ket{0 N}
+e^{i\theta_2}\ket{N 0}\right)+\sqrt{\frac{\bar{n}-N}{N}}\ket{N N}. 
\end{equation}
Then the reduced density matrix can be written as
\begin{align}
\rho = & \frac{2N-\bar{n}}{2N}\left[\sum_{k=0}^N \binom{N}{k}\rho_{2,k}+\rho_3\right]\nonumber\\
& + \sqrt{\frac{(2N-\bar{n})(\bar{n}-N)}{2 N^2}} \sum_{k=0}^N \binom{N}{k}\left(\rho_{4,k}+\rho_{5,k}\right) \nonumber\\
& +\frac{\bar{n}-N}{N}\!\!\sum_{k,l=0}^N 
\!\! \binom{N}{k}\!\binom{N}{l}\rho_{6,kl},
\label{eq:apx_linear_rho_lossgeq}
\end{align}
where 
\begin{eqnarray}
\rho_{4,k} &=& T_2^{N-k}R_2^k T_1^{\frac{N}{2}}\left(e^{i\theta_1}\ket{0,N-k}\bra{N,N-k}\right.\nonumber\\
& & \left.+e^{-i\theta_1}\ket{N,N-k}\bra{0,N-k}\right),    
\end{eqnarray}
and 
\begin{eqnarray}
\rho_{5,k} &=& T_1^{N-k}R_1^k T_2^{\frac{N}{2}}\left(e^{i\theta_2}\ket{N-k,0}\bra{N-k,N}\right.\nonumber\\
& & \left.+e^{-i\theta_2}\ket{N-k,N}\bra{N-k,0}\right),  
\end{eqnarray}
and 
\begin{equation}
\rho_{6,kl} = T_1^{N-k}R_1^k T_2^{N-l}R_2^l\ket{N-k,N-l}\bra{N-k,N-l}.   
\end{equation}

In the nonlinear case, the OFPS is the same as the counterpart in the linear case when $\bar{n}\leq N$,  
thus, the corresponding reduced density matrix is also in the form of Eq.~(\ref{eq:apx_linear_rho_lossleq}). 
When $\bar{n}\geq N$, we consider a simple case of the OFPS  
\begin{equation}
\frac{1}{\sqrt{2}}\left(\ket{\bar{n}-N, N}+e^{i\theta}\ket{N, \bar{n}-N}\right)    
\end{equation}
with $\bar{n}$ an integer in the regime $[N,\left\lfloor\frac{4N+1}{3}\right\rfloor]$. 
In this case, the reduced density matrix reads 
\begin{align}
\rho = & \frac{1}{2}\sum_{k=0}^{\bar{n}-N} \sum_{l=0}^N\binom{\bar{n}-N}{k}\binom{N}{l}\rho_{7,kl}
\nonumber\\
&+\frac{1}{2}\sum_{k,l=0}^{\bar{n}-N} \binom{\bar{n}-N}{k}^{\frac{1}{2}}\binom{N}{k}^{\frac{1}{2}}
\binom{\bar{n}-N}{l}^{\frac{1}{2}}\binom{N}{l}^{\frac{1}{2}}\rho_{8,kl},
\label{eq:apx_nonlinear_rho_lossgeq}
\end{align}
where 
\begin{eqnarray}
& &\rho_{7,kl} \nonumber \\
&=& T_1^{\bar{n}-N-k}R_1^kT_2^{N-l}R_2^l \nonumber \\
& &\times \ket{\bar{n}-N-k,N-l}\bra{\bar{n}-N-k,N-l} \nonumber \\
& &+T_1^{N-l}R_1^l T_2^{\bar{n}-N-k}R_2^k \nonumber \\
& &\times\ket{N-l,\bar{n}-N-k}\!\bra{N-l,\bar{n}-N-k},     
\end{eqnarray}
and 
\begin{eqnarray}
& & \rho_{8,kl} \nonumber \\
&=& T_1^{\frac{\bar{n}}{2}-k}R_1^kT_2^{\frac{\bar{n}}{2}-l}R_2^l \nonumber \\
& &\times\left(e^{-i\theta}\!\ket{\bar{n}\!-\!N\!-\!k,N\!-\!l}\!\bra{N\!-\!k,\bar{n}\!-\!N\!-\!l}\right. \nonumber\\
& &\left. +e^{i\theta}\ket{N\!-\!k,\bar{n}\!-\!N\!-\!l}\bra{\bar{n}\!-\!N\!-\!k,N\!-\!l}\right)\!.     
\end{eqnarray}

The QFIs for these reduced density matrices are calculated numerically via QuanEstimation~\cite{Zhang2022}. All the 
scripts for these calculations will be integrated into the SU(2) interferometer module of QuanEstimation, and will be 
announced as soon as possible. 

\subsection{Conditional probabilities for parity and particle-counting measurements}

In this section, we provide the expression of the conditional probability for parity and particle-counting measurements 
in both linear and nonlinear cases. 

\subsubsection{Parity measurement}

We first discuss the linear case. When the particle loss exists, the state before going through the phase shifts is in 
the form of Eq.~(\ref{eq:apx_linear_rho_lossleq}), thus, the expectation of the parity operator reads 
\begin{align}
\expval{\Pi_a} & = \mathrm{Tr}\left(\Pi_a e^{i\frac{\pi}{2}J_x} e^{i\phi J_z}\rho
e^{-i\phi J_z} e^{-i\frac{\pi}{2}J_x} \right) \nonumber \\
& =  \mathrm{Tr}\left(e^{i\frac{\pi}{2}n} e^{-i\pi J_y}e^{i\phi J_z}\rho e^{-i\phi J_z}\right) \nonumber\\
& = 1-\frac{\bar{n}}{N}\left[\Omega-\left(T_1 T_2\right)^{\frac{N}{2}}\cos \beta_1 \right],
\label{eq:apx_parity1_loss}
\end{align}
where 
\begin{equation}
\Omega:= 1-\frac{1}{2}\left(R_1^N+R_2^N\right), 
\label{eq:apx_Omega}
\end{equation}
and $\beta_1$ is given by Eq.~(\ref{eq:apx_beta1}). According to the conditions $\expval{\Pi_a} = P_+ - P_-$ and 
$P_+ + P_- = 1$, the probability can be calculated as 
\begin{align}
P_+ &= 1-\frac{\bar{n}}{2N}\left[\Omega-\left(T_1 T_2\right)^{\frac{N}{2}}\cos \beta_1 \right],  \\
P_-  &= \frac{\bar{n}}{2N}\left[\Omega-\left(T_1 T_2\right)^{\frac{N}{2}}\cos \beta_1\right],
\label{eq:apx_parity_LinProleq_loss}
\end{align}
and the CFI can be written as
\begin{align}
\frac{\bar{n}N^2 \left(T_1 T_2\right)^N \sin^2\beta_1}
{\left[\Omega\!-\!\left(T_1 T_2\right)^{\frac{N}{2}}\!\cos\beta_1\!\right]
\!\!\left\{2N\!-\!\bar{n}\!\left[\Omega\!-\left(T_1 T_2\right)^{\frac{N}{2}}\!\cos \beta_1\!\right]\!\right\}}.
\end{align}

Based on the expression above, the maximum CFI ($I_{\max}$) with respect to $\beta_1$ reads 
\begin{align}
& \bar{n}N\Omega-\frac{1}{2}\bar{n}\bigg\{\bar{n}\left[\Omega^2-(T_1T_2)^N\right] \nonumber\\
& +\sqrt{\left[(2N-\bar{n}\Omega)^2-(T_1 T_2)^N\bar{n}^2\right]\left[\Omega^2-(T_1 T_2)^N\right]}\bigg\}, 
\end{align}
which can be attained when $\cos\beta_1=0$ for $N = \bar{n}\Omega$, and $\cos\beta_1$ equals to 
\begin{align}
& \frac{1}{2(T_1T_2)^{\frac{N}{2}}(N-\bar{n}\Omega)}
\bigg\{2N\Omega-\left[(T_1 T_2)^N+\Omega^2\right]\bar{n} \nonumber \\
& -\sqrt{\left[\Omega^2-(T_1 T_2)^N\right]\bar{n}^2-4\bar{n}N\Omega+4N^2} \nonumber \\
& \times\sqrt{\Omega^2-(T_1 T_2)^N}\bigg\}
\end{align}
for $N \neq \bar{n}\Omega$. Then the optimal points of the true values of $\phi$ can be located accordingly. 

In the case that $\bar{n}\geq N$, the reduced density matrix is in the form of Eq.~(\ref{eq:apx_linear_rho_lossgeq}),
and the expectation of $\Pi_a$ is 
\begin{equation}
\expval{\Pi_a} = \kappa + \frac{2N-\bar{n}}{N}\left(T_1 T_2\right)^{\frac{N}{2}}\!\cos \beta_1,
\end{equation}
where 
\begin{align}
\kappa:=&\frac{\bar{n}-N}{N}\sum_{k=0}^{N}\binom{N}{k}^2\left(T_1 T_2\right)^{N-k}\left(R_1R_2\right)^k \nonumber\\
& + \frac{2N-\bar{n}}{N}\left(1-\Omega\right),
\label{eq:apx_kappa}
\end{align}
which further gives the expressions of $P_+$ and $P_-$ as follows:
\begin{align}
P_+ & = \frac{1}{2}(1+\kappa)+\frac{2N-\bar{n}}{2N}\left(T_1 T_2\right)^{\frac{N}{2}}\cos \beta_1, \\
P_- & = \frac{1}{2}(1-\kappa)-\frac{2N-\bar{n}}{2N}\left(T_1 T_2\right)^{\frac{N}{2}}\cos \beta_1.
\label{eq:apx_parity_LinProgeq_loss}
\end{align}
The CFI then reads
\begin{equation}
\frac{(2N-\bar{n})^2\left(T_1 T_2\right)^N\sin^2 \beta_1}{1-\left[\kappa\!+\!\frac{2N-\bar{n}}{N}(T_1 T_2)^{\frac{N}{2}}
\cos \beta_1 \right]^2}.
\end{equation}
The maximum CFI ($I_{\max}$) with respect to $\beta_1$ reads 
\begin{align}
&\frac{1}{2}\bigg\{N^2(1-\kappa^2)+(2N-\bar{n})^2(T_1T_2)^N\nonumber\\
&-\!\!\sqrt{\left[(2N-\bar{n})^2(T_1 T_2)^N-N^2(1+\kappa^2)\right]^2-4N^4\kappa^2}\bigg\},
\end{align}
which can be attained when $\cos\beta_1$ equals to 
\begin{align}
& \frac{1}{2(T_1T_2)^{\frac{N}{2}}N(2N-\bar{n})\kappa}\bigg\{N^2(1-\kappa^2)  \nonumber\\
& -\sqrt{\left[(2N-\bar{n})^2(T_1 T_2)^N-N^2(1+\kappa^2)\right]^2-4N^4\kappa^2}\bigg\} \nonumber \\
& -\frac{(2N-\bar{n})(T_1T_2)^\frac{N}{2}}{2N\kappa}. 
\end{align}
Then the optimal points of the true values of $\phi$ can be located accordingly. 

\begin{figure}[tp]
\centering\includegraphics[width=8.cm]{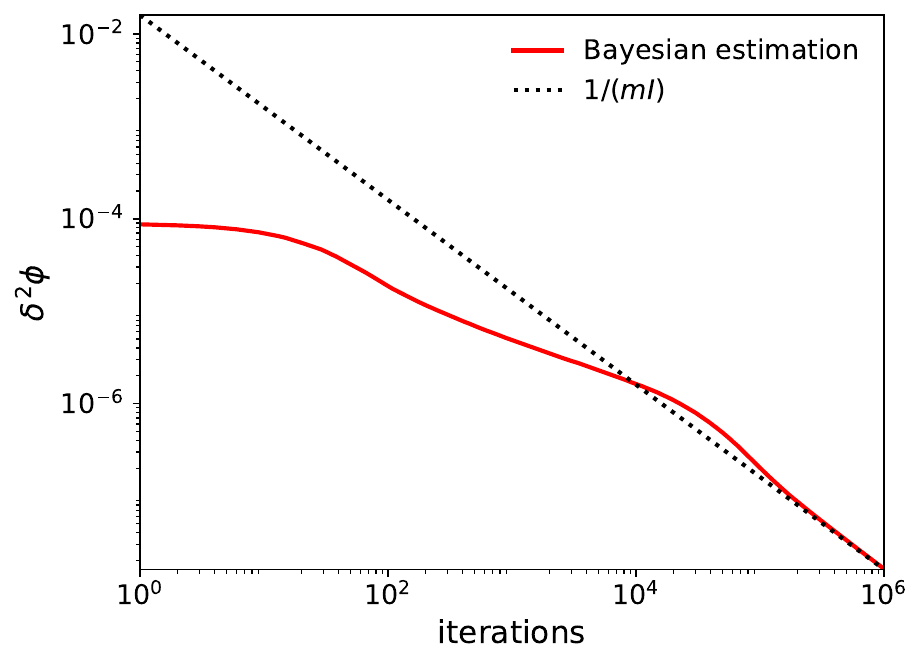}
\caption{Noisy performance of Bayesian estimation for parity measurement in the nonlinear case. 
The average particle number $\bar{n}=12$, $N=10$, and the transmission rates $T_1=T_2=0.9$. }
\label{fig:apx_Bayes}
\end{figure}

In the nonlinear case, the reduced density matrix is given by Eq.~(\ref{eq:apx_linear_rho_lossleq}) when 
$\bar{n}\leq N$. For this state, the expectation of the parity operator is
\begin{equation}
\expval{\Pi_a}=1-\frac{\bar{n}}{N}\left[\Omega-(T_1 T_2)^{\frac{N}{2}}\cos\beta_2\right],
\end{equation}  
where $\beta_2$ is given by Eq.~(\ref{eq:apx_beta2}). The corresponding probabilities $P_{\pm}$ are 
\begin{align}
P_+ & = 1 - \frac{\bar{n}}{2N}\left[\Omega-(T_1 T_2)^{\frac{N}{2}}\cos \beta_2\right], \\
P_- & =  \frac{\bar{n}}{2N}\left[\Omega-(T_1 T_2)^{\frac{N}{2}}\cos \beta_2\right].
\label{eq:apx_parity_NonProleq_loss}
\end{align}
The CFI is
\begin{equation}
\frac{\bar{n}N^4(T_1 T_2)^{N}\sin^2 \beta_2}{\left[\Omega-\!(T_1 T_2)^{\frac{N}{2}}\!\cos \beta_2\right]\!\!
\left\{2N\!-\!\bar{n}\!\left[\Omega-\!(T_1 T_2)^{\frac{N}{2}}\!\cos \beta_2\right]\!\right\}}.
\end{equation}
In this case, the maximum CFI ($I_{\max}$) with respect to $\beta_2$ is
\begin{align}
& \bar{n}N^3\Omega-\frac{1}{2}\bar{n}N^2\bigg\{\bar{n}\left[\Omega^2-(T_1T_2)^N\right]\nonumber\\
&+\sqrt{\left[(2N-\bar{n}\Omega)^2-(T_1 T_2)^N\bar{n}^2\right]\left[\Omega^2-(T_1 T_2)^N\right]}\bigg\}, 
\end{align}
where $\Omega$ is defined in Eq.~(\ref{eq:apx_Omega}). $I_{\max}$ can be attained when $\cos\beta_2 = 0$ for 
$N=\bar{n}\Omega$, and $\cos\beta_2$ equals to 
\begin{align}
& \frac{1}{2(T_1T_2)^{\frac{N}{2}}(N-\bar{n}\Omega)} \bigg\{ 2N\Omega-\left[(T_1 T_2)^N
+\Omega^2\right]\bar{n} \nonumber \\
& -\!\sqrt{\left[\Omega^2\!-\!(T_1 T_2)^N\right]\left\{\left[\Omega^2\!-\!(T_1 T_2)^N\right]\bar{n}^2
\!-\!4\bar{n}N\Omega+4N^2\right\}}\bigg\} 
\end{align}
for $N\neq \bar{n}\Omega$. Then the optimal points of the true values of $\phi$ can be located accordingly. 
In the case that $\bar{n} \geq N$, we also consider the simple case that $\bar{n}$ is an integer in the regime 
$[N,\left\lfloor\frac{4N+1}{3}\right\rfloor]$. The corresponding reduced density matrix is 
given in Eq.~(\ref{eq:apx_nonlinear_rho_lossgeq}). For this state, the value of $\expval{\Pi_a}$ reads
\begin{align}
\expval{\Pi_a} = & \sum^{\bar{n}-N}_{k=0} \binom{\bar{n}-N}{k}
\binom{N}{k}(T_1 T_2)^{\frac{\bar{n}}{2}-k}(R_1 R_2)^k \cos\gamma_k \nonumber\\
& +\frac{1}{2}\sum^{\bar{n}-N}_{k=0} \binom{\bar{n}-N}{k}\binom{N}{\bar{n}-N-k}(T_1 T_2)^{\bar{n}-N-k}\nonumber\\
& \times (R_1 R_2)^{k} \left(R^{2N-\bar{n}}_1+R^{2N-\bar{n}}_2\right),
\end{align}
where $\gamma_k:=\gamma-2k(2N-\bar{n})\phi$ with $\gamma$ given by Eq.~(\ref{eq:apx_gamma}). 
$P_{\pm}=(1\pm \expval{\Pi_a})/2$ can be calculated via the equation above correspondingly. 

\begin{figure}[tp]
\centering\includegraphics[width=8.5cm]{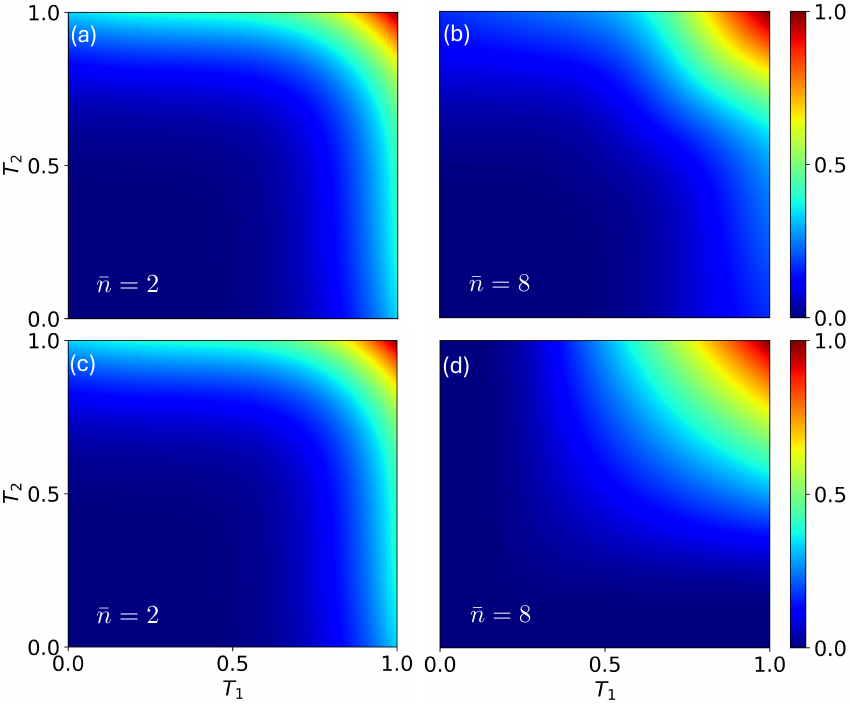}
\caption{Noisy behaviors of the QFI as a function of $T_1$ and $T_2$ in the case of (a) linear 
phase shifts with $\bar{n}<N$ ($\bar{n}=2$), (b) linear phase shifts with $\bar{n}>N$ ($\bar{n}=8$), 
(c) nonlinear phase shifts with $\bar{n}<N$ ($\bar{n}=2$), and (d) nonlinear phase shifts with 
$\bar{n}>N$ ($\bar{n}=8$). In the figure $N=6$. }
\label{fig:apx_lossyQFI}
\end{figure}

With all the expressions of the conditional probabilities, the adaptive measurement can be performed and simulated. 

\subsubsection{Particle-counting measurement}

Here we provide the expressions of the conditional probabilities for the particle-counting measurement in the case 
that particle loss exists. Recall that the reduced density matrix before going through the phase shifts is given in 
Eq.~(\ref{eq:apx_linear_rho_lossleq}) for $\bar{n}\leq N$. Then the probability $P_m$ is 
\begin{align}
P_m = &  \mathrm{Tr}\left(e^{i\phi J_z}\rho e^{-i\phi J_z}
e^{-i\frac{\pi}{2}J_x}\sum^{2N}_{j=0}\ket{m j}\bra{m j}e^{i\frac{\pi}{2}J_x}\right). \nonumber \\
= & \left(1-\frac{\bar{n}}{N}\right)\delta_{0m} +\frac{\bar{n}}{N}\Lambda +h(m-N)2^{-N}\frac{\bar{n}}{N} \nonumber \\
& \times \binom{N}{m} \left(T_1 T_2\right)^{\frac{N}{2}}(-1)^m \cos\beta_1,
\label{eq:apx_PhoCount_LinProleq_loss}
\end{align}
where $h(m-N)$ is the step function defined in Eq.~(\ref{eq:apx_step}), and $\Lambda$ is defined by 
\begin{equation}
\Lambda:=\sum^{N-m}_{k=0}2^{k\!-\!N\!-\!1}\binom{N}{k}\!\binom{N\!-\!k}{m}
\left(T^{N-k}_1 R^k_1\!+\!T^{N-k}_2 R^k_2\right).     
\end{equation}

\begin{figure*}[tp]
\centering\includegraphics[width=16.8cm]{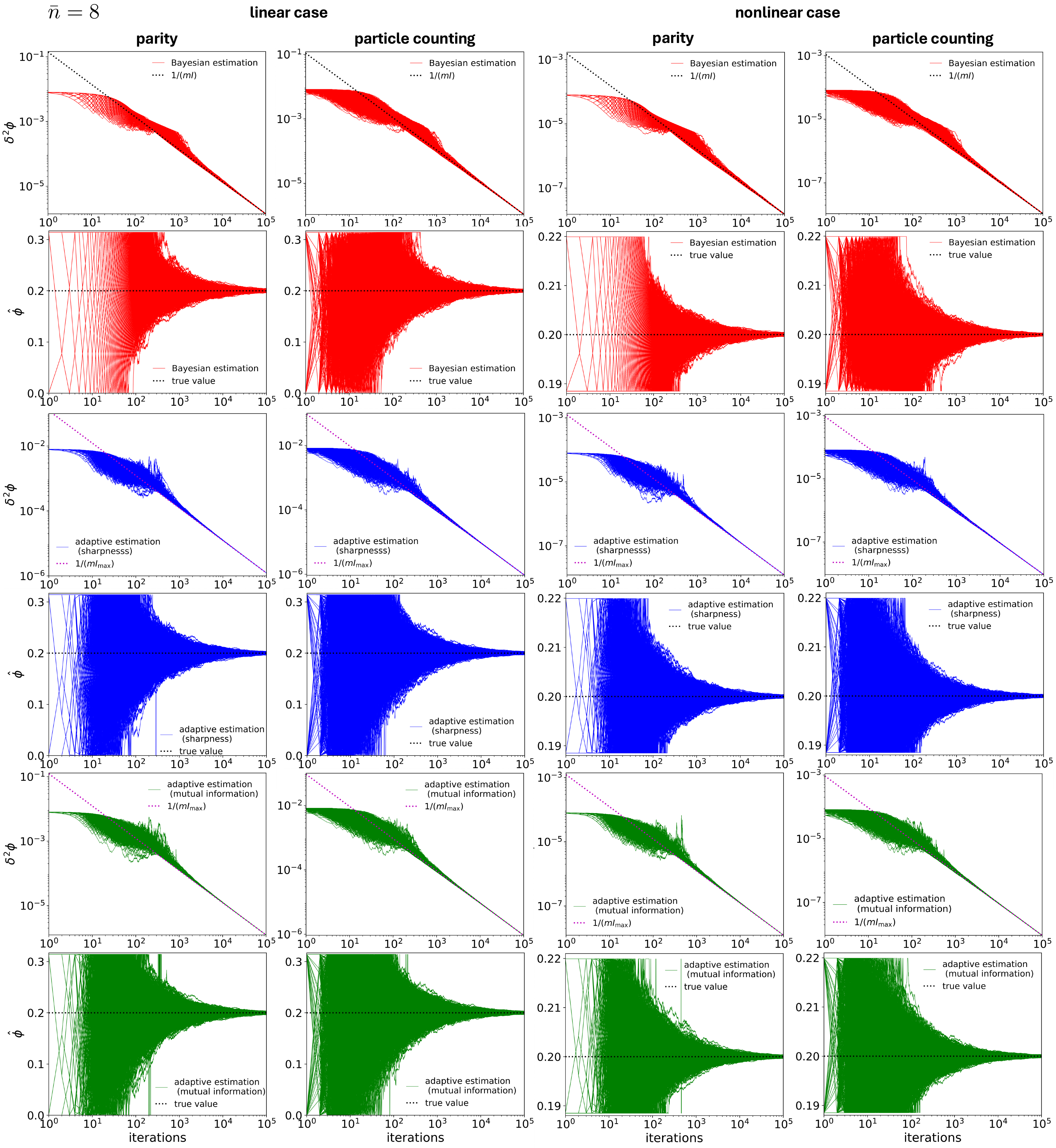}
\caption{Noisy performance of $\hat{\phi}$ and $\delta^2 \phi$ of 2000 rounds simulations 
for the adaptive measurement in the case of $\bar{n}=8$. The true value of $\phi$ is taken as 
$0.2$. The transmission rates are taken as $T_1=T_2=0.9$ and $N=10$.}
\label{fig:apx_n8loss}
\end{figure*}

\begin{figure*}[tp]
\centering\includegraphics[width=16.8cm]{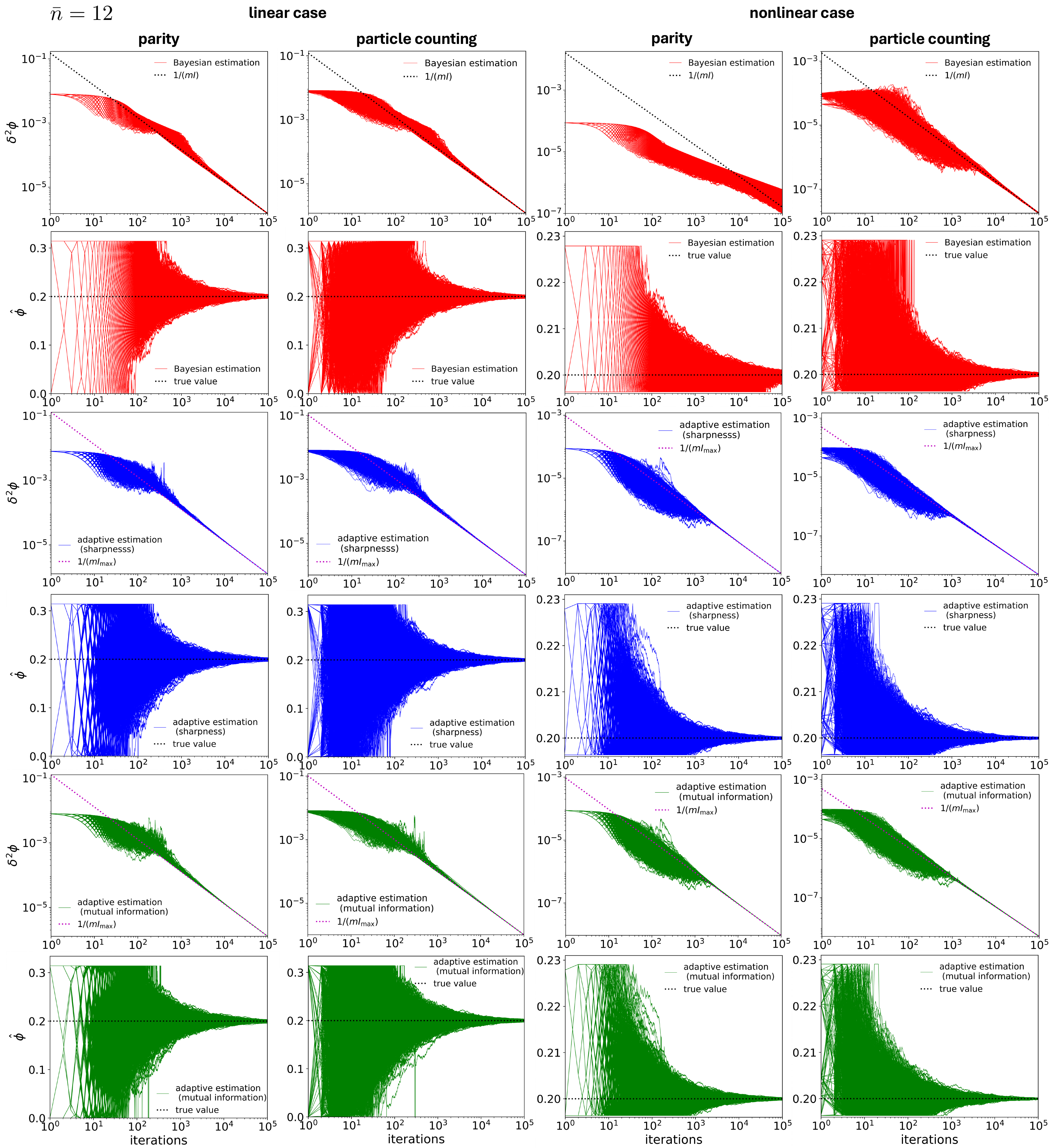}
\caption{Noisy performance of $\hat{\phi}$ and $\delta^2 \phi$ of 2000 rounds simulations 
for the adaptive measurement in the case of $\bar{n}=12$. The true value of $\phi$ is taken 
as $0.2$. The transmission rates are taken as $T_1=T_2=0.9$ and $N=10$.}
\label{fig:apx_n12loss}
\end{figure*}

In the case that $\bar{n}\geq N$, the reduced density matrix is in the form of Eq.~(\ref{eq:apx_linear_rho_lossgeq}), 
and $P_m$ then reads
\begin{widetext}
\begin{align}
P_m = & \left(2-\frac{\bar{n}}{N}\right)\Lambda + h(m-N)\left(2-\frac{\bar{n}}{N}\right)2^{-N}(-1)^m\binom{N}{m}(T_1 T_2)^{\frac{N}{2}}
\cos \beta_1 + 2^{-2N}\left(\frac{\bar{n}}{N}-1\right) \nonumber \\
& \times \sum^{N}_{k,l=0}\! 2^{k+l}\frac{m!(2N-m-k-l)!}{(N-k)!(N-l)!}T^{N-k}_1 R^k_1 T^{N-l}_2 R^l_2 \binom{N}{k}\binom{N}{l}\!\!  
\left[\!\sum^{\min \{N-k,m\}}_{s=\max\{0,m-N+l\}}\!\!(-1)^s \binom{N\!-\!k}{s}\!\binom{N\!-\!l}{m\!-\!s} \right]^2.
\label{eq:apx_PhoCount_LinProgeq_loss}
\end{align}

In the nonlinear case, the reduced density matrix is the same as that in the linear case for $\bar{n}\leq N$, 
namely, Eq.~(\ref{eq:apx_linear_rho_lossleq}). The probability $P_m$ is then calculated as 
\begin{equation}
P_m = \left(1-\frac{\bar{n}}{N}\right)\delta_{0m}+\frac{\bar{n}}{N}\Lambda
+h(m-N)2^{-N}\frac{\bar{n}}{N} \binom{N}{m} (T_1 T_2)^{\frac{N}{2}}(-1)^m \cos \beta_2. 
\end{equation}

When $\bar{n}\geq N$, the reduced density matrix is in the form of Eq.~(\ref{eq:apx_nonlinear_rho_lossgeq}) for 
the simple case that $\bar{n}$ is an integer in the regime $[N,\left\lfloor\frac{4N+1}{3}\right\rfloor]$. 
Hence, the probability can be expressed by 
\begin{align}
P_m = & \sum^{\bar{n}-N}_{k=0}\sum^N_{l=0}2^{k+l-\bar{n}-1} \frac{m!(\bar{n}\!-\!m\!-\!k\!-\!l)!}
{(\bar{n}\!-\!N\!-\!k)!(N\!-\!l)!}\binom{\bar{n}\!-\!N}{k}\!\binom{N}{l}\left[\sum^{\min\{N-l,m\}}_{s=\max\{0,N+m-\bar{n}+k\}}
(-1)^s\binom{N-l}{s}\binom{\bar{n}\!-\!N\!-\!k}{m\!-\!s}\!\right]^2 \nonumber \\
& \times\!\! \left(T^{\bar{n}-N-k}_1 R^k_1 T^{N-l}_2 R^l_2+ T^{N-l}_1 R^l_1 \!T^{\bar{n}-N-k}_2 R^k_2\right) 
+ \sum^{\bar{n}-N}_{k,l=0}\!\!2^{k+l-\bar{n}}\frac{m!(\bar{n}-m-k-l)!k!l!}{(\bar{n}-N)!N!} \binom{\bar{n}-N}{k}
\nonumber \\
&\times\!\binom{N}{k}\!\binom{\bar{n}-N}{l}\!\binom{N}{l}T^{\frac{\bar{n}}{2}-k}_1 R^k_1 
T^{\frac{\bar{n}}{2}-l}_2 R^l_2 \cos\left(\gamma-(k+l)(2N-\bar{n})\phi\right) \nonumber \\
&\times \sum^{\min\{\bar{n}-N-l,m\}}_{s=\max\{0,m-N+k\}}\sum^{\min\{N-l,m\}}_{t=\max\{0,N+m-\bar{n}+k\}}
\!\!\!(-1)^{s+t} \binom{\bar{n}-N-l}{s}\!\binom{N-k}{m-s}\!\binom{\bar{n}-N-k}{m-t}\!\binom{N-l}{t}
\end{align}
for $m\leq\bar{n}$ and zero for $m>\bar{n}$. 
\end{widetext}

The CFIs for these conditional probabilities are calculated numerically via QuanEstimation~\cite{Zhang2022}. 
The average performance of Bayesian estimation for parity measurement in the nonlinear case under noise is 
given in Fig.~\ref{fig:apx_Bayes}. The convergence speed is significantly lower than that in the noiseless 
case, which is reasonable since the actually used particles in the estimation are less than the noiseless case 
in the same time duration. 

Moreover, the noisy behaviors of the QFI as a function of $T_1$ and $T_2$ have been illustrated in 
Fig.~\ref{fig:apx_lossyQFI} for both linear and nonlinear phase shifts. In each plot, the area proportion of 
the ratio $F_{\mathrm{loss}}/F$ that is larger than a given threshold is used to reflect the robustness. Here 
$F_{\mathrm{loss}}$ and $F$ are the QFI for the OFPSs with and without loss, respectively. In this 
paper, two values of the threshold, 0.6 and 0.8, are used to make sure that the result does not rely on the 
choice of this value. 

With all the aforementioned expressions of the conditional probabilities, the adaptive measurement can be performed 
and simulated. 2000 rounds of experiments are simulated and the corresponding performance of $\hat{\phi}$ and 
$\delta^2 \phi$ are shown in Fig.~\ref{fig:apx_n8loss} for $\bar{n}=8$ and Fig.~\ref{fig:apx_n12loss} for $\bar{n}=12$. 
The average performance of 2000 rounds is given in the main text. The true values of $\phi$ in these figures are 
taken as $0.2$, and the transmission rates are taken as $T_1=T_2=0.9$.


\begin{thebibliography}{99}

\bibitem{Caves1980}
C. M. Caves,
Quantum-Mechanical Radiation-Pressure Fluctuations in an Interferometer,
\href{https://doi.org/10.1103/PhysRevLett.45.75}
{Phys. Rev. Lett. \textbf{45}, 75 (1980).}

\bibitem{Caves1981}
C. M. Caves, 
Quantum-mechanical noise in an interferometer, 
\href{https://doi.org/10.1103/PhysRevD.23.1693}
{Phys. Rev. D \textbf{23}, 1693 (1981).}

\bibitem{Yurke1986}
B. Yurke, S. L. McCall, and J. R. Klauder, 
SU(2) and SU(1,1) interferometers, 
\href{https://doi.org/10.1103/PhysRevA.33.4033}
{Phys. Rev. A \textbf{33}, 4033 (1986).}

\bibitem{Giovannetti2004}
V. Giovannetti, S. Lloyd, and L. Maccone,
Quantum-Enhanced Measurements: Beating the Standard Quantum Limit,
\href{https://doi.org/10.1126/science.1104149}
{Science \textbf{306}, 1330 (2004).}

\bibitem{Giovannetti2006}
V. Giovannetti, S. Lloyd, and L. Maccone,
Quantum Metrology,
\href{https://doi.org/10.1103/PhysRevLett.96.010401}
{Phys. Rev. Lett. \textbf{96}, 010401 (2006).}

\bibitem{Monras2007}
A. Monras and M. G. A. Paris,
Optimal Quantum Estimation of Loss in Bosonic Channels,
\href{https://doi.org/10.1103/PhysRevLett.98.160401}
{Phys. Rev. Lett. \textbf{98}, 160401 (2007).}

\bibitem{Pezze2007}
L. Pezz\'e, A. Smerzi, G. Khoury, J. F. Hodelin, and D. Bouwmeester,
Phase Detection at the Quantum Limit with Multiphoton Mach-Zehnder Interferometry,
\href{https://doi.org/10.1103/PhysRevLett.99.223602}
{Phys. Rev. Lett. \textbf{99}, 223602 (2007).}

\bibitem{Pezze2008}
L. Pezz\'e and A. Smerzi,
Mach-Zehnder Interferometry at the Heisenberg Limit with Coherent and Squeezed-Vacuum Light,
\href{https://doi.org/10.1103/PhysRevLett.100.073601}
{Phys. Rev. Lett. \textbf{100}, 073601 (2008).}

\bibitem{Boixo2008}
S. Boixo, A. Datta, M. J. Davis, S. T. Flammia, A. Shaji, and C. M. Caves,
Quantum Metrology: Dynamics versus Entanglement,
\href{https://doi.org/10.1103/PhysRevLett.101.040403}
{Phys. Rev. Lett. \textbf{101}, 040403 (2008).}

\bibitem{Pezze2009}
L. Pezz\'e and A. Smerzi,
Entanglement, Nonlinear Dynamics, and the Heisenberg Limit,
\href{https://doi.org/10.1103/PhysRevLett.102.100401}
{Phys. Rev. Lett. \textbf{102}, 100401 (2009).}

\bibitem{Paris2009}
M. G. A. Paris, Quantum estimation for quantum technology,
\href{https://doi.org/10.1142/S0219749909004839}
{Int. J. Quantum Inf. \textbf{7}, 125 (2009).}

\bibitem{Genoni2011}
M. G. Genoni, S. Olivares, and M. G. A. Paris,
Optical Phase Estimation in the Presence of Phase Diffusion,
\href{https://doi.org/10.1103/PhysRevLett.106.153603}
{Phys. Rev. Lett. \textbf{106}, 153603 (2011).}

\bibitem{Peter2011}
N. Thomas-Peter, B. J. Smith, A. Datta, L. Zhang, U. Dorner, and I. A. Walmsley,
Real-World Quantum Sensors: Evaluating Resources for Precision Measurement,
\href{https://doi.org/10.1103/PhysRevLett.107.113603}
{Phys. Rev. Lett. \textbf{107}, 113603 (2011).}

\bibitem{Spagnolo2012}
N. Spagnolo, C. Vitelli, V. G. Lucivero, V. Giovannetti, L. Maccone, and F. Sciarrino,
Phase Estimation via Quantum Interferometry for Noisy Detectors,
\href{https://doi.org/10.1103/PhysRevLett.108.233602}
{Phys. Rev. Lett. \textbf{108}, 233602 (2012).}

\bibitem{Genoni2013}
M. G. Genoni, M. G. A. Paris, G. Adesso, H. Nha, P. L. Knight, and M. S. Kim,
Optimal estimation of joint parameters in phase space,
\href{https://doi.org/10.1103/PhysRevA.87.012107}
{Phys. Rev. A \textbf{87}, 012107 (2013).}

\bibitem{Humphreys2013}
P. C. Humphreys, M. Barbieri, A. Datta, and I. A. Walmsley,
Quantum Enhanced Multiple Phase Estimation,
\href{https://doi.org/10.1103/PhysRevLett.111.070403}
{Phys. Rev. Lett. \textbf{111}, 070403 (2013).}

\bibitem{Lang2013}
M. D. Lang and C. M. Caves,
Optimal Quantum-Enhanced Interferometry Using a Laser Power Source,
\href{https://doi.org/10.1103/PhysRevLett.111.173601}
{Phys. Rev. Lett. \textbf{111}, 173601 (2013).}

\bibitem{Toth2014}
G. T\'{o}th and I. Apellaniz, Quantum metrology from a quantum information science perspective, 
\href{https://doi.org/10.1088/1751-8113/47/42/424006}
{J. Phys. A: Math. Theor. \textbf{47}, 424006 (2014).} 

\bibitem{Zhang2015}
L. Zhang, A. Datta, and I. A. Walmsley, 
Precision Metrology Using Weak Measurements, 
\href{https://doi.org/10.1103/PhysRevLett.114.210801}
{Phys. Rev. Lett. \textbf{114}, 210801 (2015)}

\bibitem{Pezze2017}
L. Pezz\'e, M. A. Ciampini, N. Spagnolo, P. C. Humphreys, A. Datta, I. A. Walmsley, M. Barbieri, F. Sciarrino, and A. Smerzi,
Optimal Measurements for Simultaneous Quantum Estimation of Multiple Phases,
\href{https://doi.org/10.1103/PhysRevLett.119.130504}
{Phys. Rev. Lett. \textbf{119}, 130504 (2017).}

\bibitem{Degen2017}
C. L. Degen, F. Reinhard, and P. Cappellaro, 
Quantum sensing,
\href{https://doi.org/10.1103/RevModPhys.89.035002}
{Rev. Mod. Phys. \textbf{89}, 035002 (2017).}

\bibitem{Gagatsos2017}
C. N. Gagatsos, B. A. Bash, S. Guha, and A. Datta, 
Bounding the quantum limits of precision for phase estimation with loss and thermal noise, 
\href{https://doi.org/10.1103/PhysRevA.96.062306}
{Phys. Rev. A \textbf{96}, 062306 (2017).}

\bibitem{Liu2020}
J. Liu, H. Yuan, X.-M. Lu, and X. Wang, 
Quantum Fisher information matrix and multiparameter estimation, 
\href{https://doi.org/10.1088/1751-8121/ab5d4d}
{J. Phys. A: Math. Theor. \textbf{53}, 023001 (2020).}

\bibitem{Rafal2020}
R. Demkowicz-Dobrza\'{n}ski, W G\'{o}recki, and M Gu\c{t}\u{a}, 
Multi-parameter estimation beyond quantum Fisher information, 
\href{https://doi.org/10.1088/1751-8121/ab8ef3}
{J. Phys. A: Math. Theor. \textbf{53}, 363001 (2020).}

\bibitem{Pezze2021}
L. Pezz\'e and A. Smerzi,
Quantum Phase Estimation Algorithm with Gaussian Spin States,
\href{https://doi.org/10.1103/PRXQuantum.2.040301}
{PRX Quantum \textbf{2}, 040301 (2021).}

\bibitem{Qiu2022}
Y. Qiu, M. Zhuang, J. Huang, and C. Lee, 
Efficient Bayesian phase estimation via entropy-based sampling, 
\href{https://doi.org/10.1088/2058-9565/ac74db}
{Quantum Sci. Technol. \textbf{7}, 035022 (2022).}

\bibitem{Holland1993}
M. J. Holland and K. Burnett, 
Interferometric detection of optical phase shifts at the Heisenberg limit, 
\href{https://doi.org/10.1103/PhysRevLett.71.1355}
{Phys. Rev. Lett. \textbf{71}, 1355 (1993).}

\bibitem{Birrittella2012}
R. Birrittella, J. Mimih, and C. C. Gerry, 
Multiphoton quantum interference at a beam splitter and the approach to Heisenberg-limited interferometry, 
\href{https://doi.org/10.1103/PhysRevA.86.063828}
{Phys. Rev. A \textbf{86}, 063828 (2012).}

\bibitem{Demkowicz2012} 
R. Demkowicz-Dobrza\'{n}ski, J.Ko{\l}ody\'{n}ski, and M. Gu\c{t}\v{a}, 
The elusive Heisenberg limit in quantum-enhanced metrology,
\href{https://doi.org/10.1038/ncomms2067}
{Nat. Commun. \textbf{3}, 1063 (2012).}

\bibitem{Berry2012}
D. W. Berry, M. J. W. Hall, M. Zwierz, and H. M. Wiseman,
Optimal Heisenberg-style bounds for the average performance of arbitrary phase estimates,
\href{https://link.aps.org/doi/10.1103/PhysRevA.86.053813}
{Phys. Rev. A \textbf{86}, 053813 (2012).}

\bibitem{Rivas2012} 
\'{A}. Rivas and A. Luis,
Sub-Heisenberg estimation of non-random phase shifts,
\href{https://dx.doi.org/10.1088/1367-2630/14/9/093052}
{New J. Phys. \textbf{14}, 093052 (2012).}

\bibitem{Demkowicz2015}
R. Demkowicz-Dobrza\'{n}ski, M. Jarzyna, and J. Ko{\l}ody\'{n}ski, 
Quantum Limits in Optical Interferometry,
\href{https://doi.org/10.1016/bs.po.2015.02.003}
{Prog. Optics \textbf{60}, 345-435 (2015).}

\bibitem{Calsamiglia2016}
J. Calsamiglia, B. Gendra, R. Mu\~{n}oz-Tapia, and E. Bagan,
Probabilistic metrology or how some measurement outcomes render ultra-precise estimates,
\href{https://dx.doi.org/10.1088/1367-2630/18/10/103049}
{New J. Phys. \textbf{18}, 103049 (2016).}

\bibitem{Branford2021}
D. Branford and J. Rubio,
Average number is an insufficient metric for interferometry,
\href{https://dx.doi.org/10.1088/1367-2630/ac3571}
{New J. Phys. \textbf{23}, 123041 (2021).}

\bibitem{Mitchell2004}
M. W. Mitchell, J. S. Lundeen, and A. M. Steinberg,
Super-resolving phase measurements with a multiphoton entangled state,
\href{http://dx.doi.org/10.1038/nature02493}
{Nature \textbf{429}, 161 (2004).}

\bibitem{Xiao1987}
M. Xiao, L.-A. Wu, and H. J. Kimble,
Precision measurement beyond the shot-noise limit,
\href{https://doi.org/10.1103/PhysRevLett.59.278}
{Phys. Rev. Lett. \textbf{59}, 278 (1987).}

\bibitem{Grangier1987}
P. Grangier, R. E. Slusher, B. Yurke, and A. LaPorta,
Squeezed-light-enhanced polarization interferometer, 
\href{https://doi.org/10.1103/PhysRevLett.59.2153}
{Phys. Rev. Lett. \textbf{59}, 2153 (1987).}

\bibitem{Higgins2007}
B. L. Higgins, D. W. Berry, S. D. Bartlett, H. M. Wiseman, and G. J. Pryde,
Entanglement-free Heisenberg-limited phase estimation,
\href{http://dx.doi.org/10.1038/nature06257}
{Nature \textbf{450}, 393 (2007).}

\bibitem{Nagata2007}
T. Nagata, R. Okamoto, J. L. O'Brien, K. Sasaki, and S. Takeuchi,
Beating the Standard Quantum Limit with Four-Entangled Photons,
\href{http://dx.doi.org/10.1126/science.1138007}
{Science \textbf{316}, 726 (2007).}

\bibitem{Kacprowicz2010}
M. Kacprowicz, R. Demkowicz-Dobrza\'{n}ski, W. Wasilewski, K. Banaszek, and I. A. Walmsley, 
Experimental quantum-enhanced estimation of a lossy phase shift, 
\href{https://doi.org/10.1038/nphoton.2010.39}
{Nat. Photon. \textbf{4}, 357 (2010).}

\bibitem{Berni2015}
A. A. Berni, T. Gehring, B. M. Nielsen, V. H\"andchen, M. G. A. Paris, and U. L. Andersen,
Ab initio quantum-enhanced optical phase estimation using real-time feedback control,
\href{https://doi.org/10.1038/nphoton.2015.139}
{Nat. Photon. \textbf{9}, 577 (2015).}

\bibitem{Xu2020}
L. Xu, Z. Liu, A. Datta, G. C. Knee, J. S. Lundeen, Y.-q. Lu, and L. Zhang, 
Approaching Quantum-Limited Metrology with Imperfect Detectors by Using Weak-Value Amplification, 
\href{https://doi.org/10.1103/PhysRevLett.125.080501}
{Phys. Rev. Lett. \textbf{125}, 080501 (2020).}

\bibitem{Liu2021}
L.-Z. Liu, Y.-Z. Zhang, Z.-D. Li, R. Zhang, X.-F. Yin, Y.-Y. Fei, L. Li, N.-L. Liu, F. Xu, Y.-A. Chen, and J.-W. Pan,
Distributed Quantum Phase Estimation with Entangled Photons,
\href{https://doi.org/10.1038/s41566-020-00718-2}
{Nat. Photon. \textbf{15}, 137 (2021).}

\bibitem{Liu2023}
L.-Z. Liu, Y.-Y. Fei, Y. Mao, Y. Hu, R. Zhang, X.-F. Yin, X. Jiang, L. Li,
N.-L. Liu, F. Xu, Y.-A. Chen, and J.-W. Pan,
Full-Period Quantum Phase Estimation,
\href{https://doi.org/10.1103/PhysRevLett.130.120802}
{Phys. Rev. Lett. \textbf{130}, 120802 (2023).}

\bibitem{Jarzyna2012}
M. Jarzyna and R. Demkowicz-Dobrza\'{n}ski, 
Quantum interferometry with and without an external phase reference, 
\href{https://doi.org/10.1103/PhysRevA.85.011801}
{Phys. Rev. A \textbf{85}, 011801(R) (2012).}

\bibitem{Pasquale2015}
A. De Pasquale, P. Facchi, G. Florio, V. Giovannetti, K. Matsuoka, and K. Yuasa, 
Two-mode bosonic quantum metrology with number fluctuations, 
\href{https://doi.org/10.1103/PhysRevA.92.042115}
{Phys. Rev. A \textbf{92}, 042115 (2015).}

\bibitem{Helstrom1976}
C. W. Helstrom, \emph{Quantum Detection and Estimation Theory}
(Academic, New York, 1976).

\bibitem{Holevo1982}
A. S. Holevo, \emph{Probabilistic and Statistical Aspects of Quantum Theory} 
(North-Holland, Amsterdam, 1982).

\bibitem{Sanders1989}
B. C. Sanders, 
Quantum dynamics of the nonlinear rotator and the effects of continual spin measurement, 
\href{https://doi.org/10.1103/PhysRevA.40.2417}
{Phys. Rev. A \textbf{40}, 2417 (1989).}

\bibitem{Boto2000}
A. N. Boto, P. Kok, D. S. Abrams, S. L. Braunstein, C. P. Williams, and J. P. Dowling, 
Quantum Interferometric Optical Lithography: Exploiting Entanglement to Beat the Diffraction Limit, 
\href{https://doi.org/10.1103/PhysRevLett.85.2733}
{Phys. Rev. Lett. \textbf{85}, 2733 (2000).}

\bibitem{Vogel1993}
K. Vogel, V. M. Akulin, and W. P. Schleich, 
Quantum state engineering of the radiation field, 
\href{https://doi.org/10.1103/PhysRevLett.71.1816}
{Phys. Rev. Lett. \textbf{71}, 1816 (1993).}

\bibitem{Park2017}
J. Park, Y. Lu, J. Lee, Y. Shen, K. Zhang, S. Zhang, M. S. Zubairy, K. Kim, and H. Nha, 
Revealing nonclassicality beyond Gaussian states via a single marginal distribution, 
\href{https://doi.org/10.1073/pnas.1617621114}
{Proc. Natl. Acad. Sci. \textbf{114}, 891-896 (2017).}

\bibitem{Boas2019}
C. J. Villas-Boas and D. Z. Rossatto, 
Multiphoton Jaynes-Cummings Model: Arbitrary Rotations in Fock Space and Quantum Filters, 
\href{https://doi.org/10.1103/PhysRevLett.122.123604}
{Phys. Rev. Lett. \textbf{122}, 123604 (2019).}

\bibitem{Lee2019}
C. Lee, C. Oh, H. Jeong, C. Rockstuhl, and S.-Y. Lee, 
Using states with a large photon number variance to increase quantum Fisher information in single-mode phase estimation, 
\href{https://doi.org/10.1088/2399-6528/ab524a}
{J. Phys. Commun. \textbf{3}, 115008 (2019).}

\bibitem{Jarzyna2015}
M. Jarzyna and R. Demkowicz-Dobrza\'{n}ski,
True precision limits in quantum metrology,
\href{https://doi.org/10.1088/1367-2630/17/1/013010}
{New. J. Phys. \textbf{17}, 013010 (2015).}

\bibitem{Dorner2009}
U. Dorner, R. Demkowicz-Dobrza\'{n}ski, B. J. Smith, J. S. Lundeen, W. Wasilewski, K. Banaszek, and I. A. Walmsley,
Optimal Quantum Phase Estimation,
\href{http://doi.org/10.1103/PhysRevLett.102.040403}
{Phys. Rev. Lett. \textbf{102}, 040403 (2009).}

\bibitem{Dobrzanski2009}
R. Demkowicz-Dobrza\'{n}ski, U. Dorner, B. J. Smith, J. S. Lundeen, W. Wasilewski, K. Banaszek, and I. A. Walmsley, 
Quantum phase estimation with lossy interferometers,
\href{https://doi.org/10.1103/PhysRevA.80.013825}
{Phys. Rev. A \textbf{80}, 013825 (2009).}

\bibitem{Liu2013}
J. Liu, X. Jing, and X. Wang,
Phase-matching condition for enhancement of phase sensitivity in quantum metrology,
\href{https://doi.org/10.1103/PhysRevA.88.042316}
{Phys. Rev. A \textbf{88}, 042316 (2013).}


\bibitem{Taylor2013}
M. A. Taylor, J. Janousek, V. Daria, J. Knittel, B. Hage, H.-A. Bachor, and W. P. Bowen, 
Biological measurement beyond the quantum limit, 
\href{https://doi.org/10.1038/nphoton.2012.346}
{Nat. Photon. \textbf{7}, 229-233 (2013).}

\bibitem{Lu2022a}
C.-Y. Lu, Y. Cao, C.-Z. Peng, and J.-W. Pan, 
Micius quantum experiments in space, 
\href{https://doi.org/10.1103/RevModPhys.94.035001}
{Rev. Mod. Phys. \textbf{94}, 035001 (2022).}

\bibitem{Stokowski2023}
H. S. Stokowski, T. P. McKenna, T. Park, A. Y. Hwang, D. J. Dean, O. T. Celik, V. Ansari, M. M. Fejer, and A. H. Safavi-Naeini, 
Integrated quantum optical phase sensor in thin film lithium niobate, 
\href{https://doi.org/10.1038/s41467-023-38246-6}
{Nat. Commun. \textbf{14}, 3355 (2023).}

\bibitem{Lu2022}
W. Lu, L. Shao, X. Zhang, Z. Zhang, J. Chen, H. Tao, and X. Wang, 
Extreme expected values and their applications in quantum metrology, 
\href{https://doi.org/10.1103/PhysRevA.105.023718}
{Phys. Rev. A \textbf{105}, 023718 (2022).}

\bibitem{Lee2016}
S.-Y. Lee, C.-W. Lee, J. Lee, and H. Nha,
Quantum phase estimation using path-symmetric entangled states,
\href{https://doi.org/10.1038/srep30306}
{Sci. Rep. \textbf{6}, 30306 (2016).}

\bibitem{Horodecki2009}
R. Horodecki, P. Horodecki, M. Horodecki, and K. Horodecki, 
Quantum entanglement, 
\href{https://doi.org/10.1103/RevModPhys.81.865}
{Rev. Mod. Phys. \textbf{81}, 865 (2009).}

\bibitem{Eltschka2014}
C. Eltschka and J. Siewert,
Quantifying entanglement resources,
\href{http://doi.org/10.1088/1751-8113/47/42/424005}
{J. Phys. A: Math. Theor. \textbf{47}, 424005 (2014).}

\bibitem{Ma2011}
J. Ma, X. Wang, C. P. Sun, and F. Nori, 
Quantum spin squeezing,
\href{https://doi.org/doi:10.1016/j.physrep.2011.08.003}
{Phys. Rep. \textbf{509}, 89 (2011)}, and references therein.  

\bibitem{Powell1994} 
M. J. D. Powell, 
\emph{in Advances in Optimization and Numerical Analysis} (Springer, New York, 1994).

\bibitem{Powell1998}
M. J. D. Powell, 
Direct search algorithms for optimization calculations,
\href{https://doi.org/10.1017/S0962492900002841}
{Acta Numer. \textbf{7}, 287 (1998).}

\bibitem{Powell2007}
M. J. D. Powell, 
A view of algorithms for optimization without derivatives, 
Math. Today-Bull. Inst. Math. Appl. \textbf{43}, 170 (2007). 

\bibitem{Qin2025}
J.-F. Qin and J. Liu, Optimal noisy quantum phase estimation with finite-dimensional states, in preparation. 

\bibitem{Gerry2001}
C. C. Gerry and R. A. Campos,
Generation of maximally entangled photonic states with a quantum-optical Fredkin gate,
\href{https://doi.org/10.1103/PhysRevA.64.063814}
{Phys. Rev. A \textbf{64}, 063814 (2001).}

\bibitem{Gerry2002}
C. C. Gerry and A. Benmoussa,
Heisenberg-limited interferometry and photolithography with nonlinear four-wave mixing,
\href{https://doi.org/10.1103/PhysRevA.65.033822}
{Phys. Rev. A \textbf{65}, 033822 (2002).}

\bibitem{Joo2011}
J. Joo, W. J. Munro, and T. P. Spiller, 
Quantum Metrology with Entangled Coherent States, 
\href{https://doi.org/10.1103/PhysRevLett.107.083601}
{Phys. Rev. Lett. \textbf{107}, 083601 (2011).}

\bibitem{Andersen2016} 
U. L. Andersen, T. Gehring, C. Marquardt, and G. Leuchs,
30 years of squeezed light generation,
\href{https://dx.doi.org/10.1088/0031-8949/91/5/053001}
{Phys. Scr. \textbf{91}, 053001 (2016).}

\bibitem{Zhang2021}
Y. Zhang, M. Menotti, K. Tan, V. D. Vaidya, D. H. Mahler, L. G. Helt, 
L. Zatti,  M. Liscidini,  B. Morrison, and Z. Vernon, 
Squeezed light from a nanophotonic molecule,
\href{https://doi.org/10.1038/s41467-021-22540-2}
{Nat. Commun. \textbf{12}, 2233 (2021).}

\bibitem{LIGO2023}
D. Ganapathy et al. (LIGO O4 Detector Collaboration),
Broadband Quantum Enhancement of the LIGO Detectors with Frequency-Dependent Squeezing,
\href{https://link.aps.org/doi/10.1103/PhysRevX.13.041021}
{Phys. Rev. X \textbf{13}, 041021 (2023).}

\bibitem{Deng2024}
X. Deng, S. Li, Z.-J. Chen, Z. Ni, Y. Cai, J. Mai, L. Zhang, P. Zheng, H. Yu, C.-L. Zou, S. Liu, F. Yan, Y. Xu, and D. Yu, 
Quantum-enhanced metrology with large Fock states, 
\href{https://doi.org/10.1038/s41567-024-02619-5}
{Nat. Phys. (2024).}

\bibitem{Kurdzialek2023}
S. Kurdzia{\l}ek, W. G\'{o}recki, F. Albarelli, and R. Demkowicz-Dobrza\'{n}ski, 
Using Adaptiveness and Causal Superpositions Against Noise in Quantum Metrology, 
\href{https://doi.org/10.1103/PhysRevLett.131.090801}
{Phys. Rev. Lett. \textbf{131}, 090801 (2023).}

\bibitem{Dobrzanski2017} 
R. Demkowicz-Dobrza\'{n}ski, J. Czajkowski, and P. Sekatski, 
Adaptive Quantum Metrology under General Markovian Noise, 
\href{https://doi.org/10.1103/PhysRevX.7.041009}
{Phys. Rev. X \textbf{7}, 041009 (2017).}

\bibitem{Hentschel2010}
A. Hentschel and B. C. Sanders,
Machine Learning for Precise Quantum Measurement,
\href{https://doi.org/10.1103/PhysRevLett.104.063603}
{Phys. Rev. Lett. \textbf{104}, 063603 (2010).}

\bibitem{Holevo1984}
A. S. Holevo,
Quantum Probability and Applications to the Quantum Theory of Irreversible Processes, 
\href{https://link.springer.com/book/10.1007/BFb0071705}
{Springer Lecture Notes Math. \textbf{1055}, 153 (1984).}

\bibitem{Berry2000}
D. W. Berry and H. M. Wiseman,
Optimal States and Almost Optimal Adaptive Measurements for Quantum Interferometry,
\href{https://doi.org/10.1103/PhysRevLett.85.5098}
{Phys. Rev. Lett. \textbf{85}, 5098 (2000).}

\bibitem{Berry2001}
D. W. Berry, H. M. Wiseman, and J. K. Breslin,
Optimal input states and feedback for interferometric phase estimation,
\href{https://doi.org/10.1103/PhysRevA.63.053804}
{Phys. Rev. A \textbf{63}, 053804 (2001).}

\bibitem{Huang2017}
Z. Huang, K. R. Motes, P. M. Anisimov, J. P. Dowling, and D. W. Berry, 
Adaptive phase estimation with two-mode squeezed vacuum and parity measurement, 
\href{https://doi.org/10.1103/PhysRevA.95.053837}
{Phys. Rev. A \textbf{95}, 053837 (2017).}

\bibitem{DiMario2020}
M. T. DiMario and F. E. Becerra,
Single-Shot Non-Gaussian Measurements for Optical Phase Estimation,
\href{https://doi.org/10.1103/PhysRevLett.125.120505}
{Phys. Rev. Lett. \textbf{125}, 120505 (2020).}

\bibitem{Garcia2022}
M. A. Rodr{\'{\i}}guez-Garc{\'{\i}}a, M. T. DiMario, P. Barberis-Blostein, and F. E. Becerra, 
Determination of the asymptotic limits of adaptive photon counting measurements 
for coherent-state optical phase estimation,
\href{https://doi.org/10.1038/s41534-022-00601-8}
{npj Quantum Inf. \textbf{8}, 94 (2022).}

\bibitem{Zhang2022}
M. Zhang, H.-M. Yu, H. Yuan, X. Wang, R. Demkowicz-Dobrza\'{n}ski, and J. Liu,
QuanEstimation: An open-source toolkit for quantum parameter estimation,
\href{https://doi.org/10.1103/PhysRevResearch.4.043057}
{Phys. Rev. Res. \textbf{4}, 043057 (2022).}

\bibitem{Bargatin2005}
I. Bargatin, 
Mutual information-based approach to adaptive homodyne detection of quantum optical states, 
\href{https://doi.org/10.1103/PhysRevA.72.022316}
{Phys. Rev. A \textbf{72}, 022316 (2005).}

\bibitem{Liu2022}
J. Liu, M. Zhang, H. Chen, L. Wang, and H. Yuan, 
Optimal Scheme for Quantum Metrology, 
\href{https://doi.org/10.1002/qute.202100080}
{Adv. Quantum Technol. \textbf{5}, 2100080 (2022).}

\bibitem{Miao2023}
M. Liu, L. Zhang, and H. Miao, 
Adaptive protocols for SU(1,1) interferometers to achieve ab initio phase estimation at the Heisenberg limit, 
\href{https://doi.org/10.1088/1367-2630/ad042f}
{New J. Phys. \textbf{25}, 103051 (2023).}

\bibitem{Tsang2012} 
M. Tsang, 
Ziv-Zakai Error Bounds for Quantum Parameter Estimation,
\href{https://link.aps.org/doi/10.1103/PhysRevLett.108.230401}
{Phys. Rev. Lett. \textbf{108}, 230401 (2012).}

\bibitem{Rubio2018}
J. Rubio,  P. Knott, and J. Dunningham, 
Non-asymptotic analysis of quantum metrology protocols beyond the Cram{\'e}r-Rao bound,
\href{https://dx.doi.org/10.1088/2399-6528/aaa234}
{J. Phys. Commun. \textbf{2}, 015027 (2018).}

\bibitem{Rzadkowski2017}
W. Rz\k{a}dkowski and R. Demkowicz-Dobrza\'{n}ski,
Discrete-to-continuous transition in quantum phase estimation, 
\href{https://doi.org/10.1103/PhysRevA.96.032319}
{Phys. Rev. A \textbf{96}, 032319 (2017).}

\bibitem{Cover1991}
T. M. Cover and J. A. Thomas, \emph{Elements of information theory}
(John Wiley \& Sons, New York, 1991).

\bibitem{Barnett1998}
S. M. Barnett, J. Jeffers, A. Gatti, and R. Loudon, 
Quantum optics of lossy beam splitters, 
\href{https://doi.org/10.1103/PhysRevA.57.2134}
{Phys. Rev. A \textbf{57}, 2134 (1998).}

\bibitem{Gardiner2004}
C. W. Gardiner and P. Zoller, \emph{Quantum Noise}, 3rd ed.
(Springer, Berlin, 2004).

\bibitem{Rubin2007}
M. A. Rubin and S. Kaushik, 
Loss-induced limits to phase measurement precision with maximally entangled states, 
\href{https://doi.org/10.1103/PhysRevA.75.053805}
{Phys. Rev. A \textbf{75}, 053805 (2007).}

\bibitem{Huver2008}
S. D. Huver, C. F. Wildfeuer, and J. P. Dowling, 
Entangled Fock states for robust quantum optical metrology, imaging, and sensing, 
\href{http://doi.org/10.1103/PhysRevA.78.063828}
{Phys. Rev. A \textbf{78}, 063828 (2008).}

\bibitem{Zhang2013}
X.-X. Zhang, Y.-X. Yang, and X.-B. Wang, 
Lossy quantum-optical metrology with squeezed states,
\href{http://doi.org/10.1103/PhysRevA.88.013838}
{Phys. Rev. A \textbf{88}, 013838 (2013).}

\bibitem{Knott2014}
P. A. Knott, T. J. Proctor, K. Nemoto, J. A. Dunningham, and W. J. Munro, 
Effect of multimode entanglement on lossy optical quantum metrology,
\href{https://doi.org/10.1103/PhysRevA.90.033846}
{Phys. Rev. A \textbf{90}, 033846 (2014).}

\bibitem{Lang2014}
M. D. Lang and C. M. Caves, 
Optimal quantum-enhanced interferometry, 
\href{https://doi.org/10.1103/PhysRevA.90.025802}
{Phys. Rev. A \textbf{90}, 025802 (2014).}

\bibitem{Datta2011}
A. Datta, L. Zhang, N. Thomas-Peter, U. Dorner, B. J. Smith, and I. A. Walmsley, 
Quantum metrology with imperfect states and detectors, 
\href{https://doi.org/10.1103/PhysRevA.83.063836}
{Phys. Rev. A \textbf{83}, 063836 (2011).}

\bibitem{Popviciu1935}
T. Popviciu, Sur les \'{e}quations alg\'{e}briques ayant toutes leurs racines r\'{e}elles, 
Mathematica \textbf{9}, 129 (1935).

\bibitem{Boyd2004}
S. Boyd and L. Vandenberghe, \emph{Convex Optimization} 
(Cambridge University Press, Cambridge, England, 2004).

\bibitem{zenodo2025}
J.-F. Qin, Y.-Q. Xu, and J. Liu, 
Optimal finite-dimensional probe states for quantum phase estimation (Part I), Zenodo (2025), 
\href{https://doi.org/10.5281/zenodo.16518656}{10.5281/zenodo.16518656}; 
Optimal finite-dimensional probe states for quantum phase estimation (Part II), Zenodo (2025), 
\href{https://doi.org/10.5281/zenodo.16668895}{10.5281/zenodo.16668895}.

\end{thebibliography}
\end{document}